\documentclass[preprints,article,accept,moreauthors,pdftex]{mdpi}

\usepackage{float}
\usepackage{hyperref}
\usepackage{upgreek}
\usepackage{pdflscape}
\usepackage{rotating}
\usepackage{epsfig}
\usepackage{booktabs}
\firstpage{1} 
\pubvolume{9}
\issuenum{2}
\articlenumber{24}
\pubyear{2021}
\copyrightyear{2021}
\externaleditor{Academic Editor: {Phil Edwards}}
\datereceived{24 February 2021} 
\dateaccepted{24 March 2021} 
\datepublished{31 March 2021} 
\hreflink{https://doi.org/10.3390/\linebreak galaxies9020024} 

\Title{History of Astronomy in Australia: Big-Impact Astronomy from World War II until the Lunar Landing (1945--1969)}

\TitleCitation{History of Astronomy in Australia: Big-Impact Astronomy from World War II until the Lunar Landing (1945--1969)}

\Author{Alister W.\ Graham $^{1,}$* \orcidA{}, 
Katherine H.\ Kenyon $^{1,\dagger}$, 
Lochlan J. Bull $^{2,\ddagger}$\orcidB{},           
Visura C.\ Lokuge Don $^{3,\mathsection}$      
and Kazuki~Kuhlmann $^{4,\|}$}

\AuthorNames{Graham, A.W.; Kenyon, K.H.; Bull, L.J.; Don, V.C.L.; Kuhlmann, K.}

\AuthorCitation{Graham, A.W.; Kenyon, K.H.; Bull, L.J.; Don, V.C.L.; Kuhlmann, K.}

\address{
$^1$ \quad Centre for Astrophysics and Supercomputing, Swinburne 
University of Technology, \mbox{Hawthorn, VIC 3122, Australia}; katherinehope95@hotmail.com\\
$^2$ \quad Horsham College, PO Box 508, Horsham, VIC 3402, Australia; lochie.bull@gmail.com\\ 
$^3$ \quad Glen Waverley Secondary College, 39 O'Sullivan Rd, Glen Waverley, VIC 3150, Australia; vlokuge@gmail.com \\
$^4$ \quad Marcellin College, 160 Bulleen Road, Bulleen, VIC 3105, Australia; kazjgk@gmail.com
}

\corres{Correspondence: AGraham@astro.swin.edu.au}

\firstnote{Current address: Central Clinical School, The Alfred Centre, Monash University, \mbox{Melbourne, VIC 3004,~Australia}.} 
\secondnote{Current address: Faculty of Business and Economics, University of Melbourne, Carlton, VIC 3053, Australia.}
\thirdnote{Current address: Faculty of Engineering, Monash University, Clayton, VIC 3800, Australia.} 
\fourthnote{Current address: School of Engineering, RMIT, GPO Box 2476, Melbourne, VIC 3001, Australia.}

\abstract{Radio astronomy commenced in earnest after World War II, with
Australia keenly engaged through the Council for Scientific and Industrial
Research.  At this juncture, Australia's Commonwealth Solar Observatory 
expanded its portfolio from primarily studying solar phenomena to 
conducting stellar and extragalactic research.  Subsequently, in the 1950s and
1960s, astronomy gradually became taught and researched in Australian
universities.  However, most scientific publications from this era of growth
and discovery have no country of affiliation in their header information,
making it hard to find the Australian astronomy articles from this
period.  In 2014, we used the then-new Astrophysics Data System (ADS) tool
Bumblebee to overcome this challenge and track down the Australian-led
astronomy papers published during the quarter of a century after World War II,
from 1945 until the lunar landing in 1969.  This required knowledge of the
research centres and facilities operating at the time, which are briefly
summarised herein.  Based on citation counts---an objective,
universally-used measure of scientific impact---we report on the Australian
astronomy articles which had the biggest impact.  We have identified the
top-ten most-cited papers, and thus also their area of research, from five
consecutive time-intervals across that blossoming quarter-century of
astronomy.  Moreover, we have invested a substantial amount of time
researching and providing a small tribute to each of the 62 scientists
involved, including several trail-blazing women.  Furthermore, we provide an
extensive list of references and point out many interesting historical
connections and anecdotes.}

\keyword{history and philosophy of astronomy; publications, bibliography; obituaries, biographies; sociology of astronomy; instrumentation: miscellaneous; methods: observational; atmospheric effects; Sun: activity; stars: general; (galaxies:) quasars: general}

\begin{document}

\section{Introduction}\label{sec:intro}



Inspired by Toner Stevenson's article \citep{2014PASA...31...18S} 
in the Publications of the Astronomical Society
of Australia\footnote{This followed a (poster) presentation at the
  Astronomical Society of Australia's 
  Annual General Meeting \citep{2013:Stevenson}.} regarding women's role 
in constructing Australia's ``Astrographic Catalogue''~\citep{1960AJ.....65..189W}, 
we identify other big-impact astronomy
publications from Australia's past while  using this as an opportunity to
showcase additional women who contributed to the legacy of
Australian astronomy.  For example, while in Sydney from 1890 to 1895, Mary
Acworth Orr had recognised the need for a smaller catalogue providing the
general public with a more informal guide than the ``Astrographic
Catalogue'', and she published just such a book in 1896 \citep{1896egss.book.....O}. 
Due to demand, this was reprinted in 1911, and Mary Acworth Orr went on
to have a career in astronomy, publishing many scientific articles \citep{2002amcr.book.....B}. 

 
The period of Australian history that we have focussed on is the quarter of a
century after World War II, from 1945 to 1969.  This era was a boom time for
astronomy. Adding to Australia's proficiency at observing the heavens at
optical wavelengths, it is well known that immediately after World War II, 
Australia became a world leader in conducting astronomy at radio wavelengths
\citep{1953JRASC..47..137P, 2006JAHH....9...35O, 2009cnhe.book.....S}. 
At the same time, Mount Stromlo Observatory, which had produced optical
munitions for the war effort, grew into one of the largest optical
observatories in the Southern hemisphere.  
The year 2020 marked the `Diamond, gold' jubilee (75th anniversary)
since the end of World War II (1st of September 1939--2nd of September
1945).\footnote{Technically, Britain and France declared a state of war with
  Germany on the 3rd of September 1939, after Germany had invaded
  Poland on the 1st of September.  The war in Europe ended on the 8th of May
  1945, and the war in Asia ended on the 15th of August that year, with a formal
  surrender by Imperial Japan signed on the 2nd of September 1945.}
Here we investigate and commemorate Australia's biggest impact astronomy
papers and the people who wrote them, from the quarter of a century after
this war until the first lunar-landing half a century ago in 1969.  
As we shall see, this involved scientists from the UK, USA, and elsewhere. To
address this objectively, we have split these 25 years into five 
consecutive time intervals and identified the ten most-cited publications from
each interval.  This approach helps to negate the growing number of citations
with time and allows one to see the evolution in research topics.


As we shall see, 
over the quarter-century covered, exploration ventured from predominantly
studying solar-related phenomena to researching objects far out into space.  Indeed, in
1963 the optical counterpart to the radio quasar 3C~273 was determined using
lunar eclipsing at the Parkes radio telescope and its cosmological redshift
(which was immediately obtained by colleagues in the USA) dramatically and
forever changed our understanding of the scale of the Universe.  In 1969 the
iconic live footage of Neil Armstrong stepping onto the Moon was received at,
and sent around the world from, the Honeysuckle Creek radio antenna near
Canberra, with the Parkes antenna relaying the latter footage of this historic
lunar landing. The famous pulsar paper by Goldreich \& Julian was additionally written
in 1969 while Goldreich was at the University of Sydney.
Continuing Australia's long
tradition of exploring the southern skies, in 1969 we also encounter Lindsey
Fairfield Smith, who mapped the positions of many southern Wolf-Rayet stars, and
Betty `Louise' Webster, who mapped the positions of numerous southern planetary nebulae
around aging~stars.\footnote{Both Smith and Webster obtained their Ph.D.\ degrees at Mount
Stromlo Observatory during Bart Bok's directorship, while Bart's wife
Priscilla Bok worked there as a successful and well-respected astronomer.
The Bok's themselves had arrived at Mount Stromlo Observatory after another
very prolific astronomy couple, Antoinette and G\'erard de Vaucouleurs, who had been
there the preceding six years (1951--1957).}

From this same quarter-century, the most cited astronomy paper from 
Australia and the world pertains to the ``initial mass function'' of
stars.  It was written by E.\ Salpeter (a refugee from war-torn Europe) while
he was at Mount Stromlo Observatory, after having attended high school in Sydney
and obtained his B.Sc.\ and M.Sc.\ degrees from the University of Sydney.
Impressively, Salpeter authored another three highly-ranked articles in our
compilation, one of which was with T.\ Hamada (a postdoc from Japan working on nuclear
reactions with Salpeter in Australia just 11 years after World War II had ended
via the dropping of nuclear bombs on Japan). 
J.P.\ Wild (radio astronomy) and 
A.E.\ Ringwood (geophysics \& geochemistry, and whose subsequent studies 
of lunar rocks from the Apollo missions led to the 
Moon-from-Earth theory) 
also have several highly-ranked articles spanning 
the first four and the last three of our five time-intervals, respectively. 
Australian-based research from other well-known astronomers, such as C.W.\ Allen,
A.R.\ Sandage, and G.H.\ de Vaucouleurs, additionally feature here, as does work
by Father D.J.K.\ O'Connell (who went on to become the Director of the Vatican
Observatory for 18 years) and Ronald Bracewell (who is known for his
hypothesised alien ``Bracewell probes'' and his important
contributions to radio imaging and medical tomography). 


Moreover, several of the astronomers we will encounter had interesting
roles during WWII, which may not be well known.  For example, optical
astronomers in Australia were asked to design and construct optical munitions 
(`Walter' Stibbs, `Ben' Gascoigne), while the proto-(radio astronomers) 
established early-warning radar which detected and prevented Japanese bombing
raids over Australia (Jack H.\ Piddington, Ruby Payne-Scott, with significant
contributions from other notable women such as Joan Freeman and Rachel
E.B.\ Makinson).  Their stories, touched on within this article (and with
references to further information), tempt one to draw 
parallels with the recent movie ``The Imitation Game'', 
and makes one wonder if Australia may have its own movie waiting to be made. 

In Section~\ref{Sec_Method}, we describe the methodology adopted to
find the high-impact Australian astronomy articles from 1945 to 1969.
This involved knowledge of the facilities that were operational at the time. 
In Section~\ref{Sec-Tables}, we tabulate the ten most popular articles
from each of our five time-intervals over the quarter of a century
from 1945 to 1969.  While the titles of the articles are (usually)
self-explanatory, and one may wish to merely peruse the tables herein,
we have also gone to considerable effort to provide a paragraph or two
about the authors of these articles that have had a significant and
enduring impact.  A concluding summary reflecting on the astronomers,
including the female astronomers encountered herein, 
their work, and the facilities which enabled much of that work,
is provided in Section~\ref{Sec_Con}, along with connections to the present. 

This project was a substantial undertaking and hopefully represents a
detailed yet concise, unbiased view of Australia's astronomical
endeavours during the quarter-century after World War II, which have
had the biggest impact in the scientific literature.  By its nature,
it is, of course, not a complete picture of Australia's astronomical
activities during this time, and readers may enjoy delving further
into the topic through the many enjoyable books we have listed in
Appendix~\ref{AppA}.

\section{Methodology}\label{Sec_Method}

Identifying the Australian articles was a rather time-consuming task.
Many of the older articles have no address
in their header information, and thus there is no `affiliation
tag'\footnote{The affiliation tag, which appears at the top of
  articles today, identifies the author with a place and country of
  work.} upon which to search for the keyword ``Australia''.  However,
thanks to the availability of scanned articles in the Smithsonian
Astrophysical Observatory's ``Astrophysics Data System'' 
(ADS)\footnote{\url{http://adsabs.harvard.edu} (accessed 15 February 2015).} now stretching back
decades and centuries, coupled with their text recognition software
and advanced web-based query form, investigations such as ours have
become possible.  In particular, the ``ADSLabs Full Integrated
  Search'' (which became ``Bumblebee'' in mid-2015 and the 
``User Interface'' in 2018) had recently, in 2013, opened up a new frontier for
historical research by enabling one to instantly search the full text
of scanned articles dating back to well before Australia was reached
by the British in 1770.  It should be noted that while this wonderful
new system is adequate for our objective, as with all bibliographic
databases, the ADS is not 100\%
complete\footnote{\url{http://doc.adsabs.harvard.edu/abs\_doc/faq.html\#complete} (accessed 15 February 2015).},
and it becomes increasingly less so the further back in time one
samples.  Indeed, the citations record is grossly incomplete during
and before World War II (1939--1945), and as such, it is not easy to
obtain useful/complete quantitative information from before 1945.

As noted above and in Section 2 of \citet{2007IAUSS...5....9H}, a
critical issue that needed overcoming was identifying the Australian
research papers.  In the past, authors would simply include their name and a
very brief address at the end of their article.  Moreover, this `signature'
information is not stored neatly in ADS.  If it was, we could have
quickly found the Australian-led papers in ADS with the
`affiliation position' search command: pos(aff:Australia,1).  One could then
include the years of interest, and with the click of a button, rank the results
according to how many citations they had accrued.  However, given that
Australian articles did not always include the word ``Australia'', we needed 
to uncover a set of keywords that authors did use to identify
their place of work, and we do that in this section.  ADS then automatically
searched through the full text of the scanned articles looking for these
keywords pertaining to Australian places and facilities. We then manually 
checked the papers found by ADS to confirm/deny their Australian heritage.  For example,
references to Australian eclipses or someone else's work in a named part of Australia do
not just appear in Australian articles, consequently requiring us to filter
those out.  This contamination made our task somewhat 
laborious but possible by individually inspecting the most
well-cited papers to check if they were affiliated with Australia.

As for the journals, 
in 1947, the Monthly Notices of the Royal Astronomical Society (MNRAS) 
introduced the {\sc REFERENCE} section at the end of their articles,
abandoning the previous (and often vague) footnote referencing style on
individual pages.  The scientific journals Nature, Science and
The Astronomical Journal (AJ) were already using this format before
1945, and it was later adopted by CSIRO's Australian Journal of Scientific
  Research A (AuSRA: 1948--1952) and CSIRO's Australian Journal of
  Physics (AuJPh: 1953--2001).
%
%
However, the The Astrophysical 
  Journal (ApJ) did not start doing this until 1954, the year that the ApJ
  Supplement began publication.  As such, citations given in ApJ articles
published before 1954 are typically not included in ADS.  
For reference, we note that the European journal Astronomy and Astrophysics did not
commence until 1969, two years after the Publications of the Astronomical
  Society of Australia (PASA) began. 

It should be recognised that (pre-internet) papers in Australian journals, 
without the same worldly distribution as some other journals, would have had a smaller
global reach and thus a lower impact.  Similarly, 
the Australian Journal of Physics 
likely had a smaller astronomy 
audience than the mainstream astronomy journals commanded. Therefore, an understandable CSIRO policy
to support and grow the local journals may have restricted CSIRO research's impact 
if these journals were not distributed widely. 
Additionally, like X-ray astronomy in the late 1960s
(e.g., \citep{1965P&SS...13..565P, 1968Natur.217...43E}), 
radio astronomy in the late 1940s was a new field of research and 
would have had a smaller group of participants than optical astronomy. 
It would, however, give rise to over 20 field stations in the 1950s and 1960s, 
and today many astronomers continue to use and cite the
early radio astronomy publications, over-coming the early 
participation-bias and thereby generating a broad impact. 
This ongoing trend, bolstered by a new wave of radio astronomers, 
is evident in Appendix~\ref{AppB}. 

From 1945 to 1969, we split the period into five time-intervals. 
To help nullify or offset the increased number of publications and citations over the
years (for example, \citep{2012PLoSO...746428P}), 
we searched within a 6-year interval (1945--1950), three 5-year intervals (1951--1955,
1956--1960, 1961--1965) and then within a 4-year interval (1966--1969). 
We identified the ten most-cited articles within these five time periods, 
with the final citation data collected in February 2015. 
As such, we may have captured some articles that were ahead of their time and not
immediately well-cited, and we may have avoided articles that were a flash-in-the-pan and
not sustained by a long period of citation.

\subsection{Post WWII Observing Facilities}

In what follows, we identify Australia's main astronomical observing
facilities from 1945 to 1969.  Besides simply listing them in
Table~\ref{Tab1}, we provide a brief historical context for each, as this
helps in part to better connect them with the research and researchers whom we
later identify.
\end{paracol}
\clearpage
\newpage
\begin{specialtable}[H]
\setlength{\tabcolsep}{3.6mm}
\widetable
\tablesize{\scriptsize}
\caption{Post 1945 Keywords used for full-text search in ADS.\label{Tab1}}
\begin{tabular}{ll}
\toprule
\textbf{Keyword}      &  \textbf{Associated Place(s)}  \\ 
\midrule
Commonwealth Solar Observatory    &  Mount Stromlo Observatory (1924--1950) \\ 
Commonwealth Observatory          &  Mount Stromlo Observatory  (1950--1957) \\ 
Mount Stromlo                     &  Mount Stromlo Observatory (1957-- )  \\ 
Australian National University    &  The Australian National University (1946--) \\
Uppsala Southern Station          &  Mount Stromlo Observatory (1957--1982) \\
Yale-Columbia Southern Station    &  Mount Stromlo Observatory (1952--1992, burnt in 2003) \\
Mount Bingar                      &  Mount Bingar field station, N.S.W.\ (1959--1962) \\
Siding Spring                     &  Siding Spring Observatory, N.S.W.\ (1964--) \\ 
\midrule
CSIR   & Council for Scientific and Industrial Research (1926--1949) \\ 
CSIRO  & Commonwealth Scientific and Industrial Research Organization (1949--) \\ 
Chippendale &  National Standards Laboratory (1939--1973) \\
            &  Division of Radiophysics HQ, (Sydney) University Grounds (1939--1968) \\
            &  Division of Physics HQ (1945--1973) \\
Dover Heights & Division of Radiophysics, key field station (1946--1954) \\ 
Georges Heights & Division of Radiophysics, Middle Head, Sydney (1947--1948) \\
Hornsby Valley & Division of Radiophysics, (1947--1952) \\
Potts Hill  & Division of Radiophysics, key field station (1948--1962) \\
Penrith & Division of Radiophysics, (1949--1950) \\
Badgery's Creek & Division of Radiophysics, on a cattle research station (1949--1956) \\
Murraybank & Division of Radiophysics, Orchard of astronomer John Murray (1956--1961)  \\
Dapto   & Division of Radiophysics, Dapto Dairy's Radio Spectrograph (1952--1965) \\
Culgoora OR Narrabri & Division of Radiophysics, Culgoora Solar Observatory and \\
                     & Radio Heliograph at the Paul Wild Observatory (1967--1984) \\
Parkes               & Parkes Observatory (1961--) \\
Molonglo             & Molonglo Radio Observatory (1965--) \\
Buckland             & Buckland Park Aerial Array (1969--) \\
\midrule 
Riverview  & Saint Ignatius' College; Riverview College Observatory (1909--) \\
N.S.W.\ OR New South Wales  &  Many\\
Sydney  &  Sydney Observatory; Sydney University Grounds; etc. \\
Canberra &  Mount Stromlo Observatory; The Australian National University \\
Melbourne  & Melbourne Observatory; The University of Melbourne \\
Adelaide  &  Adelaide Observatory; The University of Adelaide \\
Perth OR Western Australia  & The Univ.\ of Western Australia; Perth Observatory \\
Hobart OR Tasmania  &  The Univ.\ of Tasmania; Comm.\ Obs.\ Ionospheric Prediction Service \\
Brisbane OR Queensland  & The University of Queensland  \\
Darwin   & RAAF Radar Station 59 \\
\bottomrule
\end{tabular}
\end{specialtable}

\begin{paracol}{2}
\switchcolumn
\vspace{-6pt}
\subsubsection{Mount Stromlo Observatory}

By the start of the 20th century, astronomers around the world were very
interested in the sunspot cycle, the structure of the convection zone, solar
flares, and the heating of the outer layers of the Sun.  However, they
realised how little they knew about the Sun.  Starting in 1905, Walter
G.\ Duffield (1879--1929), from Adelaide but at the time undertaking an
M.Sc.\ at the University of Manchester in the United Kingdom, campaigned to
establish a {\bf Commonwealth Solar Observatory} in Australia 
\citep{Allen:1979, Love:1984}.  Once it was eventually approved, he helped
ensure that it would be incorporated into the plans for the new Australian
Capital Territory (ACT).  To his credit, in 1911, the very first Commonwealth
building\footnote{This building would also later serve as a quiet office for
  Ph.D.\ students, such as Ronald (Ron) David Ekers (b.1941), who obtained his
  Ph.D.\ from The ANU in 1967 before going on to become the Director of the
  Very Large Array, in New Mexico, from 1980 to 1987, and the Director of the
  Australia Telescope National Facility (1988--2003).} in the new Australian Capital Territory had a
dome at one end: it housed the donated 9-inch Oddie Telescope\footnote{In
  1910, {\bf James Oddie} (1824--1911) donated an excellent 9-inch refracting
  telescope which he had imported from Thomas Grubb of Dublin and kept in its
  box since 1888.}  to test Mount Stromlo's suitability and ``seeing''
conditions.  Although the site checked out\footnote{Victorian Government
  astronomer {\bf Pietro P.G.E.\ Baracchi} (1851--1926) and J.M.\ Baldwin
  reported that Mount Stromlo never had less than 4 fine nights out of 7,
  after week-long samples taken every month over a year.}, due to delays,
including World War I, the Commonwealth Solar Observatory did not `open' until
1924, 13 years after the Oddie Telescope had been erected there.  In that
year, British engineer and amateur astronomer, J.H.\ Reynolds donated a
30-inch reflector, following on from Lord Farnham's 1914 donation of a 6-inch
refractor.  Even then, from late 1924 to 1926, the founding Director,
W.G.\ Duffield---who had been at the University of Reading in the UK from
1910 to 1923---was based in what is now the main bar of the Hotel Canberra.
He had to wait until the buildings on Mount Stromlo were finally fit for
occupancy.

During WWII, the Commonwealth Solar Observatory converted their
workshop\footnote{A bushfire destroyed the workshop in February 1952
  \citep{1952PASP...64...62E}, and another bushfire destroyed the re-built
  workshop in January 2003.} into something of a factory to produce `gun
sights', and prisms for binoculars, telescopes, and periscopes for the
military campaign.  After WWII, the Commonwealth Solar Observatory switched
from just solar and atmospheric observations (for
  example, \citep{1939ApJ....89..555G, 1940MNRAS.100..635A, 1944MNRAS.104...13A,
  1942MNRAS.102...24H})
%
%
to stellar and galactic work.\footnote{Over the ensuing years, much of
  Australia's solar and meteorological research steadily transitioned to the
  Commonwealth Scientific and Industrial Research Organisation (CSIRO).}  In
1944, the Victorian State Government closed the Melbourne Observatory---parts
of which became a psychiatric clinic---and in 1945, the Commonwealth Solar
Observatory in Canberra became the {\bf Commonwealth Observatory}, having purchased in
1944 the Great Melbourne telescope (which had effectively been obsolete since the
1890s, \citet{1953ASPL....6..346H}).\footnote{\citet{1953ASPL....6..346H} 
outlined the historical development of 
optical observatories in Australia.}  The Commonwealth 
Observatory grew to become one of the largest optical observatories in the
Southern Hemisphere from 1945 to 1969. As such, it attracted and produced many of the
world's best observers at that time.  The observatory's 74-inch optical
telescope (1955--2003) was a replica of the 74-inch telescope built by the
British in Pretoria, and as such, it was the equal second-largest optical
telescope in the Southern hemisphere from 1955 to 1974.  {\bf The Australian
  National University} (ANU) opened in 1946, and in 1957 the observatory
became known as {\bf Mount Stromlo
  Observatory}, officially transitioning from the Commonwealth Department of the
Interior to become a part of The ANU.


In searching for the location of a new dark-sky site, in 1957, Mount Stromlo
Observatory staff, together with Isadore Epstein from Yale University (she was
funded 
by the US National Science Foundation to find a new site in Australia), 
searched as far south as Horsham, just north of the Victorian Grampians, and
as far west as Geraldton in Western Australia.  While travelling with Harley
Wood\footnote{The first President of the Astronomical Society of Australia (ASA).}, 
Ben Gascoigne\footnote{The first Vice-President of the ASA.}, 
and Richard Twiss (who was looking for a site for a new
optical interferometer for The University of Sydney), 
Epstein inspected {\bf Siding Spring} Mountain\footnote{Furthermore, known as Mount
  Woorat.} in 1957. 
In 1962 the site was chosen, edging out {\bf Mount Bingar}\footnote{From 1959 to
  1962, The ANU had a `field station' on Mount Bingar, with a 26-inch 
reflector used for photoelectric photometry.} in the district
of Griffith, and it officially opened in 1965.  
While The ANU owns {\bf Siding Spring Observatory}, several telescope
sites are leased to other organisations, such as the Australian
Astronomical Observatory (formerly the Anglo Australian Observatory). 
The Anglo Australian Telescope (AAT) keywords were, however, not required 
because the AAT did not open at Siding Spring Observatory until 1974.
It was co-funded by the British, thanks largely to Richard
vdR.\ Woolley, former Director of Mount Stromlo Observatory and at that time the
Astronomer Royal in England.  Similarly, the Royal Observatory of Edinburgh's
1.2 m UK Schmidt telescope at Siding Spring Observatory did not commence
operations until 1973; it was handed over to the Anglo Australian
Observatory in~1988.  

Foreign-funded facilities in Australia were nothing new, and
naturally the British still had an interest in (supporting) astronomy from
Australia after Australia became a Federation in 1901.  In addition, due to
the Swedish astronomer Karl Gunnar Malmquist
(1893--1982: \citet{1982ATi....15..176E}, who is known for ``Malmquist
bias'' \citep{1922MeLuF.100....1M, 1925MeLuF.106....1M}, the University of
Uppsala had located a 0.65 m reflecting telescope at the Commonwealth
Observatory in Canberra for their {\bf Uppsala Southern Station}.  The Yale
and Columbia Universities had also previously relocated their enormously-long
{\bf Yale-Columbia} 26-inch refractor from Johannesburg, South Africa, to the
Commonwealth Observatory to continue their campaign of southern
stellar parallaxes. Their interest in 1957 pertained to installing a 20-inch
astrograph at a better site.


\subsubsection{University of Sydney}

In June 1940, physics departments around Australia were asked to assist in the
war effort.  At the University of Sydney, Rachel E.B.\ Makinson, born Kathleen
Rachel White in London (1917--2014: \citet{Rachel-Makinson}), 
had just arrived before the war 
with a Double First Bachelor degree in Physics from Cambridge.  She had also
just married Richard (Dick) Makinson\footnote{Richard Makinson (1913--1979) was a member of
  the Communist Party of Australia (CPA) whose physics career suffered once identified
  as a member.  Not surprisingly, ASIO also kept close surveillance on Rachel
  Makinson.}, an Australian physicist whom she met at Cambridge.  If not for
the war (with so many men away fighting overseas), she may well have struggled
to find employment because of her marital status. However, Professor Victor
A.\ Bailey took her on as a research assistant at the University of Sydney,
where she tutored radio physics to RAAF airmen and others (known as the
  ``Bailey Boys'', \citep{Fielder-Gill:1998}) involved in secret radar work within
the radiophysics lab \citep{Wild:1971, 1987PASAu...7...95W, Macleod:1998,
  1998:Minnett, 2005JAHH....8...11S}. Many of them went on to become
Australia's first radio astronomers.  After the war, the University of Sydney
would not offer R.\ Makinson a permanent job in the physics department because
her husband had one there.  In fact, for 20 years after the war, she worked
for the {\bf Council for Scientific and Industrial Research} (CSIR,
1926--1949), later known as the {\bf Commonwealth Scientific and Industrial
  Research Organisation} (CSIRO, 1949--), on annual contracts because married
women were denied permanent positions.  She studied the physics of wool fibres
until she was finally granted a permanent position.  She became CSIRO's first
female Chief Research Scientist in 1977 and the first female Assistant Chief
of the Division of Textile Physics (1979--1982).\footnote{A pioneer of Australian
  science in more ways than one, R.\ Makinson has a connection with the
  coloniser Elizabeth Macarthur who studied astronomy in Sydney with the First
  Fleet's official astronomer Lieutenant William Dawes.  (Lieutenant Dawes'
  son, born William Rutter Dawes in England, is himself well known for ``Dawes
  limit''.)  Together with her husband, Elizabeth and John Macarthur are
  largely responsible for establishing the wool industry in Australia
  (including Tasmania); indeed, for 8 years, Elizabeth managed the merino
  flocks in Sydney while her husband was overseas.}

Having headquartered CSIRO's Division of Radiophysics and their Division of
Physics---and trained many of their staff---The University of Sydney was well
placed to expand into astronomy and astrophysics. 
Their Astrophysics Department was formed in 1962 and charged with building the
{\bf Molonglo Radio Observatory}. Headed by B.Y.\ Mills, it subsequently
opened in November 1965.  Primarily funded by the USA National Science
Foundation (USD \$846,000 initially, and later US \$107,500 extra), the radio
telescope originally had cross-shaped arms 12-metres-wide by one-mile-long.  
In addition to this grand venture, in 1961, the Chatterton\footnote{Stanley
  Chatterton, one of the five founders of Woolworths in Australia, provided a
  grant of 50,000 pounds which enabled Prof.\ Harry Messel, Head of the School
  of Physics since 1952, to establish the new Astronomy Department.  By this
  time, Messel had raised more than \pounds3M for the School of Physics, with
  more than \pounds2 M from overseas.  The School's `theoretical physics'
  department, for example, was named after Sir `Frank' Packer (1906--1974), the
  media magnate, father of Kerry F.B.\ Packer (1937--2005) and granfather of
  James D.\ Packer (1967--).  Given the Fairfax family's prior connections with
  astronomy, encompassing the 1874 transit of Venus and friendship with George
  D.\ Hirst (1846--1915: \citep{1916MNRAS..76R.261.}), it is perhaps
  surprising that the Astrophysics Department was not supported by and thus
  named after the Fairfax family---although one may speculate that Messel
  likely tried.  In 1872, John Fairfax established the Fairfax Prize for the
  greatest proficiency among the female candidates of the Junior and Senior
  Public Examinations for matriculation honours and certificates---effectively providing two student scholarships to The University of Sydney.}
Astronomy Department was established within the School of Physics at the
University of Sydney.  It was co-located in the same building and floor as the
Astrophysics Department.  The Astronomy Department would commission and
operate the {\bf Narrabri Stellar Intensity Interferometer}.  R.\ Hanbury
Brown and R.Q.\ Twiss had come to Sydney in 1962 from the University of
Manchester, with Hanbury Brown leading this new department.  He and Twiss had
previously built an optical intensity interferometer in the UK, but the cloudy
UK weather meant it was hard to acquire data.  The University of Manchester,
together with the University of Sydney's School of Physics, therefore
co-funded the installation and maintenance of the Narrabri Stellar Intensity~Interferometer.


\subsubsection{CSIRO}

During and after WWII, the CSIR's and CSIRO's {\bf Division of Radiophysics}, and 
the Australian Postmaster General (PMG) Research 
Laboratories (now Telstra) developed more than 20 different radar systems or `field stations'.
While too numerous to describe here, most are listed 
in Table~\ref{Tab1} or Section~\ref{Sec_Other}, with more detailed discussions found in
\citet{2009cnhe.book.....S}, \citet{2019JAHH...22..266W}, 
and other references listed in Appendix~\ref{AppA}. 
The Division of Radiophysics (not to be confused with the {\bf Division of
  Physics}, which also conducted astronomical work and was similarly
headquartered on the {\bf University of Sydney Grounds} in Chippendale)
commenced their radio astronomy research programs in 1945.  Ron Giovanelli led
one group that built a solar optical observatory at {\bf Culgoora} in New South
Wales, while Paul Wild led the group that built a solar radioheliograph there.
CSIRO's Division of Radiophysics had also received a \pounds0.63 M grant from
the Ford Foundation of America to pay for half the 
radioheliograph cost, which opened in 1968 with CSIRO paying for the land.  
Radio astronomy was booming in the 1960s.  Construction of the {\bf Parkes radio
  telescope}\footnote{The telescope recently received a 
  traditional Indigenous name, `Murriyang', representing the `skyworld'
  where a prominent creator spirit is said to~live.} 
was 50\% paid for by the US-based Carnegie Corporation
(USD \$250 k) and the Rockefeller Foundation (USD \$380 k) with the Australian
Government contributing the remaining 50\% and operational
costs \citep{1981PASAu...4..267B}.   
The telescope is, of course, located in the regional NSW town called Parkes,
named after Sir Henry Parkes (1815--1896), the ``Father of Federation''
who appears on the Centenary of Federation commemorative AUD \$5 note (issued
in 2001), on the AUD \$1 coin from 1996, and a
3 pence Australian postage stamp from 1951 (for the Golden Jubilee).  The 
Parkes telescope featured on the first AUD \$50 note (along with other
astronomy-related images), and on an Australia Post stamp from 1986 (in
connection with Comet Halley), and also on an AUD \$1 coin from 2009  
(issued to celebrate the International Year of Astronomy by the Royal
Australian Mint, marking 400 years since Galileo used a telescope in 1609 to observe
the heavens). 
In 1988 the Culgoora radioheliograph was replaced by the Australia Telescope
Compact Array {ATCA} 
%
%
of radio antennas.  In an unusual break from tradition, this was fully-funded by the
Australian Government (AUD\$50 million).  This was perhaps achieved because the
project was badged successfully as an official Bicentennial project
celebrating the 200th anniversary of European colonisation in Australia, which 
had its origins in astronomical endeavours.  Spending only 
20\% of the funds offshore and calling it the {\em Australia} Telescope also
likely helped its cause, although one would be forgiven for expecting to have
seen a collaboration with, and contribution from, the British and/or~Americans. 


\subsubsection{The University of Adelaide}

By 1969 the Department of Physics at The University of Adelaide had developed
the {\bf Buckland Park Aerial
  Array}\footnote{\url{http://www.physics.adelaide.edu.au/atmospheric/facilities.htm} (accessed 29 March 2021).} 
 \citep{1969Natur.223.1321B}. 
Some 40 km north of Adelaide and 1 km square in size, it was built for
studying the ionosphere, including its drift overhead, and meteors\footnote{The disturbing effects of meteors on the ionosphere and thus long radio transmissions were first noted in Japan \citep{Nagaoka1929} and studied in the 1930s \citep[e.g.,][and references therein]{1937RSPTA.236..191A, 1947PPS....59..858H, 1982VA.....26..325H}.  Radar would later detect aircraft, meteors, and missiles.}.  The
Buckland Park Research Station's air shower array, built to detect radio and Cerenkov
emission from high energy cosmic ray showers in our atmosphere,
commenced operations in 1972.  As with the AAT and ATCA,
this facility came online post-1969. 


The University of Adelaide and the Weapons Research Establishment (WRE) at
Salisbury used rockets and satellites to observe the Sun at UV and X-ray
wavelengths.  The University of Adelaide also worked with The University of
Tasmania to conduct X-ray astronomy in the late 1960s using sounding rockets
and balloon payloads. 

Due to our focus on astronomy, we do not include Australia's military
involvement in space which began in 1946 through the establishment of an
Anglo-Australian Joint Project, which resulted in Australia becoming the 4th
nation in the world to place a satellite into orbit from home soil on the 29th
of November 1967.  Over the dozen or so years before the launch of WRESAT,
the Project employed thousands of people in Woomera.  WRESAT contained a suite
of instruments designed and built by the former Department of Supply's 
WRE, together with The University of Adelaide,
to probe, among other things, Earth's upper atmosphere and the temperature of
the Sun's corona.

\subsubsection{The University of Tasmania}

Grote Reber (1911--2002: \citet{2003ITPS...31.1112P, 2003PhT....56h..63T,
  2004PASP..116..703K, 2005ASSL..334...43K} 
moved from the USA to Tasmania in 1954 to observe the southern skies (the
centre of the Milky Way) and determine the lowest radio frequencies that one
could observe \citep{1982CosSe...4...14K, 1988JRASC..82..107K}. Together with
Graeme R.A.\ `Bil' Ellis (1921--2011: \citet{Delbourgo:2013} from the Ionospheric
Prediction Service in Tasmania\footnote{The IPS was headquartered at the
  Commonwealth Observatory in Canberra.}, they built a receiver the following
year which operated at frequencies as low as 0.52 MHz.  In the 1960s, Reber
built another telescope, and The University of Tasmania constructed several
others, including the 609 m $\times$ 609 m Llanherne Low-Frequency
  Array \citep{1972PASAu...2..135E} near Hobart's Llanherne Airport.
Construction on this started in 1967, but it did not become fully operational
until 1972.

\subsubsection{Monash University}

The Mount Burnett Observatory\footnote{\url{https://mbo.org.au/} (accessed 29 March 2021).} did
not commence operations until the 1970s and was thus not included in our
search.


\subsubsection{Other}\label{Sec_Other} 

The following keywords were additionally checked, but this did not yield any
previously missed articles: Fleurs\footnote{The 18 m dish at Fleurs, a key
  field station near Badgery's Creek (1954--1963), was referred to as the
  `Kennedy Dish'.  It was relocated to Parkes and formed a part of the Parkes
  Interferometer \citep{2012JAHH...15...96O}.}, including the Fleurs Synthesis
Telescope; Cumberland Park; Rodney Reserve; 
Freeman's Reach; Llandilo; Rossmore; Seacliff Observatory; Long Reef; 
Wallacia; North Head; West Head; Collaroy; Llanherne; Bothwell; Kempton; and
Penna.  In addition to the early radar stations, we checked for publications
from Australian-based Tracking Stations, such as Honeysuckle Creek
(1967--1981), Tidbinbilla (1965--), Island Lagoon, Muchea, Carnarvon, Cooby
Creek, and Orroral Valley.\footnote{In 1986, the 26 m satellite-tracking
  telescope at Orroral Valley near Canberra was given to the University of
  Tasmania.}


\section{The Publications and the People}\label{Sec-Tables}

In this Section, five tables display the top-ten Australian publications from
five sequential time intervals.  In the first table (1945--1950), 
we increased this count to show the top-15 publications.
Doing so allowed us to capture additional interesting 
material from the earlier years in our study.  

In terms of world-ranking, 
as opposed to the Australian-ranking shown in the first column of the tables, the 
articles are among the top 1 to 2 percent most cited astronomy publications from their
respective era.  Over the years 1945--1950, 1951--1955, 1956--1960, 1961--1965 and
1965--1969, the ADS records a total of 13,765, 12,487, 17,536, 30,577 and
41,220 astronomy-related bibliographical sources.  To a good approximation, the
first half of the articles appearing in Tables~\ref{Tab45-50}--\ref{Tab61-65}
are ranked in the top 1 percent globally, while all of the articles appearing
in Table~\ref{Tab66-69} (1966--1969) 
are ranked in the top 1 percent. 
Of particular note is the first entry in Table~\ref{Tab51-55}, ranked
number one in the world, and the first entry in Table~\ref{Tab66-69}, 
ranked number five in the world, from their respective eras.

\subsection{1945--1950}\label{Sec_45-50}

As seen in Table~\ref{Tab45-50}, studies related to the Sun dominate the
popular articles from 1945 to 1950.  Below we provide information about the
scientists who researched and wrote these articles.

Sydney-born {\bf Douglas `Walter' Noble Stibbs} (1919--2010:
Lloyd Evans~\cite{2010AG....51d..41L}, Stickland~\cite{2010Obs...130..272S}) obtained the 1942 University
of Sydney Medal in physics for his B.Sc.\ Honours degree, before obtaining in
1943 his M.Sc.\ degree and becoming an elected Fellow of the Royal
Astronomical Society that same year.  During this time, he worked at Mount
Stromlo's Commonwealth Observatory, designing a folded optical system for
gunsights and a sun-compass for desert navigation (both of which were used
during the War).  In 1950, before leaving the Commonwealth Observatory in
1951, he wrote the influential paper~\citep{1950MNRAS.110..395S} explaining
the variable magnetic star HD~125,248\footnote{The original Henry Draper (HD)
  Catalogue contained spectra from, and classifications for, 225,300 stars.
  HD~125,248 (CS Virginis; BD-18 3789), the variable star of $\alpha^2$ CVn
  type, was included in the 1920 volume \citep{1920AnHar..95....1C} by Annie
  Jump Cannon and Edward Charles Pickering, with the spectra obtained from the
  Bache Telescope, mounted at Arequipa, in Peru.}, which is Australia's
most-cited astronomy-paper from 1945 to 1950.  HD~125,248 reverses its magnetic
field every 9.3 days, with the radial velocity of the star's elements moving
in synchronisation with this.  As the elements move outward along lines of
magnetic force, their flow velocity responds according to the magnetic field
strength.  \citet{1950Natur.165..195S} wrote an accompanying Letter to Nature
explaining this phenomenon.  He then spent time at Oxford and worked in South
Africa on the radial velocities of southern cepheids for tracing Galactic
rotation~\citep{1955MNRAS.115..363S, 2019Galax...7...68A}---research which
earned him his Ph.D.  Stibbs is, however, perhaps best known as the Napier
Professor of Astronomy and Director of the Observatory at the University of
St.\ Andrews, Scotland (1959--1989), where he introduced the largest computer
in Europe in 1961.  Upon retiring, he spent the next 20-odd years as a
`Visiting Fellow' back at Mount Stromlo Observatory and The ANU's Mathematical
Sciences Institute, giving occasional lectures in statistical methods, and
since 2013 he is remembered through the ``Professor Walter Stibbs Lectures''
organised by the University of Sydney.

In 1929, Scottish-born {\bf David Forbes Martyn} (1906--1970: \citet{1970SSRv...11....3P,
  Piddington:1971}) 
came from Imperial College London to
join the Australian Radio Research Board (now the Australian
Telecommunications and Electronics Research Board).  A part of the CSIR at the
University of Melbourne, the Board was established in 1927 to improve
Australia's radio procedures \citep{White:1975}.  In 1932, Martyn transferred to
the Board's group at the University of Sydney, which was directed by Sir John
Percival Vissing Madsen (1879--1969: \citet{White:1971}).  Martyn helped Australia's
broadcasting services with their radio transmissions while also studying the Earth's
ionosphere. 
Martyn is known for ``Martyn's theorem'' regarding the attenuation of radio
waves~\citep{1935PPS....47..323M} 
and the ``Bailey-Martyn Theory''~\citep{Bailey.Martyn:1934} 
which solved the mystery of the ``Luxembourg effect''.\footnote{In England, a
  Luxembourg radio station playing popular music could be heard intermittently
  between the BBC's stronger signals. The modulation of a passing radio
  signal by a powerful intervening source had some relevance for the war
  effort regarding the jamming of (an enemy's) radio~signals.}  In 1939, Martyn
was on the short-list---along with Subrahmanyan Chandrasekhar (1910--1995, 
co-recipient of the 1983 Nobel Prize for Physics)---for the Directorship of
the Commonwealth Solar Observatory in Canberra.  This position was
ultimately awarded to R.\ Woolley (see later).  In February 1939,
Martyn was invited by the British Government to spend six months in England
learning the secrets of radar.  The Australian Prime Minister Lyons had, 
at the time, been 
asked by the Australian High Commission in London to send the best-qualified
physicist for a secret mission concerning radar.  When Martyn returned to
Australia, he was made the Chief of the CSIR's new Radiophysics Division and
Laboratory, back at the University of Sydney, for secret WWII radio direction
finding and distance-ranging \citep{1958:Mellor, 1970hrab.book.....E,
  1973hrab.book.....E, 1998:Minnett}. 
Martyn did, however, eventually move to the Commonwealth Solar Observatory at
the end of 1944.  In 1947, he published his award-winning modified dynamo
theory connecting the Earth's magnetic field with its ionised atmosphere and
the influence of solar and lunar tides \citep{1947RSPSA.189..241M,
  1947RSPSA.190..273M}. He successfully expanded this research into the
1950s with three more articles \citep{1951Natur.167...92M, 1953RSPTA.246..281B, 
  1953RSPTA.246..306M}, see Table~\ref{Tab51-55}.  Sadly, after much service to 
Australia, including President of the Australian Academy of Science
(1969--1970), 
and starting the Radiophysics Division which developed Australia's
radar surveillance during WWII, he was given electric shock `treatment' 
for a mental illness and eventually committed suicide in 1970. 

{\bf John Paul Wild} (1923--2008: \citet{2008SpReT.172...40., Frater:2012, 2017fpra.book.....F}) was
born in Sheffield, England.  A gunnery and radar officer in the Royal Navy
(1943--1947), he was also involved in research trying to understand why
early radar was sometimes jammed, later found to be due to radio interference
from the active Sun\footnote{In 1942, James Stanley Hey (1909-2000) recognised the cause was due to sunspots and solar flares
 \citep{1946Natur.157...47H, 1946PMag...37...73A}, as did George Clark Southworth (1890-1972: \cite{Southworth2010}) at the Bell Telephone Laboratories.}.
Wild became a CSIR radio astronomer in 1947,
working for L.L.\ McCready (mentioned below) in building a radio-spectrograph.
Using this to monitor the emission of solar bursts from 70 to 130 MHz, in 1950, 
they classified three types of spectral burst \citep{1950AuSRA...3..399W,
  1950AuSRA...3..541W, 1950AuSRA...3..387W}.  Over the ensuing years, the
frequency coverage and technological complexity of their radio-spectrographs
were steadily increased.  Wild subsequently worked his way up the ranks and
became chief of CSIRO's Division of Radiophysics in 1971 and then chairman of
CSIRO from 1978 to 1985.  His team built the 3 km diameter Culgoora
radioheliograph (1967--1982), producing spatially-resolved, real-time images of
activity in the Sun's corona.  The ``Paul Wild Observatory'' at Culgoora,
now home to the Australia Telescope Compact Array, the Sydney University
Stellar Interferometer, a node of the Birmingham Solar Oscillations Network,
and the Ionospheric Prediction Service\footnote{\url{http://www.ips.gov.au/} (accessed 29 March 2021).}, is
named in his honour.

Further afield, during the 1970s, Wild received financial support from the
Commonwealth Department of Transport for the ``Interscan'' microwave landing
system (MLS) for aircraft.  However, the global positioning system (GPS) was
ultimately preferred by the International Civil Aviation Organisation.  During
the years 1986--1991, immediately after Wild had resigned from CSIRO, he tried
to establish the ``Very Fast Train joint venture''.  After failed talks with
the NSW State Rail Authority, this was to be a privately-run, high-speed
railway between Melbourne-Canberra-Sydney.  Somewhat unrelated, but of
interest, is that during WWII, just after Wild arrived at CSIR, the NSW
Government Railway\footnote{Curiously, prior to becoming the first President of the
  Astronomical Society of Victoria, Charles J.\ Merfield (1866--1931) from
  Ararat, Victoria, had distinguished himself with the Victorian and NSW State
  Rail Authority for his mathematical knowledge of track laying.} had been
constructing (Radiophysics Laboratory)-designed aerials and
transmitters/receivers for radar detection.  J.G.\ Worledge engineered these 
at the Eveleigh workshops---where many indigenous Australians
had been working for the war effort.

While working with J.P.\ Wild, J.L.\ Pawsey, R.\ Payne-Scott, and others, {\bf
  Lindsay Leslie McCready} (1910--1976), from Leeton, New South Wales (NSW),
led much of the engineering effort at CSIR(O).  This was due in part to his previous
employment with Amalgamated Wireless (Australasia).  In so doing, he helped the
budding astronomers observe a correlation between sunspot coverage and radio
noise, and they also deduced that the temperature of the Sun [or rather
  its corona] is far over 6000 degrees and as high as one million
degrees~\citep{1946Natur.157..158P, 1946Natur.158..633P, 1946Natur.158..632M}. 



Noted below, Pawsey and Payne-Scott developed radio interferometry 
in 1946 (as did the British radio astronomers Martin Ryle\footnote{M.\ Ryle (1918--1984), a founding father of radio astronomy, 
shared the 1974 Nobel Prize in Physics with Antony Hewish (1924--) for their pioneering radioastrophysics research.}
and Derek Vonberg\footnote{D.\ Vonberg (1921--2015) became a medical research scientist.})  to enable higher resolution
imagery.  The first use for astronomy was by McCready, Pawsey \& Payne-Scott
using the single dish (WWII radar antenna) sea-cliff interferometer in
Sydney. This built on the unpublished, classified work of J.C.\ Jaeger
(1907--1979: \citet{Paterson:1980}) in 1943 
to determine the elevations of incoming aircraft \citep{2010ASSL..363.....G} (p. 277). 
They did this using reflections of the Sun, at sunrise, off the sea
which acted as a reflecting surface to produce the interference pattern, akin
to the well-known Lloyd's mirror interferometer \citep{Lloyd:1834}. 
\startlandscape
\begin{specialtable}[H] 
\widetable
\setlength{\tabcolsep}{6.8mm}
\caption{The ten most cited articles from 1945 to 1950
  are numbered in column 1, and involve ten distinct authors.\label{Tab45-50}} 
\begin{tabular}{llllclc}
\toprule
\textbf{\#} & \textbf{Title}  &  \textbf{Affiliation}  &  \textbf{Author(s)}  &  \textbf{Year}  &  \textbf{Journal}   &  \textbf{Cites}  \\  
   &        &               &             &        & \textbf{Vol.\ Page} &         \\
\midrule
1. & A study of the spectrum and magnetic variable                            & MSO                   &  Stibbs, D.W.N.                    & 1950a & MNRAS     & (251) \\  
   &  \hspace{5mm} star HD 125248                                             &                       &                                    &       & 110, 395  &       \\  
2. & Atmospheric Tides in the Ionosphere.\ I.\                                & CSIR-Radiophysics   &  Martyn, D.F.                      & 1947a & RSPSA    & (134) \\  
   &  \hspace{5mm} Solar Tides in the F$_2$ Region                            &  \hspace{5mm} \& MSO  &                                    &       & 189, 241  &       \\  
3. & ... Spectrum of High-Intensity Solar Radiation at                        & CSIRO-Radiophysics  &  Wild, J.P.                        & 1950b & AuSRA     & (126) \\   
   & \, Metre Wavelengths.\ III.\ Isolated Bursts                             &                       &                                    &       & 3, 541    &       \\   
4. & Magnetic and Electric Phenomena in the Sun's                              & CSIR - Physics        &  Giovanelli, R.G.                  & 1947  & MNRAS    & (103) \\  
   &  \hspace{5mm} Atmosphere associated with Sunspots                         &                       &                                    &       & 107, 338 &       \\  
5. & A Theory of Chromospheric Flares                                          & CSIR - Physics        &  Giovanelli, R.G.                  & 1946  & Nature   & (101) \\ 
   &                                                                           &                       &                                    &       & 158, 81  &       \\
6. & Observations of the Spectrum...\ I.\ The Apparatus                        & CSIRO-Radiophysics  &  Wild, J.P.                        & 1950  & AuSRA    & (100) \\
   &  \hspace{5mm} and Spectral Types of Solar Burst Observed                  &                       &  \& McCready, L.L.                 &       & 3, 387   &       \\
7. & Interpretation of Electron Densities from Corona                          & MSO                   &  Allen, C.W.                       & 1947  & MNRAS    &  (97) \\ 
   &   \hspace{5mm} Brightness                                                 &                       &                                    &       & 107, 426 &       \\ 
8. & Positions of Three Discrete Sources of Galactic                           & CSIRO-Radiophysics  &  Bolton, J.G.,                     & 1949  & Nature   &  (92) \\
   & \hspace{5mm} Radio-Frequency Radiation                                    &                       &  \hspace{2mm} \&  Stanley, G.J.    &       & 164, 101 &       \\
   &                                                                           &                       &  \hspace{2mm} \& Slee, O.B.        &       &          &       \\
9. & The Coronal Emission Spectrum                                             & MSO                   &  Woolley, R.D.V.R.\                & 1948  & MNRAS    &  (72) \\
   &                                                                           &                       &  \hspace{2mm} \& Allen, C.W.       &       & 108, 292 &       \\
10. & Atmospheric Tides ...\ II.\ Lunar Tidal Variations                      & CSIR-Radiophysics   &  Martyn, D.F.                      & 1947b & RSPSA     & (63)  \\    
    & \hspace{5mm} in the F Region Near the Magnetic Equator                  &  \hspace{5mm} \& MSO  &                                    &       & 190, 273  &       \\    
\midrule
\multicolumn{7}{c}{Extending the list to the 15 most cited articles from 1945 to 1950 involves 15 distinct authors.}\\
\midrule
11.& Observations [] at Metre Wavelengths.\ II.\                               & CSIRO-Radiophysics  &  Wild, J.P.                        & 1950a & AuSRA    & (55)  \\
   & \hspace{5mm} Outbursts                                                    &                       &                                    &       &  3, 399  &       \\
11.& The spectrum of the Corona at the eclipse of                              & MSO                   &  Allen, C.W.                       & 1946  & MNRAS    &  (55) \\
   &   \hspace{5mm} 1940 October 1                                             &                       &                                    &       & 106, 137 &       \\
12.& Radio-Frequency Radiation from the Quiet Sun                              & CSIRO-Radiophysics  &  Smerd, S.F.                       &  1950 &  AuSRA   &  (54) \\  
   &                                                                           &                       &                                    &       & 3, 34    &       \\ 
   12.& Solar Radiation at Radio Frequencies and Its                              & CSIR-Radiophysics   &  McCready, L.L.,                   &  1947 &  RSPSA   &  (54) \\  
   & \hspace{5mm} Relation to Sunspots                                         &                       &  \hspace{2mm} \& Pawsey, J.L.      &       & 190, 357 &       \\  
   &                                                                           &                       &  \hspace{2mm} \& Payne-Scott, R.   &       &          &       \\  
   \bottomrule
\end{tabular}  \end{specialtable}
\begin{specialtable}[H]\ContinuedFloat
\widetable
\setlength{\tabcolsep}{8.5mm}
 \small
\caption{{\em Cont.}} \label{xx}
\begin{tabular}{llllclc}
\toprule
\textbf{\#} & \textbf{Title}  &  \textbf{Affiliation}  &  \textbf{Author(s)}  &  \textbf{Year}  &  \textbf{Journal}   &  \textbf{Cites}  \\  
   &        &               &             &        & \textbf{Vol.\ Page} &         \\ \midrule
13.& Microwave Thermal Radiation from the Moon                                 & CSIR-Radiophysics   &  Piddington, J.H.\                 & 1949  &  AuSRA   &  (49) \\  
   &                                                                           &                       &  \hspace{2mm} \& Minnett, H.C.     &       &  2, 63   &       \\  
\bottomrule
\end{tabular} 
\end{specialtable}
\unskip
\begin{specialtable}[H] 
\widetable
\setlength{\tabcolsep}{7.15mm}
\caption{The ten (plus 2) most cited articles from 1951 to 1955 are numbered
  in column 1.  The 12 most cited articles shown here involve 14 distinct authors.\label{Tab51-55}}
\begin{tabular}{llllclc}
\toprule
\# & Title  &  Affiliation  &  Author(s)  &  Year  &  Journal  &  Cites \\  
   &        &               &             &        & Vol.\ page &         \\
\midrule
1. & The Luminosity Function and Stellar               &  MSO                                  &  Salpeter, E.E.                   &  1955  &   ApJ    & (4645)  \\ 
   & \hspace{5mm}  Evolution                           &                                       &                                   &        & 121, 161 &         \\
2. & Electrons Screening and Thermonuclear             &  MSO                                  &  Salpeter, E.E.                   &  1954  & AuJPh    &  (369)  \\  
   & \hspace{5mm}  Reactions                           &                                       &                                   &        &  7, 373  &         \\
3. & On the Distribution of Mass and                   &  MSO                                  &  de Vaucouleurs, G.               &  1953b & MNRAS    &  (234)  \\  
   & \hspace{5mm} Luminosity in Elliptical Galaxies    &                                       &                                   &        & 113, 134 &        \\  
4. & Electric Currents in the Ionosphere.\ I.\         &  Amalgamated Wireless Ltd             &  Baker, W.G.                      &  1953  & RSPTA    & (208)  \\
   & \hspace{5mm}  The Conductivity                    & \hspace{2mm} \& Radio Research Board  & \hspace{2mm} \& Martyn, D.F.      &        & 246, 281 &        \\
5. & A Survey of Southern HII regions                  &  MSO                                  &  Gum, C.S.                        &  1955  &  MmRAS   & (130)   \\
   &                                                   &                                       &                                   &        &  67, 155 &         \\ 
6. & Evidence for a local supergalaxy                  &  MSO                                  &  de Vaucouleurs, G.               &  1953a &  AJ      &  (106)  \\
   &                                                   &                                       &                                   &        &  58, 30  &         \\
7. & Electric Currents in the Ionosphere.\ III.\       &  Radio Research Board                 &  Martyn, D.F.                     &  1953  &   RSPTA  & (102)  \\
   &  \hspace{5mm} Ionization Drift due to Winds       &                                       &                                   &        & 246, 306 &        \\
   &  \hspace{5mm} and Electric Fields                 &                                       &                                   &        &          &        \\

8. & Aerial Smoothing in Radio Astronomy               &  CSIRO-Radiophysics                 &  Bracewell, R.N.                  &  1954  &  AuJPh   &  (92)  \\  
   &                                                   &                                       & \hspace{2mm} \& Roberts, J.A.     &        &  7, 615  &        \\  
9.& Harmonics in the Spectra of Solar                  & CSIRO-Radiophysics                  &  Wild, J.P.                       &  1954  &  AuJPh   &  (88)  \\  
   &    \hspace{5mm} Radio Disturbances                &                                       & \hspace{2mm} \& Murray, J.D.      &        &  7, 439  &        \\  
   &                                                   &                                       & \hspace{2mm} \& Rowe, W.C.        &        &          &        \\  
10.& The Theory of Magnetic Storms                     &  MSO                                  &  Martyn, D.F.                     &  1951  & Nature   & (80)   \\  
   &   \hspace{5mm}  and Auroras                       &                                       &                                   &        & 167, 92  &        \\
10.& The so-called Periastron effect in                &  Riverview College Obs.               &  O'Connell, D.J.K.                &  1951  &  PRCO    &  (80)  \\  
   &  \hspace{5mm} close eclipsing binaries            &                                       &                                   &        &  2, 85   &        \\  
   
    \bottomrule
\end{tabular}  \end{specialtable}
\begin{specialtable}[H]\ContinuedFloat
\widetable
\setlength{\tabcolsep}{10mm}
 \small
\caption{{\em Cont.}} 
\begin{tabular}{llllclc}
\toprule
  \# & Title  &  Affiliation  &  Author(s)  &  Year  &  Journal  &  Cites \\  
   &        &               &             &        & Vol.\ page &         \\
\midrule 

10.& Red and Infrared Magnitudes for 138               &  Lick Obs.\ (USA)                     &  Kron, G.E.                       &  1953  &  ApJ     &  (80)  \\
   &  \hspace{5mm} Stars Observed as Photometric       & \hspace{2mm} \& MSO                   & \hspace{2mm} \&  White, H.S.\ \&  &        & 118, 502 &        \\
   &  \hspace{5mm}  Standards                          &                                       & \hspace{2mm} Gascoigne, S.C.B.    &        &          &        \\
%
%
\bottomrule
\end{tabular}
\end{specialtable}
\unskip
\begin{specialtable}[H] 
\widetable
\setlength{\tabcolsep}{7.1mm}
\caption{The ten most cited articles from 1956 to 1960
  are numbered in column 1, and involve 17 distinct authors. Eleven distinct
    authors can be found above the horizontal line.\label{Tab56-60}}
\begin{tabular}{llllclc}
\toprule
\textbf{\#} & \textbf{Title}  &  \textbf{Affiliation}  &  \textbf{Author(s)}  &  \textbf{Year}  &  \textbf{Journal}  &  \textbf{Cites} \\   
   &        &               &             &        & \textbf{Vol.\ Page} &         \\
\midrule
1. & A catalogue of H$\alpha$-emission regions in        &  MSO                                  &  Rodgers, A.W.                    &  1960  & MNRAS     &  (317)  \\ 
   & \hspace{5mm}  the southern Milky Way                &                                       & \hspace{2mm} \& Campbell, C.T.    &        & 121, 103  &         \\ 
   &                                                     &                                       & \hspace{2mm} \& Whiteoak, J.B.    &        &           &         \\ 
2. & Radiation Transfer and the Possibility of           &  CSIRO-Radiophysics                 &  Twiss, R.Q.                      &  1958  &  AuJPh    &  (197)  \\  
   & Negative Absorption in Radio Astronomy              &                                       &                                   &        &  11, 564  &         \\  
3. & Interferometry of Intensity Fluctuations            &  Jodrell Bank (UK)                    &  Hanbury Brown, R.                &  1957  & RSPSA     &  (170)  \\
   & \hspace{5mm} in Light. I. Basic Theory...           & \hspace{1mm} \& CSIRO-Radiophys.    & \hspace{2mm} \& Twiss, R.Q.       &        & 242, 300  &         \\
4. & Red and infrared magnitudes for 282                 &  Lick Obs.\ (USA)                     &  Kron, G.E.                       &  1957  &  AJ       &  (170)  \\
   & \hspace{5mm} stars with known trigonometric         & \hspace{1mm} \& MSO                   & \hspace{2mm} \& Gascoigne, S.C.B. &        & 62, 205   &         \\
   & \hspace{5mm} parallaxes                             &                                       & \hspace{2mm} \& White, H.S.       &        &           &         \\
5. & A Catalogue of Radio Sources between                &  CSIRO-Radiophysics                 &  Mills, B.Y.,                     &  1958  & AuJPh     &  (143) \\  
   & \hspace{5mm} Declinations +10$^\circ$ and $-$20$^\circ$   &                               &  \hspace{2mm} Slee, O.B.\ \& Hill, E.R. &     & 11, 360   &        \\  
6. & A Catalogue of Radio Sources between                &  CSIRO-Radiophysics                &  Mills, B.Y.,                      & 1960   & AuJPh     &  (130) \\ 
   & \hspace{5mm} Declinations $-$20$^\circ$ and $-$50$^\circ$   &                              & \hspace{2mm} Slee, O.B.\ \& Hill, E.R. &       & 13, 676   &      \\ 
\midrule
7. & Geomagnetic Storm Theory                            &  CSIRO-Radiophysics                &  Piddington, J.H.                  &  1960  &  JGR      &  (129)  \\
   &                                                     &                                      &                                    &        & 65, 93    &         \\  
8. & ...Speed of the Solar Disturbances                  &  CSIRO-Radiophysics                &  Wild, J.P.,                       &  1959  & AuJPh     &  (120)  \\  
   & \hspace{3mm} responsible for Type III Radio Bursts  & \hspace{2mm} \& MSO                  & \hspace{2mm} \& Sheridan, K.V.     &        & 12, 369   &    \\  
   &                                                     &                                      & \hspace{2mm} \& Neylan, A.A.       &        &           &    \\  
9. & Strip Integration in Radio Astronomy                &  CSIRO-Radiophysics                &  Bracewell, R.N.                   &  1956  &  AuJPh    &  (114)  \\  
   &                                                     &                                      &                                    &        & 9, 198    &         \\
10. & On the chemical evolution and densities            &  The ANU:                            &  Ringwood, A.E.                    &  1959  &  GeCoA    &  (112)  \\  
    & \hspace{4mm} of the planets                        & \hspace{2mm} Dept.\ of Geophysics    &                                    &        &  15, 257  &        \\  
%
%
\bottomrule
\end{tabular} 
\end{specialtable}
\finishlandscape




%
%

{\bf Ronald Gordon Giovanelli} (1915--1984: 
\citet{1984EOSTr..65..409P, Piddington:1984b, 1985PASAu...6..112S}) 
from Grafton, NSW, obtained his M.Sc.\ from the University of Sydney in 1939,
4 years before Stibbs.  During the war, within the CSIR's National Standards
Laboratory (which became the Division of Physics in 1945), Giovanelli
developed many things. These included safety goggles to protect the eyes of
anti-aircraft spotters/gunners while watching for dive-bombers attacking from
the direction of the Sun, and aircraft illumination panels that do not spoil
one's night-adapted vision.  His most important astronomical research
published in 1946--1947 was the notion of magnetic field reconnection 
for generating solar flares \citep{1946Natur.158...81G, 1947MNRAS.107..338G}. 
From 1958--1974 he was the Chief of the Division
of Physics.  His team built the solar optical observatory at Culgoora.  In
1984 at the sixth National Congress of the Australian Institute of Physics, he
was remembered through a series of Solar-Terrestrial Physics Workshops 
co-sponsored by the Australian Academy of Science, dedicated to his memory.

{\bf Clabon (Clay) Walter Allen} (1904--1987: \citet{1973QJRAS..14..311S,
  1990QJRAS..31..259M}) from Subiaco, Perth, was one of the original 1926 staff
employed by W.G.\ Duffield at the Commonwealth Solar Observatory.  Clay had
recently 
obtained his B.Sc.\ from the University of Western Australia.  While observing
the 1940 solar eclipse from South Africa, he obtained measurements (published
after the war, in 1946--1947) of the electron density in the solar corona.
These proved highly useful for radio astronomers \citep{1946MNRAS.106..137A,
  1947MNRAS.107..426A}.  Later on, working at the University College, London,
and as Director of the University of London Observatory at Mill Hill from
1951--1973, he authored the highly useful ``Astrophysical
Quantities'' (\citep{1955asqu.book.....A, 1963asqu.book.....A}, cited over 1400
times). 
The third edition \citep{1973asqu.book.....A} has been cited over 4300
times and was published as Allen retired and returned to Mount Stromlo
Observatory.  Due to the multiple editions of his book\footnote{In honour of
  C.W.\ Allen's work, the 4th edition, although prepared and edited by Cox
  2000, is called ``Allen’s Astrophysical
  Quantities'' \citep{2000asqu.book.....C}.}, C.W.\ Allen is one of Australia's
most well-known names in astronomical circles around the globe.

{\bf John Gatenby Bolton} (1922--1993: \citet{1994PhT....47d..73K,
  1996PASP..108..729K, Wild-Rad:1994}) was born in Sheffield, England, one year
before J.P.\ Wild.  While also in the Navy during World War II, Bolton visited
Australia and later started work in the CSIR's Division of Radiophysics in
1946.  From Dover Heights, he advanced the radio interferometer
using the sea as a reflecting element (see \citep{2002JAHH....5...21O}). He and his
colleagues discovered three radio sources in the sky, matched these with
optical counterparts, and thereby in their 1949 article they discovered (two)
extragalactic ``radio galaxies'' \citep{1949Natur.164..101B}. From 1955--1961, 
Bolton also helped establish Caltech's ``Owens Valley Radio Observatory'' in
the USA, before returning to CSIRO to oversee the construction of the Parkes radio
telescope which subsequently identified the location of thousands of extragalactic radio
sources and helped discover the extreme distances of quasars 
(see \citep{2007JAHH...10...79W, 2014JAHH...17..283R}). 
Arising from
the ``John G.\ Bolton Memorial Symposium'', a special issue of the Australian
Journal of Physics \citep{1994AuJPh..47..495G} is devoted to Bolton, and since
1998 he is additionally honoured through the ``The Bolton Fellowship''
administered by CSIRO Astronomy and Space Science.
    
{\bf Gordon James Stanley} (1921--2001: \citet{2005ASSL..334...43K,
  2005PASA...22...13K}) came from Cambridge, New Zealand.  From 1944 he helped
build radio receivers in Australia before relocating to the USA at the start
of 1954.  He constructed the 200 MHz receiver for Allen's solar antenna
located at the Commonwealth Solar Observatory.  Working with Bolton at Caltech
from 1954 to 1961, he found the site for, and from 1961 to 1975 was the first
Director of, the Owens Valley Radio Observatory.    
While {\bf Owen Bruce Slee} (1924--2011: \citet{2004PASA...21...23O}) of Adelaide
was the third key player in the exciting 1940s discovery of extra-galactic
radio galaxies, he is also well known for his catalogs of radio
sources produced a decade later.  A detailed and worthy tribute to Slee, and
his 60 years in radio astronomy, is given in a special issue of the Journal of
Astronomical History and Heritage~\citep{2005JAHH....8....3O}. 
Today, AGN emissions from radio to tera-electron volt (TeV) energies
offer exciting clues to understanding the 
these systems' emission processes (e.g., \citep{2019Galax...7...23R}). 

British-born {\bf Richard van der Riet Woolley, Sir} 
(1906-1986: \citet{Davies:1984, 1987MNSSA..46....4F, 1987QJRAS..28..546L,
  McCrea:1987, 1987Obs...107...99S}) 
went to South Africa at age 15, obtaining
his B.Sc.\ and M.Sc.\ (at age 19) from the University of Cape Town before
returning to England to obtain his Ph.D.\ from the University of Cambridge.  He
was (i) the second Director\footnote{From 1929-1939 the Commonwealth Solar
  Observatory was in the care of Mr W.B.\ Rimmer after the early death of
  W.G.\ Duffield.} (referred to as the
Commonwealth Astronomer at that time) of the Commonwealth Solar
Observatory, later known as the Commonwealth Observatory by 1950, 
in Canberra (1939--1955), and (ii) the UK ``Astronomer Royal''
(1956--1971), and (iii) the Director of the South African Astronomical Observatory
(1972--1976) where he retired.  Although just 33 when he started at the
Commonwealth Solar Observatory, he is recognised as having been wonderful in
the role and successfully transformed it from a solar observatory to a stellar
and extra-galactic observatory.  Together with Walter Stibbs, Woolley
co-authored the valuable textbook about radiative transfer called ``The outer
layers of a star'' \citep{1953ols..book.....V}.  He also co-wrote 
a `top-ten' article from 1945 to 1950 
with C.W.\ Allen about emission lines from the Sun's corona
\citep{1948MNRAS.108..292W}. 

Furthermore, Woolley succeeded in having the
74-inch optical telescope built at Mount Stromlo.  It was the equal largest
telescope in the 
Southern Hemisphere for two decades, along with the matching telescope in South
Africa from which it was copied.  Woolley was also an early and crucial
supporter of the 3.9 m Anglo Australian Telescope.  The ``Woolley
building'' at Mount Stromlo Observatory was opened in 1995. In addition, the
University of Cambridge in the UK host ``Woolley Conferences'' and offer
``Woolley Studentships'' for students primarily resident in the Southern
Hemisphere and preferably from South Africa.


Although we have now been briefly introduced to the authors of the first ten
papers in Table~\ref{Tab45-50}, there are many notable individuals from this
period immediately following the war.  Therefore, the additional five authors
from the following three most-cited papers are included in Table~\ref{Tab45-50} and
mentioned below.

Born in Vienna, {\bf Stefan (Steve) Friedrich Smerd}
(1916--1978: \citet{1980IAUS...86....5W}) became a world-leading solar physicist
after emigrating to Britain in 1938 and obtaining his B.Sc.\ degree from the
University of Liverpool in 1942.  Like most of the early radio astronomers, 
he had previously been recruited (in Britain) to work on secret projects
connected with radar development, which ultimately brought him to
the CSIR in 1946. Known for his theoretical work on the `quiet'
Sun \citep{1950AuSRA...3...34S}, which is his 1950 entry in
Table~\ref{Tab45-50}, he worked with J.L.\ Pawsey and applied his ideas to
practical problems and observational data from the Sun.  By 1971 he succeeded
Wild as the Director of the solar observatory at Culgoora, but sadly died in
1978 while undergoing heart surgery at the Royal Prince Alfred Hospital,~Camperdown.

{\bf Joseph (Joe) Lade Pawsey} (1908-1962: \citet{1963QJRAS...4..316.}) came from
Ararat, Victoria, Australia, the birthplace of astronomer C.J.\ Merfield. 
Universities in Australia did not offer
Ph.D.\ degrees until The University of Melbourne introduced them in
1944 \citep{Rae:2002}.  Prior to this, Australian graduates would go to Britain
for their research training.  After Pawsey received his Ph.D.\ from Cambridge
in 1935 (and after a few years with the Electric and Musical Industries
  Inc., EMI)\footnote{Sir Ernest Thomas Fisk (1886--1965: \citet{2007:Given}) 
  headed Amalgamated Wireless (Australasia) from 1917 to 1944, and was
  managing director of EMI in London from 1945 to 1951.}, D.F.\ Martyn
recruited Pawsey, who returned to Australia in 1940 to develop radar equipment
at the CSIR radiophysics laboratory. After WWII had finished, as with others
noted earlier, Pawsey and his group turned their antennae and attention to the
Sun, finding enhanced radio emission associated with sunspots.  Pawsey is
sometimes referred to as the `Father of radio astronomy in Australia'. This is in part
for having helped introduce radio interferometry (via the sea-reflection
interferometry technique using the cliff-top aerial at Dover Heights), for
his leading role in introducing the Fourier Synthesis
concept \citep{1947RSPSA.190..357M}\footnote{The article by McCready, Pawsey \&
  Payne-Scott (1947) was submitted with Pawsey as the lead-author.  However, the
  journal's editor made the order alphabetical (R.D.\ Ekers, 
  priv.\ comm.\ 2015).}, and of course for his textbook ``Radio
Astronomy'' \citep{1955radi.book.....P}. Having been the Assistant Chief of the
Division of Radiophysics from 1951, he passed away in 1962 just before taking 
up the Directorship of the US National Radio Astronomy Observatory in Green
Bank, West Virginia.  Pawsey is remembered through the annual Pawsey Memorial
Lecture and the ``Pawsey Medal'' which has been awarded since 1967 by the
Australian Academy of Science (of which he was a founding Fellow) to recognize
outstanding research in any field of physics by an Australian scientist under
40 years of age. The Pawsey crater on the Moon is named after him, as is the
Pawsey Centre, and the Pawsey Supercomputing Centre in Perth---housing
one of the most powerful petascale supercomputers in the Southern Hemisphere.
It is used for, among other things, processing and storing data from CSIRO’s
Australian SKA Pathfinder (ASKAP) Telescope.

%

%
During the 1910s, 
there must have been something in the water at Grafton, NSW, or perhaps a
particularly inspiring local teacher.  {\bf Ruby Payne-Scott}
(1912--1981) was also born in Grafton, just three years before R.G.\ Giovanelli.
In 1941, Payne-Scott (after {\bf Joan Jelly, n\'ee Freeman}\footnote{Joan
  Freeman was exceptionally smart and seemingly came first in
  everything she did. She received numerous prizes while in Australia,
  including the Fairfax Prize.  At the end of WWII (after having obtained 
  a B.Sc.\ in 1939, and an M.Sc.\ in 1943, under Victor A.\ Bailey while she 
  also worked for the CSIR), the CSIR offered her a Senior Studentship to attend 
  Cambridge for her Ph.D.  Her story from early hardship in Perth to
  international success is impressive.  She became a world-leading
  Nuclear Physicist and, in 1976, was the first woman to receive the
  Rutherford Medal (shared with Roger Blin-Stoyle).}, 1919--1998: \citet{Freeman:1991}) was
one of the first two female radio astronomers employed by CSIR in Australia.
She worked at 
the Radiophysics Laboratory at the University of Sydney, where she had
obtained a B.Sc., M.Sc., and Dip.Ed.\ over the previous decade, often as
the only girl in the class---as was the case with Joan Freeman.  
Payne-Scott used the facilities at Dover Heights, Hornsby, and Potts
Hill, and during the war she performed top-secret work investigating 
radar.  
In 1915, during WWI, women had to march through London for the right to join
the war effort, but by WWII, 
women's wartime services were recognised and encouraged (Figure~\ref{Fig_stamp}).
Payne-Scott helped monitor the positions of aircraft
flying off Australia's coast during WWII and is famously said to have
protected Australia's coastline with radars maintained with
'coathangers and sticky~tape'.  
\begin{figure}[H]
\includegraphics[height=5.0cm,angle=0]{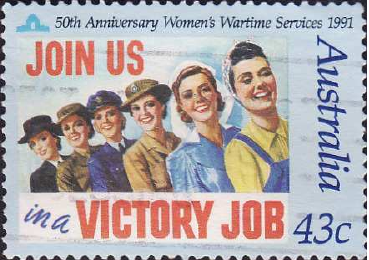}
\caption{
National Service Office poster from WWII featured on a 1991 Australian postage
stamp commemorating the 50th Anniversary of Women's Wartime Services.  
\label{Fig_stamp}
}
\end{figure}

Payne-Scott had a prominent role in the first radio astronomical
interferometer observation in 1946 (of the rising Sun over the sea).  She
openly noted (see the references in~\citep{1946Natur.157..158P}) 
that this research followed the war-time radar detection of the 
Sun by people such as {\bf Dr Frances `Elizabeth' Somerville Alexander, n\'ee Caldwell} 
(1908--1958: Sullivan~\cite{1984eyra.book.....S}, Orchiston~\cite{2005ASSL..334...71O}), who 
had a truly 
remarkable life and career\footnote{Alexander was at the time operating radar
  for the Royal 
  New Zealand Air Force and a former Captain in the British Royal Navy.}).
The Sun had also been observed at radio wavelengths by O.B.\ Slee at the RAAF
Radar Station 59 in Darwin \citep{2005JAHH....8....3O} and it was first detected in 1942 by \citet{1946PMag...37...73A} and independently by G.C.\ Southworth, who would later detect it at microwave (centimeter) wavelengths \citep{SOUTHWORTH1945285}. 
Payne-Scott went on to research Type I and III solar radio bursts and had
astronomy publications from 1946 to 1952.  Secretly married in 1944, it was kept
a secret for six years because back then, and until 1966, married women could
not hold a permanent position in the public service.  In 1950, she lost her
permanent position before `resigning' in 1951 when she had her first
child\footnote{Her son, Peter Gavin Hall, went on to become one of Australia's top mathematicians at
  The University of Melbourne.} with PMG telephone mechanic William (Bill)
Holman Hall (1911--1999: \citet{2010ASSL..363.....G}).  Since 2008, she is remembered through
CSIRO's ``Payne-Scott Award'' to support researchers who have taken extended
leave to care for a newborn child or family-related matters.  Many
articles and at least two books \citep{2010ASSL..363.....G, 2013mwsr.book.....G} have been
written about her.  With combinations of WWII intrigue and secrecy\footnote{A
  member of the Communist Party, as was Rachel Makinson's husband Richard
  Makinson, the Australian Security Intelligence Organisation (ASIO) had a
  large file on Ruby Payne-Scott's activities.}, 
pioneer intellect, and discrimination from the authorities, the story
of Payne-Scott, Freeman, Martyn, Piddington, Alexander, and others, is
in some ways Australia's version of the motion picture ``The Imitation
Game'' involving Alan Mathison Turing (1912-1954: Hodges
\cite{Hodges:1983, 2009pttt.book...13H}), the WWII codebreaker and
computer pioneer who features on the UK's new 50 pound note.


{\bf Jack Hobart Piddington}\footnote{The middle name originates from his
  great-great-grandparents, who came to Australia and started their new life in
  Hobart around 1836.} (1910--1997: \citet{Piddington:1998}) 
%
%
came from Wagga Wagga, NSW, 120-odd km south-east from
the country town Leeton, where L.L.\ McCready came.  Piddington was a highly-cited 
%
%
theoretical astrophysicist who worked on cosmic plasma physics 
and electrodynamics.  However, Piddington was much more than that.  As with Joan
Freeman, he won an astounding number of prizes and awards while studying at
school and then at the University of Sydney before obtaining his Ph.D.\ and
yet more scholarships at Cambridge, England.  Having aided the British Air
Ministry and Royal Air Force while working at Cambridge\footnote{Piddington from Australia was friends at Cambridge with Ernest Rutherford (1871--1937) from New Zealand, who received the 1908 Nobel Prize in Chemistry.}, where 
Appleton\footnote{Sir Edward Victor Appleton (1892--1965) won the 1947 Nobel Prize in Physics for his research into the ``Appleton layer'' of the atmosphere and the ionosphere more generally, which involved his development of radar.} had been his Ph.D.\ supervisor and godfather to his son, he became a tremendous asset to
Australia.  During WWII, he was instrumental in the secret development of
Australia's shore-defence radar.  He designed Sydney's first
aircraft-detecting (air-raid warning) system within five days of the bombing
of Pearl Harbour by the Japanese (7 December 1941).  These radars were constructed at
the Radiophysics Labs and by the NSW Railways and ``Her Majesty's Voice''
(HMV).  While a set was rushed to Darwin at the end of January 1942, the RAAF
was not able to get it operational before the devastating bombing of Darwin
by the Japanese on 19 February 1942.  One month later, Piddington was up
there helping with the installation, which he achieved just in time to detect
another secret Japanese raid which was then intercepted and thankfully
dispersed by US fighters just 30 km out to sea.  If not for Piddington, Darwin
would have been bombed a second time, not to mention other cities in Australia
that were spared WWII bombing raids thanks to the new radar
detection \citep{1998a:Minnett, 1998b:Minnett}. 





Piddington does not have the recognition that one might expect.
Moreover, according to Keith David Cole (1929--2010: \citet{DysonCole:2014}),
who is quoted by \citet{Piddington:1998}, 
Piddington's somewhat withdrawn nature later in life meant that he did
a lot of outstanding research in Australia but did not always receive
fair credit.  Raised by his mother since he was six, she can be proud
of his contributions to Australia.  Although Piddington and
D.F.\ Martyn worked together, 
they could not publish their war-time work.  Their 200 MHz Shore Defence (ShD)
radar, which was later used by early radio astronomers, was installed at 17
sites around Australia's coastline during WWII.  Such was its success that Air
Chief Marshal Sir Robert B.\ Popham requested Piddington to advise on suitable
sites in Singapore\footnote{Before Piddington's radar could be installed in
  Singapore, on 8 December 1941 it was bombed, and on 15
  February 1941 it fell to the might of the Japanese.  This was arguably Britain's
  worst defeat in WWII.} (where Dr `Elizabeth' Alexander was stationed) and
several other countries in Asia.  After the war, and partly due to the
subsequent dominance of observational research in Australia, as opposed to 
(Piddington's post-1956 preference for) theoretical research, Piddington was the
sole-author on almost all of his refereed publications\footnote{Not included in ADS is \citet{AppPidd1938}, regarding the ionosphere.}, and he collaborated on just 
three papers from 1954 to 1985.  Following Pawsey \& Bracewell's book 
``Radio Astronomy'' \citep{1955radi.book.....P}, 
Piddington published his own valuable book in 1961 with the same 
title~\citep{1961raas.book.....P}. 


While briefly involved with {\em observational} astronomy from 1949 to 1956,
Piddington co-authored five papers with engineer {\bf Harry Clive Minnett}
(1917--2003: \citet{Thomas:2005}) from Sydney.  While almost everyone else at that
time continued to perform metre wavelength observations of the Sun, Piddington
and Minnett followed Southworth \citet{SOUTHWORTH1945285} and pushed into the microwave regime, where they successfully studied the Moon
and correctly inferred properties of its dust layer
\citep{1949AuSRA...2...63P}.  In 1985--1986, Minnett was the
Deputy Chief Executive of Interscan International Limited, a spinoff company
of the Division of Radiophysics which started by  developing a 
microwave approach and landing system for aircraft in the 1970s.  Working with
J.H.\ Piddington, he also discovered the powerful
radio source Sagittarius A \citep{1951AuSRA...4..459P},  
now known to be powered by a 4 million solar mass 
black hole at the centre of our Galaxy.
Not only a research astronomer but
also an engineer, Minnett joined CSIR in 1940 as an assistant research officer
and was to later successfully help lead the Parkes 64 m radio telescope to
completion. He worked in the Parkes telescope control room during the Apollo 11 lunar landing,
provided significant contributions to the construction of the 3.9 m
Anglo-Australian Telescope \citep{1971PASAu...2....2M}, and eventually became
Assistant Chief (1972--1978), and then Chief, of the Division of Radiophysics
from 1978 until his retirement from CSIRO in~1981.

\subsection{1951--1955}\label{Sec_51-55}
With a remarkable 4645 citations (in 2014) and counting, the article by Austrian-born
and Australian-educated {\bf Edwin Ernest Salpeter}
(1924--2008: \citet{2009PASP..121..101H, 2009BAAS...41.1208T}) 
about ``The Luminosity Function and Stellar
Evolution'' \citep{1955ApJ...121..161S} 
is not just the most cited Australian astronomy article from
1951--1955 but is (within the ADS) the world's most cited astronomy article
from that period.  In fact, as of writing, no pre-1970 astronomy article has more citations
than this within the ADS.  Salpeter wrote this work about the initial mass
function of stars (recently reviewed by \citet{2018PASA...35...39H}) 
while he was at Mount Stromlo Observatory, on leave of
absence from Cornell University (USA).  Belonging to a Jewish family in
Austria, Salpeter had previously (in 1939) escaped to Australia as a teenager
during WWII.  After two years at Sydney Boys High School and obtaining a
B.Sc.\ and M.Sc.\ from the University of Sydney, in 1945 he left for England
to undertake his Ph.D.\ at Birmingham University before spending his career at Cornell
University.  While at Mount Stromlo in 1954--1955, he also published on refined
nuclear reaction rates inside of stars, producing the second most-cited
Australian article from 1951 to 1955 \citep{1954AuJPh...7..373S}. 
Salpeter maintained ties with 
Australia, and one of his very last articles, from the
year he died, was in the Publications of the Astronomical Society of Australia and 
reviewed the state of Nuclear Astrophysics prior to 
1957 \citep{2008PASA...25....1S}.

\textls[-15]{{\bf G\'erard Henri de Vaucouleurs} (\mbox{1918--1995: \citet{1995Natur.378..440B,
  1995GAst...22d...2C}}, Buta~\cite{1996BAAS...28.1449B}) from France joined the staff at
Mount Stromlo Observatory in 1951 as their first Research Fellow.  In 1953 he
furthered his famous $R^{1/4}$ galaxy model \citep{1948AnAp...11..247D} to
describe the projected (on the plane of the sky) radial light profiles of early-type
galaxies~\citep{1953MNRAS.113..134D}.  This model had a very successful run and
was eventually replaced by the 1963 $R^{1/n}$ model of Jos\'e Luis
S\'ersic \citep{1963BAAA....6...41S, 2005PASA...22..118G}). 
In 1954, and published in \citet{1956VA......2.1584D}, G\'erard discovered 
a Local ``supergalaxy''.  This represented a supercluster of 
nearby galaxy clusters and bright galaxies orientated in a plane which runs roughly
perpendicular to that of our Milky Way.  He is also 
known for proposing a flat, rotating disk model for the Large Magellanic Cloud
(LMC)~\citep{1955AJ.....60..126D}, which was confirmed by Kerr, McGee and others
at the Radiophysics Lab with 21-cm observations.  Combining their ideas and
results, \citet{1955AuJPh...8..508K} was the first
extragalactic 21 cm astronomy article. 
\citet{1955AJ.....60..126D} also advocated that there had been a past
interaction between the Milky Way and the LMC, a result later supported by
Kerr's observations of our galaxy's disturbed gas
disc \citep{1957AJ.....62...93K}. After solidifying their his on astronomy, de
Vaucouleurs left Mount Stromlo in 1957.  Among other things, he went on to
announce at an International Astronomical Union (IAU) meeting in Canberra that
the Milky Way galaxy is barred \citep{1964IAUS...20..195D}. He is, of course,
also known for his {\it Reference catalogues of bright
  galaxies} (for example, \citep{1991rc3..book.....D}), compiled with his astronomer wife {\bf
  Antoinette de Vaucouleurs, n\'ee Pi\'etra}
(\mbox{1921--1987: \citet{1987LAstr.101..566B, 1987JAF....31....2B,
  1988PhT....41g..92B}}) and other key contributors.  While at Mount Stromlo, Antoinette
produced the first luminosity and spectral classification of 366
southern-hemisphere B, A, and F stars on the Morgan-Keenan
system \citep{1957MNRAS.117..449D}.  When Antoinette died, a special conference
was held in her (and G\'erard's) honour \citep{1988LAstr.102..233B,
  1989AdSAC...4.....C}. }


It is interesting that \citet{1952PASP...64..185D}\footnote{{\bf Olin Jeuck Eggen}
  (1919--1998: \citet{1993ARAA..31....1E, 2000BAAS...32.1661F,
    2001PASP..113..131T}), from the USA's Lick Observatory, had an extended
  visit to Mount Stromlo Observatory in 1951, where he was later the Director
  from 1966 to 1977.} reported variability in eta Carinae (Figure~\ref{Fig_Eta}). 
Mr Francis Abbott (1799--1883: \citep{FrancisAbbott}) 
had claimed that the nebulae surrounding this star also varied 
\citep{1871Natur...4..478A,
  1873AReg...11..221A}.  However, his discovery from Tasmania was met with stern
criticism in the 1860s by the
British and European astronomical establishment~\citep{1871MNRAS..31..228H,
  1871MNRAS..31..233A}.   
Mr Abbott was an English watchmaker sent to Tasmania
  in 1845 as a convict.\footnote{We have not established if Abbott's 
wife, Mary Woolley, was a relative of English-born Richard 
vdR.\ Woolley (1906--1986), the Director of the Commonwealth Observatory in
Canberra from 1939 to 1955.} 
While using a pair of
binoculars from Mount Stromlo Observatory, de Vaucouleurs noticed that the
star was brighter than reports from almost half a century earlier.  As
discussed in the scholarly review of $\eta$ Carinae by 
\citet{1952ASPL....6..244D}, it was D.\ O'Connell (Director of the
Riverview College Observatory operated by the Saint Ignatius' College,
Riverview, Sydney) who subsequently discovered in a long series of patrol
plates that eta Carinae had brightened by 1 mag more than a decade earlier in
June 1941.  Increasing from 8.5 to 7.5 mag, it was still much fainter
than its peak in 1843, when it became the second brightest star in the sky
after Sirius, and brighter than $\alpha$ Carinae---known as Canopus---the
(usually) second brightest star in the sky.\footnote{The central star is
  expected to explode as a supernova any day now (astronomically speaking),
  hence explaining, in part, the large citation count to de Vaucouleurs and Eggen
  \cite{1952PASP...64..185D,2005ASPC..332...14H}. {\bf Bart Jan Bok}
  (1906--1983: \citet{Gascoigne:1992a}), the third Director at Mount Stromlo
  Observatory, and his astronomer wife {\bf Priscilla Fairfield Bok}
  (1896--1975), are well-known to have had a strong fascination and liking for
  the $\eta$ Carinae nebula.}
\begin{figure}[H]
\includegraphics[height=5.0cm,angle=0]{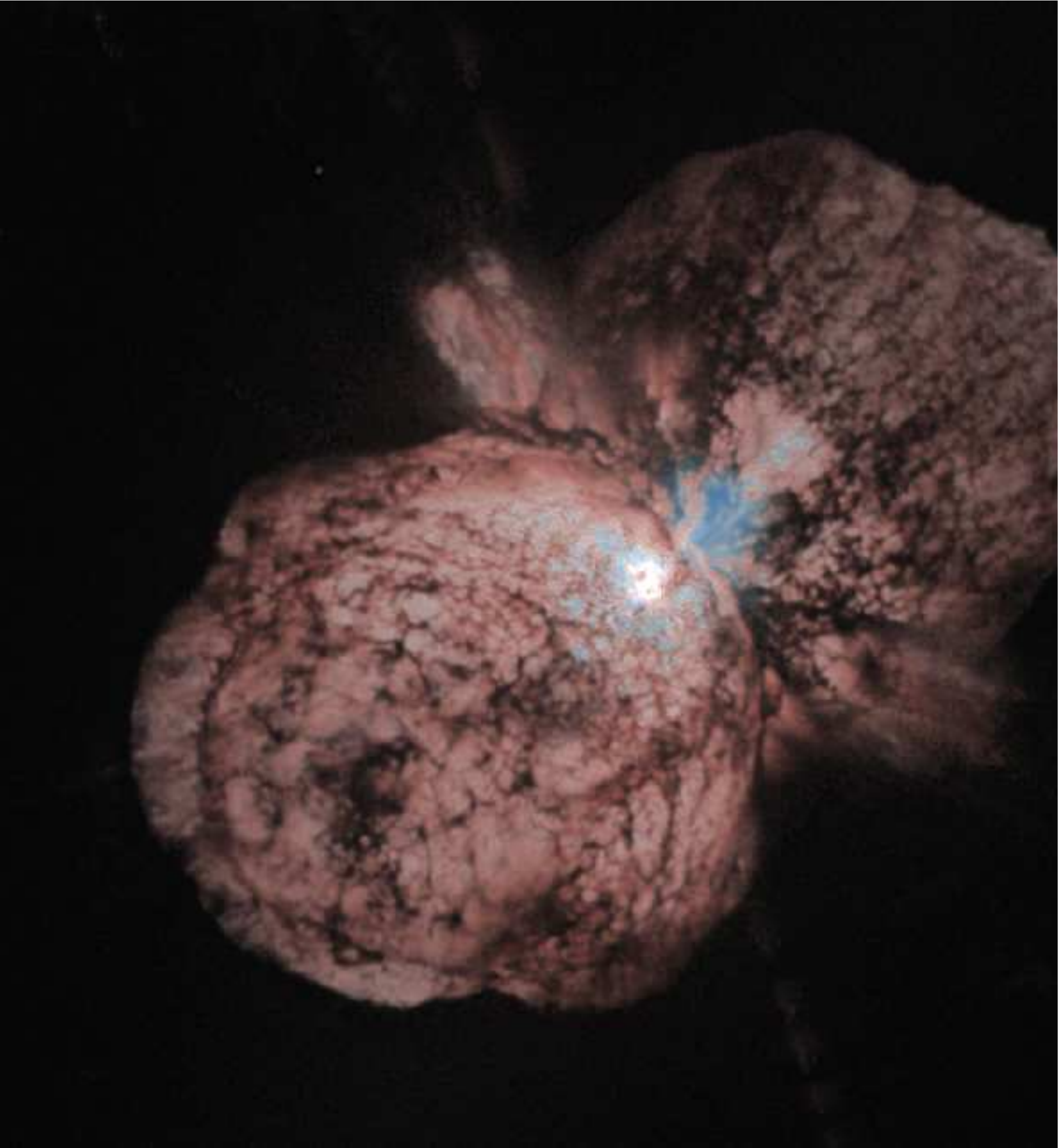}
\caption{The star $\eta$ Carinae. 
HST image credit: Jon Morse (University of Colorado), and NASA. 
\label{Fig_Eta} 
}
\end{figure}


{\bf Dr W.G.\ Baker} was not (primarily) an astronomer but rather was
affiliated with the electronics manufacturer and broadcaster Amalgamated
  Wireless (Australasia) Ltd., where L.L.\ McCready had also worked.  
  AWA Ltd.\ was Australia's largest designer and builder of radios and
televisions in the 1900s, and they still exist today as an ICT company.  As
the title of Baker's 1953 paper \citep{1953RSPTA.246..281B} 
reveals, his work pertained to the
conductivity of electric currents in the Earth's ionosphere.  Another
highly-cited Australian paper, but not listed in the Tables because by the
1960s it starts to fall under the banner of atmospheric physics rather than
astronomy, pertained to the ``excitation of atmospheric oscillations''.  It
was written by 
Butler\footnote{Stuart Thomas Butler (1926--1982: \citet{Watson-Munro:1980}).} 
\& Small \citep{1963RSPSA.274...91B} while working at The Daily Telegraph Theoretical
Department\footnote{The Daily Telegraph Theoretical Department was
later named the Sir Frank Packer Theoretical Department.} 
within the School of Physics at the University of Sydney.  This
unusual name for a Department came about because of the enabling donation by
Sir Douglas `Frank' Hewson Packer, media magnate, keen yachtsman, and
patriarch of the Packer family.  Packer ran The Daily
  Telegraph\footnote{The Daily Telegraph was sold to Rupert Murdoch
  in 1972.}, one of Australia's leading newspapers, and he started the 
  Australian Women's Weekly (which would make for another interesting, and
welcome, Departmental name).

{\bf Colin Stanley Gum} (1924--1960: \citet{1961QJRAS...2...37.}) was an
undergraduate student from The University of Adelaide who worked with
C.W.\ Allen at the Commonwealth Observatory.  In 1946, they commenced the
first 200 MHz radio map of the southern part of the
Galaxy \citep{1950AuSRA...3..224A} using equipment built by G.J.\ Stanley and
loaned by J.L.\ Pawsey from the Radiophysics laboratory.  For Gum's subsequent
Ph.D.\ thesis undertaken at the Commonwealth Observatory, he used optical 
filters to discover 40 (new) ionised hydrogen clouds glowing red due to their
H$\alpha$ emission.  While such HII regions usually trace star formation
(e.g., \citep{2019Galax...7...88C}), 
the giant ``Gum Nebulae'' 
\citep{1952Obs....72..151G} (see Figure~\ref{Fig_Gum}) was later shown by radio
astronomers to be ionised by the Vela supernova remnant.  Combining his
H$\alpha$ maps with those in the northern hemisphere \citep{1953ApJ...118..318M}
revealed the spiral structure of our Galaxy.  Gum's 1955
catalogue \citep{1955MmRAS..67..155G} was furthered by 
\citet{1960MNRAS.121..103R} at Mount Stromlo, and this is Australia's 
most-cited article in the following 1956--1960 time interval (Table~\ref{Tab56-60}).
While working with J.L.\ Pawsey, F.J.\ Kerr, and others, Gum also defined the
new IAU System of Galactic Coordinates \citep{1960MNRAS.121..123B}. 


Possibly due to the stress of his Ph.D.\ (awarded in 1955), Gum 
suffered a nervous breakdown and his thesis was initially rejected.
Tragically, Gum died aged just 36 in a skiing accident at Zermatt in
Switzerland (28 April 1960).  Employed by the University of Sydney in 1959,
after three years with CSIRO's Division of Radiophysics, he had been in Europe
looking for a 36-inch telescope design to bring to Australia.  However, the plan by
Harry Messel at the School of Physics for a new optical observatory then had
to be canned.  The Gum crater on the Moon is named after Colin Gum. 
\begin{figure}[H]
\hfill
{\includegraphics[height=5.0cm]{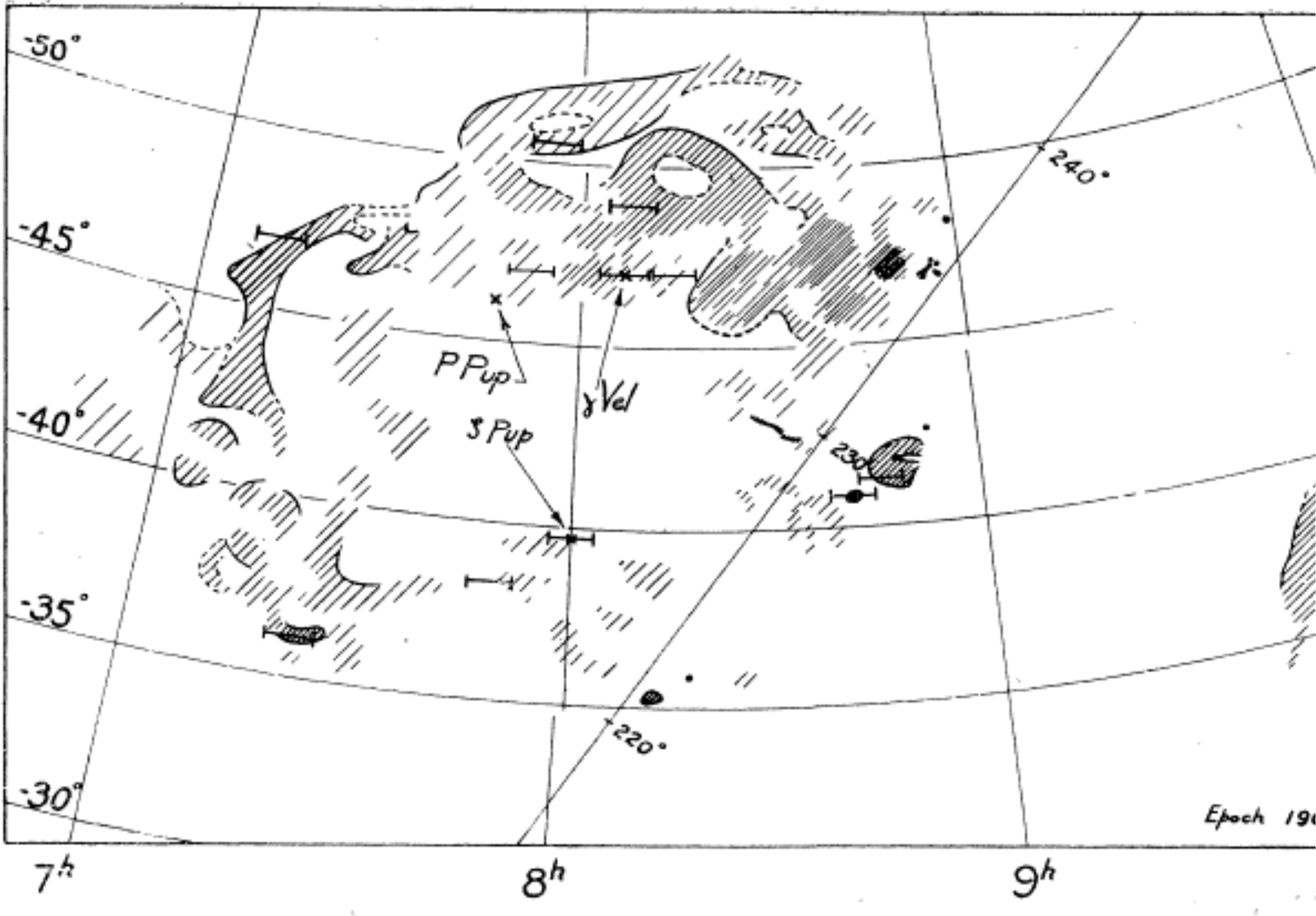}}
\hfill
{\includegraphics[height=5.0cm]{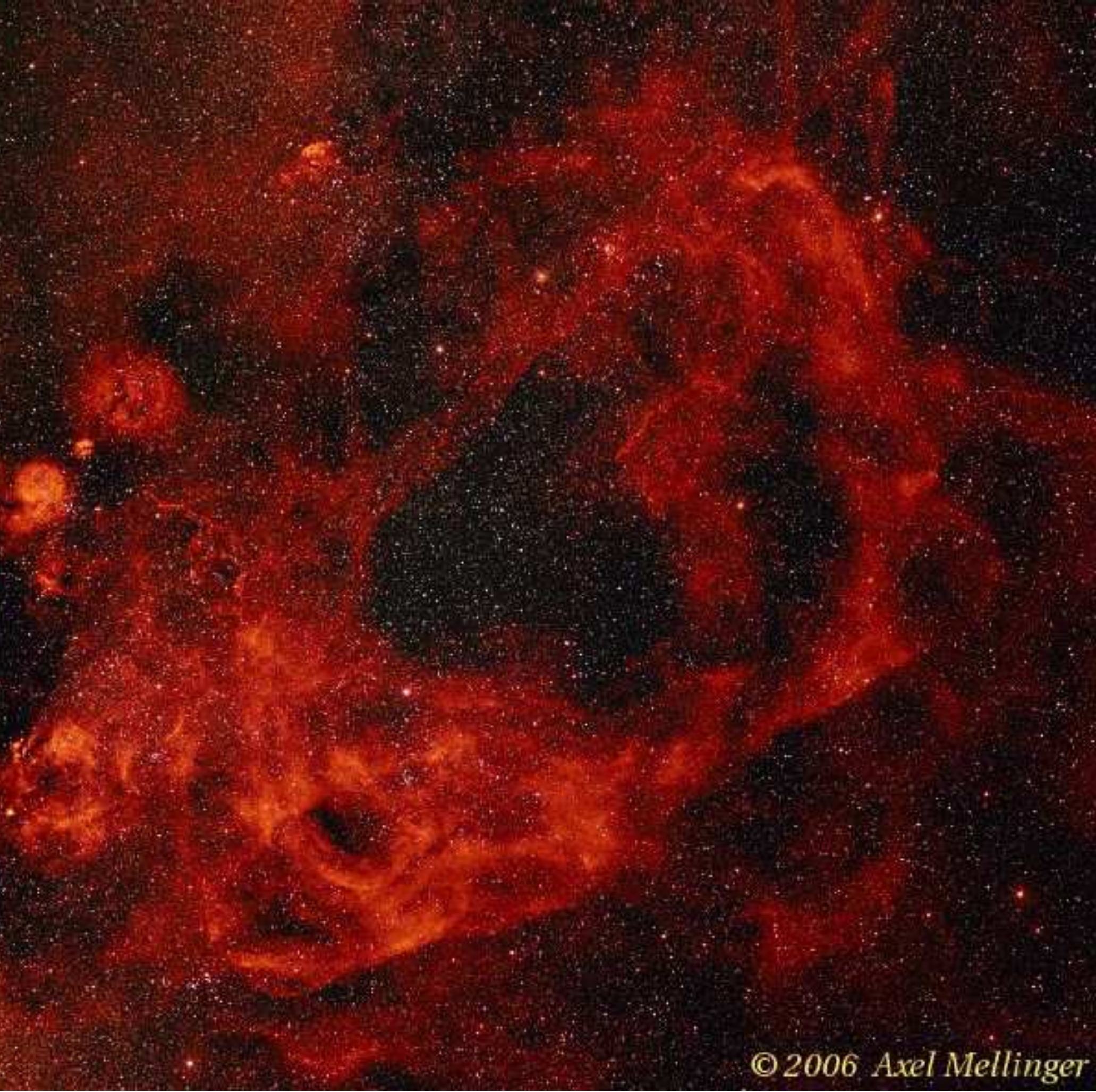}}
\hfill
\caption{The Gum Nebula is not named after a similarly-shaped Australian gum
  leaf, but rather honours its discover Colin S.\ Gum.  Some see a love heart 
  while others see a face in this image. 
Left panel: Sketch from Gum's 1952 paper. 
Right panel: $H\alpha$ image taken by Prof.\ Axel Mellinger, in 2006, with both the horizontal and vertical axis flipped.
\label{Fig_Gum}
}
\end{figure}



Sydney-born {\bf Ronald Newbold Bracewell} (1921--2007:
\citet{2017fpra.book.....F}, \mbox{\citet{2009BAAS...41..566P}}, \citet{2010JAHH...13..172T}) had the
same traditional `educational trifecta' as many of the astronomers listed
here.  He went to Sydney Boys High School, then the University of Sydney, and
then Cambridge, England, for his doctorate.  He was a major contributor to not
just radio interferometry but also medical scanning.  His signal processing
techniques and mathematical methods of image analysis are now applied to
medical images and X-ray tomography.  During WWII, and while pursuing his
engineering degrees from the University of Sydney, Bracewell also worked for
J.L.\ Pawsey and H.C.\ Minnett by helping to develop radar at microwave
wavelengths for the Royal Australian Navy before then spending 1946--1949 at
Cambridge obtaining his Ph.D.\ in ionospheric research.  Bracewell's ``strip
integration'' article \citep{1956AuJPh...9..198B} (Table~\ref{Tab56-60})
explains how Fourier Transforms \citep{1822:Fourier} can be used to create a
two-dimensional image from a series of one-dimensional scans taken with
different positional angles; a technique later adapted to
(medical\footnote{Bracewell is not the only astronomer from this era to have
  links with the medical profession. With an M.D.\ from Cornell University in
  1925, and a Ph.D.\ from Princeton University in 1927, US-born Theodore (Ted)
  Dunham Jr (1897--1984: \citet{1984ST....68R.203.}) co-discovered, through the
  use of optical spectroscopy, that the atmosphere of Venus contains large
  quantities of pressurised CO$_2$ and is thus not like Earth's atmosphere ---
  ending decades of speculation \citep{1932PASP...44..243A}). Physicist and
  physician, from 1946 to 1957 he studied the spectrophotometry of cells in
  medicine and surgery at Harvard University and the University of Rochester,
  before joining Mount Stromlo Observatory in 1957, where he helped Ted Dunham
  Junior develop the coud\'e spectrograph for their new 74-inch telescope
  \citep{1962AJ.....67..114D}.  Dunham then used this to study the spectrum of
  $\eta$ Carinae \citep{1966ApJ...146..126A} and other stars, and from 1965 to
  1970 he worked at The University of Tasmania.})  tomography.  His 1954
``aerial smoothing'' article \citep{1954AuJPh...7..615B}
(Table~\ref{Tab51-55}) also explained, using Fourier Transforms, how to
restore the true profile of a source from an antenna scan, and, importantly,
what information was lost and could not be recovered.  He was an exceptionally
inventive, capable, and active scientist who was lauded with many prizes and
medals during his career.  In 1998 he was named an officer of the Order of
Australia, in recognition of his contributions to radio astronomy and image
reconstruction.

Bracewell's successful 1955 book with Pawsey (titled ``Radio Astronomy'') 
undoubtedly contributed to his delivery of lecture courses in the USA at that time,
and he subsequently became a Faculty member at Stanford University's
Electrical Engineering department in 1955 until he retired from teaching
in 1991.  In 1965, albeit now at Stanford, he wrote a book called ``The
Fourier Transform and its Applications'' \citep{1965ftia.book.....B}, 
which has become a standard 
reference work, with updated editions published in 1986 and again in 2000.
During the lunar missions and the first landing on the Moon, Bracewell's
USA-based 32-dish solar spectroheliograph, which he built upon arriving in
Stanford, was used by NASA to keep a daily look-out for dangerous solar
outbursts.  Given the onset of the space-race, and Bracewell's observations of
Sputnik 1, Bracewell \citep{1960Natur.186..670B} famously speculated that alien probes 
might have been sent into the habitable zones of around a thousand stars nearby 
the alien civilisation's home star---to establish radio communications with
other civilisations.  Bracewell confidently wrote, over half a century ago,
that ``Our impending contact cannot be expected to be the first of its kind;
rather it will be our induction into the chain of superior communities''. 
He publicised his ideas further in his 1975 monograph titled ``The Galactic
Club: Intelligent life in outer space'' \citep{1975gcil.book.....B}. 
His hypothesised messengers have became known as ``Bracewell Probes''.

After obtaining a Ph.D.\ under Fred Hoyle at Cambridge, {\bf James (Jim)
  Alfred Roberts} (1927--: \citet{JimRoberts}) worked in the Division of
Radiophysics from 1952 to 1958.  During that time he co-wrote the important
``aerial smoothing'' paper \citep{1954AuJPh...7..615B} with R.N.\ Bracewell 
and characterised solar bursts of spectral Type II as observed from Dapto 
\citep{2009:Stewart} from
1949 to 1957 \citep{1959AuJPh..12..327R}.  Among other things, he noted that
half of the bursts contained harmonic structure, that they could be observed
from near the solar limb and not just the solar centre (with scattering and
refraction implications), and that geomagnetic activity occurred two days
later and thus implied particle travel speeds of 1000 km s$^{-1}$.  He then
spent some years at Caltech, Owens Valley, Arecibo, and as a visiting
Professor at the University of Toronto before returning to Parkes where he
continued to make significant contributions.

With an engineering degree from the University of Tasmania, {\bf John
  D.\ Murray} joined CSIR in 1948 and remained a scientific staff member until
1989.  After assisting with the Penrith radio-spectrograph in the late 1940s
\citep{2010JAHH...13....2S}, he helped technical assistant {\bf William (Bill)
  C.\ Rowe} (deceased 1954) build both the 40--240 MHz receivers and display
for the Dapto radio-spectrograph, and then shared the observing with W.C.\ Rowe
and J.P.\ Wild.  Together they observed a Type II burst with a (first and
second) harmonic structure \citep{1954AuJPh...7..439W}.  A corner of the West
Pennant Hills' orchard belonging to John Murray's father, known as {\it
  Rosebank}, was later to become the {\it Murraybank} Field Station
\citep{2011ASSP...23..433W} after increasing radio noise levels at Potts Hill
\citep{2005JAHH....8...87D, 2011ASSP...23..379W} required them to move their
receivers further out of Sydney \citep{2005ASSL..334..119O}.


\clearpage


{\bf Father Daniel Joseph Kelly O'Connell, S.J.}\footnote{S.J.\ = Society of
  Jesus} (1896--1982: \citet{1982IrAJ...15..347B}) was born in England, studied
in Ireland and the USA, and was the Director of the Jesuit Riverview College
Observatory in Sydney from 1938 to 1952, before becoming the Director at the
Vatican Observatory for the next 18 years.  He specialised in variable stars,
and his 1951 article \citep{1951PRCO....2...85O} revealed that the difference
in brightness between the maxima, on either side of the principal minimum, in
eclipsing binary stars is {\em not} a `periastron effect' as was previously
thought.  His monitoring of the sky from Sydney led to
discovering the 1941 brightening of $\eta$ Carinae.  Father O'Connell is
responsible for the large telescopes in the Barberini Gardens at Castel
Gandolfo, upgrading from the smaller telescopes on the Papal Palace's roof.

The three authors of the final article in Table~\ref{Tab51-55} are familiar 
names, and they appear again in Table~\ref{Tab56-60}.  We are introduced to
them in the following subsection.

\subsection{1956--1960}\label{Sec_56-60}
Following the 1955 catalogue of Gum \citep{1955MmRAS..67..155G}, the paper by 
\citet{1960MNRAS.121..103R} became the standard
``RCW'' reference for (182) southern HII regions.  This paper had a significant
impact in cementing Australia's reputation as a significant player in astronomy.
{\bf Alexander (Alex) William Rodgers} (1932--1997: \citet{1997SouSt..37..194F,
  1998BAAS...30.1464F}) from Newcastle, NSW, passed away just weeks before his
planned retirement from Mount Stromlo Observatory where he had presided as
Director from 1986 to 1992.  Among other works, Rodgers is also known for his
spectrophotometry of $\eta$ Carinae \citep{1967MNRAS.135...99R}. 
%
%
After obtaining an undergraduate degree from the University of Sydney, Rodgers
started at Mount Stromlo as a Ph.D.\ student of Woolley's in 1954, and he
graduated in 1958. After a short stint at the Carnegie Observatories in
Pasadena, and at the Royal Greenwich Observatory in the UK with R.\ Woolley,
from 1959 onwards, Rodgers spent his career at Mount Stromlo.  {\bf Colin
  T.\ Campbell} was one of Bart Bok's first Ph.D.\ students, although, after
two years, Campbell decided to leave.   
{\bf John Bartlett Whiteoak}\footnote{John Bartlett Whiteoak should not be
  confused with Jonathan Bartlett Z.\ Whiteoak, who obtained his Ph.D.\ from
  the University of Sydney in 1994 and published the `MOST supernova remnant
  catalogue' (MSC: \citep{1996AAS..118..329W}).} (1937--) was also an early
student of Bok's, graduating in 1962.  After a stint at Caltech, where he
first co-authored with V.\ Radhakrishnan (see later), Whiteoak joined the
Division of Radiophysics in 1965 and was the Deputy Director of the ATNF from
1989 until retirement in 2001.  In addition to his scientific discoveries,
particularly his work on radio polarisation (for
  example, \citep{1966ARAA...4..245G}), he is known for successfully campaigning for
greater protection of the high-frequency (71--275 GHz) radio bands for
astronomical~observations.

\textls[-15]{Born in India and educated in England, {\bf Richard Quentin Twiss}
(1920--2005: Tango~\cite{2006AG....47d..38.}) is known for the ``Hanbury Brown and
Twiss effect'' \citep{1956Natur.178.1046H, 1957RSPSA.242..300B}, which led to the
creation of the optical intensity interferometer.\footnote{The technique was
  previously established and used at radio wavelengths to measure the
  structure of radio sources at very high spatial
  resolution \citep{1952Natur.170.1061H}.}  They built the first in the
UK in 1954. Despite much early skepticism from the scientific community as to its
feasibility, it was famously used to measure
(the optical size of the star) Sirius A \citep{1956Natur.178.1046H}.  Mathematician and electronics
engineer Twiss later helped to construct the Narrabri Stellar Intensity
Interferometer \citep{1967MNRAS.137..375H}, which came online in 1965.  Twiss is
also highly cited for his paper \citep{1958AuJPh..11..564T} about coherent
electron cyclotron maser emission, seen in, for example, Jupiter's decimetric
radio emission.   
Like Twiss, {\bf Robert Hanbury Brown} (1916--2002: \citet{1991bpse.book.....H,
  Davis:2002}) was born in India and grew up in England, but unlike the very
English Twiss, Hanbury Brown reportedly loved Australia.  He was a physicist
and astronomer who helped develop radar and establish radio astronomy in the
UK.  After constructing the (Hanbury Brown)---Twiss intensity interferometer in
the UK, he came to Australia in 1962 (where he stayed for 27 years).  He
co-built and used the Narrabri Stellar Intensity Interferometer to measure the
diameters of stars \citep{1967MNRAS.137..393H, 1974MNRAS.167..121H}, 
enabling the calibration of the temperatures of stars hotter than the Sun.  In
1975 he began work on the Sydney University Stellar Interferometer (SUSI),
which came online shortly after he retired in 1981.  In 1986 he received the
Companion of the Order of Australia honour before returning to England in
1989.}

{\bf Sidney Charles Bartholemew (Ben) Gascoigne}
(1915--2010: \citet{2010Obs...130..274W}) from
Napier, New Zealand, came to Canberra in 1941 after undertaking a Ph.D.\ in
England.  He was the first Vice-President of the Astronomical Society of
Australia \citep{Lomb:2015}, and he remained at Mount Stromlo Observatory 
until his retirement in 1980.  Initially, 
he worked on the design of gun-sights along with Walter Stibbs at the CSO.
His 1952 paper with {\bf G.E.\ Kron}\footnote{Kron was a (highly successful)
  American-astronomer, on an extended visit to Mount Stromlo
  Observatory in 1951, and was later a Senior Research Fellow at Mount
  Stromlo from 1974 to 1976.} (1912--2012) reported the discovery of young
``populous clusters\footnote{In terms of star formation and luminosity,
  ``populous clusters'' reside between open star clusters and super star
  clusters.}'' in the Magellanic Clouds, revealing that the Clouds had not
strictly evolved in parallel with the Milky Way.  They also found blue cepheid
variables in the Small Magellanic Cloud, which led to a doubling of its
distance from us, which confirmed a doubling of the Universe's `distance
scale' \citep{1944ApJ...100..137B, 1956PASP...68....5B}. 
Together with the US-colleagues {\bf H.S.\ White} 
and {\bf G.E.\ Kron}---who had brought
over a photoelectric device from the Lick Observatory (USA) in 1953---they
measured the first near-infrared magnitudes of stars in the Southern
Hemisphere \citep{1953ApJ...118..502K, 1957AJ.....62..205K}.  ``Gascoigne's
leap'' at the AAT remembers the site where Gascoigne (former joint Project
Scientist, then senior commissioning astronomer, for the AAT) fell some six 
meters in 1974 from the interior walkway to the observation floor, breaking
his arm.\footnote{In the early 1990s, Dr Brett Wells also fell and injured
  himself badly from within the darkness of the AAT dome.}


Using the Mills cross radio telescope, from 1958 to 1961, Mills, Slee \& Hill
(MSH) published several catalogues \citep{1958AuJPh..11..360M,
  1960AuJPh..13..676M, 1961AuJPh..14..497M} of radio sources at 3.5 m
wavelengths (and they also mapped the Galactic plane at this
wavelength \citet{1958AuJPh..11..530H}).  In their useful catalogs, they
reported 1159, 892, and 219 radio sources within declinations of $+10$ to
$-20$, $-20$ to $-50$, and $-50$ to $-80$ degrees.  Like so many other
astronomers mentioned here, {\bf Bernard (Bernie) Yarnton Mills}
(1920--2011: \citet{2012AG....53b..19B, 2013BMFRS..59..215F, 2017fpra.book.....F}) from Manly obtained
his undergraduate degree from the University of Sydney.  Equipped with this,
in 1942, he started to help develop radar receivers and displays under Joe
Pawsey at CSIR, and he went on to obtain his Masters of Engineering in 1950 and
Doctor of Science in Engineering in 1959.  He is known for designing and
building the 450 m long ``Mills Cross telescope'' \citep{1956AJ.....61..167B}
that was at Fleurs---now called Badgery's Creek where Sydney's second
airport is under construction---in which the fan beams of two long antennae at
right-angles to each other were combined to produce a (49 arcmin) pencil
beam \citep{1953AuJPh...6..272M}. Left without adequate CSIRO funding---which
was committed to the Parkes radio telescope and the Culgoora Radioheliograph---in 1960, he and Wilbur Norman ``Chris'' Christiansen
(1913--2007: \citet{2011:FraterGoss, 2009JAHH...12...11O}) of the ``Chris Cross
Telescope'' left CSIRO and joined the staff at the University of
Sydney.  
In 1967 they completed the ``Super Cross'' (aka One-Mile Mills
Cross) telescope with 1.6 km arms out near Bungendore, 30 kilometres east
of Canberra.  In 1978 this telescope's north--south arm was removed, and the
east--west arm reconfigured to give us the Molonglo Observatory Synthesis
Telescope (MOST), which operates today, albeit with various
upgrades \citep{2008JAHH...11...63M, 2017PASA...34...45B}.  The Molonglo Reference
Catalogue \citep{1981MNRAS.194..693L} identified 12,000 radio sources, a
five-fold increase on the MSH 
survey.  
As we have already encountered O.B.\ Slee in Section \ref{Sec_45-50}, we
quickly note here that {\bf Eric R.\ Hill} obtained his physics degree from
the University of Melbourne in 1951.  He then joined the CSIRO group in
Sydney. From 1957 to 1971, he published some 20 articles after a traineeship at
the Leiden Observatory (1953--1956), which resulted in his derivation of our
Galaxy's gravitational field perpendicular to the plane
\citep{1960BAN....15....1H}. 


We have now met the first eleven authors from Table~\ref{Tab56-60}, 
and having already met Piddington, Wild, and Bracewell, this leaves
Sheridan, Neylan, and Ringwood. 

In 1959 Wild, Sheridan \& Neylan revealed that the coronal plasma oscillations
responsible for Type III radio bursts were excited by flares moving at half
the speed of light, and faster \citep{1959AuJPh..12..369W}.  These flares
launched protons and electrons up into, and out of, the corona, with some of
the electrons trapped by the corona's magnetic field lines subsequently
emitting `Type IV' synchrotron continuum 
radiation.
{\bf Kevin V.\ Sheridan} (1918--2010) was born in Brisbane.
He joined CSIR in 1945, working on the Mills Cross radio telescope, and later 
engaged in radio astronomy research in 1953 after 
obtaining his B.Sc.\ degree from the University of Queensland.  At Dapto, he
was the chief electronics engineer and widely acknowledged by J.P.\ Wild as
the ``man who put it all together''. 
Sheridan later became the solar group leader at Culgoora from 1975 until
retiring in 1984.
%
%
As a Mount Stromlo Ph.D.\ student, {\bf Tony Neylan} spent much of 1958 at
Dapto, thanks to a CSIRO Research Grant, helping with the observations and
analysis of the radio-interferometer data concerning the coronal heights
of Type III solar radio bursts and the associated Type IV emission.  He later
dropped out of the Ph.D.\ program to join the Diplomatic
Service \citep{1984PASAu...5..608W}, and apparently~\citep{2009:Stewart} after
that became a Jesuit Priest like Father O'Connell.

{\bf Alfred Edward (Ted) Ringwood} (1930--1993: \citet{1993Natur.366..509G,
  1994Metic..29..290W, 1996PEPI...96...77I}) was born in Kew, Victoria, 
just up the road from what was then the Swinburne Technical College. 
He 
obtained his Ph.D.\ in Geology from Melbourne University in 1956.  He was a
geochemist and geophysicist who spent most of his career at The ANU's
Department of Geophysics, which in 1964 became the Department of Geophysics
and Geochemistry while still under the Directorship of J.C.\ Jaeger, and then
became the Research School of Earth Sciences in 1973 (with Ringwood the
Director from 1978 to 1983).  Ringwood has an impressive list of honours and
awards \citep{1998:Green}, and it is not surprising that his (self-explanatory
titled) works appear in not just the table for 1956--1960, but also for
1961--1965 and then again for 1966--1969.  In addition to his 1959 article about
the chemical evolution and densities of the planets \citep{1959GeCoA..15..257R},
his 1961 article about the chemistry of meteorites
\citep{1961GeCoA..24..159R}, 
and his 1966 article about the chemistry of the
planets \citep{1966GeCoA..30...41R}, Ringwood has two other equally well-cited
articles from the 1960s which we have not included because, pertaining to the
Earth's mantle, they are more of a geological than astronomical nature.
Ringwood's substantial international recognition helped our nation's
reputation as a key contributor in the field of high pressure petrology,
geochemistry, planetary and meteoric mineralogy.  His analysis of Moon rocks
returned from the Apollo missions led to his recognition of their similarity
with Earth and the idea that the Moon was in some way derived from the
Earth's mantle \citep{1970JGR....75.6453R, 1979oem..book.....R}.  The mineral
``ringwoodite'', discovered in fragments of the Tenham meteorite, contains a
specific structure previously predicted to exist by Ringwood and, as such, now
reflects his name.  Since 2012, Ted Ringwood is additionally honoured through
the Ringwood Medal, administered by the Geological Society of Australia.




\subsection{1961--1965}
From 1956 to 1961, Japanese {\bf Tetsuo Hamada} was a postdoctoral fellow at
the University of Sydney's School of Physics, where he authored the
important paper with (the then visiting) astronomer E.E.\ Salpeter,
introducing the ``Hamada--Salpeter White Dwarf
Model'' \citep{1961ApJ...134..683H}.  Little more than a decade after WWII had
ended with the dropping of nuclear bombs on Japan, Hamada's appointment to
work on matters related to nuclear reactions revealed how science can often
transcend politics.  Properly allowing for inverse $\beta$ decays at high
densities, where electrons tunnel into nuclei and create neutrons, Hamada
and Salpeter developed an improved mass-radius relation for white dwarf stars:
dead stars supported by the degeneracy pressure of the free electrons.
Salpeter further explored what happens once the thermal energy in a plasma/star is zero \citep{1961ApJ...134..669S}.  Working in the Daily Telegraph
Theoretical Department, Hamada was as much a nuclear physicist (studying the
interior of neutron stars and white dwarfs) as an astronomer.  He is 
widely known for the ``Hamada-Johnston potential model''.  This analytical
expression for an energy-independent nucleon-nucleon
potential \citep{1962NucPh..34..382H} was developed using Australia's second
electronic computer\footnote{Following CSIRAC \citep{Willis:2006}, SILLIAC was
  officially opened in 1956 by Adolph Basser who had previously donated 50,000
  pounds (the prize money of his 1951 Melbourne Cup winner, Delta) for its
  construction, and then another 50,000 pounds in 1956 for its upgrade.}, with
vacuum tubes that apparently filled a room, at a time when Japan had no
electronic computers.  This latter Australian-based
paper \citep{1962NucPh..34..382H} has been cited 900 times.  In 1962 Hamada
relocated to Ibaraki University, Mito, Japan, where he eventually became
University President from 1988 to 1992.
\startlandscape

\begin{specialtable}[H] 
\widetable
\setlength{\tabcolsep}{7mm}
\caption{The ten (plus 1) most cited articles from 1961 to 1965
  are numbered in column 1, and involve 19 distinct authors. Ten distinct
  authors can be found above the horizontal line.\label{Tab61-65}}
\begin{tabular}{llllclc}
\toprule
\textbf{\#} & \textbf{Title}  &  \textbf{Affiliation}  &  \textbf{Author(s)}  &  \textbf{Year}  &  \textbf{Journal}  &  \textbf{Cites} \\  
   &        &               &             &        & \textbf{Vol.\ Page} &         \\
\midrule
1. & Models for zero-temperature stars                         &  Sydney Univ., Cornell Univ.\            &  Hamada, T.                     &  1961  &  ApJ      & (595)  \\  
   &                                                           & \hspace{2mm} \& Cambridge (UK)           & \hspace{2mm} \& Salpeter, E.E.  &        & 134, 683  &        \\  
2. & Solar Bursts                                              &  CSIRO-Radiophysics                    &  Wild, J.P., Smerd, S.F.        &  1963  &  ARA\&A   & (382)  \\  
   &                                                           &                                          & \hspace{2mm} \& Weiss, A.A.     &        & 1, 291    &        \\  
3. & Energy and Pressure of a Zero-                            &  Cornell Univ., Sydney Univ.             & Salpeter, E.E.                  &  1961  &  ApJ      & (327)  \\
   & \hspace{5mm} Temperature Plasma                           & \hspace{2mm} \& Cambridge (UK)           &                                 &        & 134, 669  &        \\
4. & Variation of Radio Star and Satellite                     &  Univ.\ of Adelaide:                     & Briggs, B.H.\                   &  1963  &  JATP     & (172)  \\
   & \hspace{5mm} Scintillations with zenith angle             & \hspace{2mm} Dept.\ of Physics           & \hspace{2mm} \& Parkin, I.A.    &        & 25, 339   &        \\
5. & Investigation of the Radio Source                         &  Univ.\ of Sydney                        & Hazard, C.                      &  1963  & Nature    & (136)  \\
   & \hspace{5mm} 3C~273 by the Method of                      & \hspace{2mm} \& CSIRO-Radiophysics     & \hspace{2mm} \& Mackey, M.B.    &        & 197, 1037 &        \\
   & \hspace{5mm} Lunar Occultations                           &                                          & \hspace{2mm} \& Shimmins, A.J.  &        &           &        \\
\midrule 
6. & A study of the decimetric emission                        &  CSIRO-Radiophysics                    &  Cooper, B.F.C.                 &  1965  &  AuJPh    &  (118)  \\  
   & \hspace{5mm} and polarization of Centaurus A              & \hspace{2mm}\& Mount Stromlo             & Price, R.M., Cole, D.J.         &        & 18, 589   &         \\  
6. & The Parkes catalogue of radio sources,                    &  CSIRO-Radiophysics                    & Bolton, J., Gardner, F.         &  1964  &  AuJPh    &  (118)  \\  
   & \hspace{5mm} declination zone $-$20$^\circ$ to $-$60$^\circ$  &                                         & \hspace{2mm} \& Mackey, M.B..   &        & 17, 340   &         \\  
7. & Galactic Velocity Models and the                          &  CSIRO-Radiophysics                    &  Kerr, F.J.                     &  1962  &  MNRAS    &  (110)  \\  
   & \hspace{5mm} interpretation of 21-cm surveys              &                                          &                                 &        & 123, 327  &         \\  
8. & Population I in the Large Magellanic                      &  MSO \&                                  & Westerlund, B.                  &  1961a &  UppAn    &  (100) \\
   & \hspace{5mm} cloud                                        & Uppsala Southern Station                 &                                 &        & 5, 1      &        \\
9. & Observations of Jupiter's radio                           &  CSIRO-Radiophysics                    & Roberts, J.A.\ \&               &  1965  &  Icarus   & (96) \\
   & \hspace{5mm} spectrum and polarization in                 &                                          & Komesaroff, M.M.                &        & 4(2), 127 &       \\
   & \hspace{5mm} the range from 6 cm to 100 cm                &                                          &                                 &        &           &        \\
10.& Chemical and genetic relationships                        & The ANU:                                 & Ringwood, A.E.                  &  1961  &  GeCoA    &  (87)  \\  
   & \hspace{5mm} among meteorites                             &  \hspace{2mm} Dept.\ of Geophysics       &                                 &        &  24, 159  &         \\  
\bottomrule
\end{tabular}
\end{specialtable}
 
\finishlandscape

Having transferred from The University of Adelaide, where he had made radar
observations of meteors and published about our ionosphere since 1953,
Captain\footnote{During WWII, Captain Weiss was enlisted on 17 November 1942.}
{\bf Alan Austin Weiss} (1917--1964), born in Monreith, South Australia, came to
Sydney around 1960 to supervise the analysis of the Dapto solar data.  He
wrote papers regarding Type II, III, and IV radio bursts, and working with
J.P.\ Wild and S.F.\ Smerd, in 1963, he co-authored a detailed review article
about the properties of solar bursts \citep{1963ARAA...1..291W} before he died
from leukaemia, as had W.C.\ Rowe roughly a decade earlier at Dapto.
His work built on the 1942 discovery of radio bursts from sunspots 
made during WWII when such radar findings 
had to be kept secret \citep[e.g.,][and references therein]{1946Natur.157...47H}. 

British-born {\bf Basil Hugh Briggs} (1923--1994: \citet{1994JATP...56.1533E}) 
obtained his Ph.D.\ from Cambridge in 1952 and worked with the Cavendish
Laboratory's Radio Research Group from 1946 to 1961.  While there, he explored
the fading of radio waves reflected from an irregular ionosphere and developed
a formula to determine the horizontal motion of the lower ionosphere using
radar reflections measured by separated ground-based
antennas \citep{1950PPSB...63..106B}. In 1962 he joined The University of
Adelaide, where he published papers for the next three decades, including
work on the scatter and scintillation of radio waves due to irregularities in
the ionosphere \citep{1963JATP...25..339B}. In the early 60s, Briggs faced the
same problem as B.Y.\ Mills (who built the Molonglo telescope).  CSIRO 
was committed to building two radio facilities (Parkes and the Culgoora
Radioheliograph) and could not support others.  However, The University of
Adelaide was able to fund the construction of the Buckland Park Array, a
$1\times1$ square kilometre array of 89 crossed dipoles that became the
world's largest low-frequency radio telescope.  Rather than studying the
heavens, it provided, and provides, detailed observations of the lower
ionosphere.   
In 1968, co-author {\bf I.A.\ Parkin} published a second paper on radio
scintillation through the ionosphere before apparently leaving the University
of Adelaide to join the Radio and Space Research Station of the Science
Research Council in the UK.  
As his papers' titles reveal, his work
had implications for satellite communications (military defense,
navigational, and civilian~purposes).

Using the Parkes telescope and the Moon as a screen, in 1963, {\bf Hazard,
  Mackey \& Shimmins} reported \citep{1963Natur.197.1037H} an accurate position
for the 'radio star' 3C~273.  This method was not the first time something like this
had been done.  For example, \citet{1962MNRAS.124..343H}
had previously reported the position of 3C~212 to an accuracy of
3$^{\prime\prime}$, and \citep{1963ApJ...138...30M} had already 
taken optical spectra of other 3C `radio
stars' whose optical counterpart had been identified.\footnote{A.R.\ Sandage
  did not believe that these objects were galaxies, but instead stars, because
  their brightness varied on short time scales.}  However, 
\citet{1963Natur.197.1037H} were fortunate in that Maarten Schmidt became
involved with 3C~273.  In an adjoining Nature article, 
\citet{1963Natur.197.1040S} reported on his realisation that his optical spectra
associated with the star-like object 3C~273 had a cosmologically-significant
redshift of 0.158, that is, 3C~273 was receding at 1/6th the speed of light.
The first quasar\footnote{The term `quasar' appears to have been introduced in
  1964 by \citet{1964PhT....17e..21C} to refer to ``quasi-stellar radio
  sources''.  The term QSO was later introduced to refer to
  optically-identified `quasi-stellar objects' that were not detected at
  radio wavelengths.} had been identified, the steady-state model of the
Universe was dead, and our known size of the Universe had dramatically
increased.  

\textls[-15]{Furthermore, the path to the discovery of supermassive black
holes\footnote{\citet{2014JAHH...17..267K} and \citet{2016ASSL..418..263G} extensively review the 
historical development of how we came to believe in black 
  holes.} had begun~\citep{2013BASI...41....1K,
  2018PASA...35....6H, 2019Galax...7...20B}.  
In 1963, the spectra of the previously studied 'radio
stars' were also realised to be highly redshifted, and much later work 
continued to push this frontier (for example, \citep{1982ApJ...260L..27P}). While more than
half a century on, the \citet{1963Natur.197.1037H} Nature article 
is still badged as a wholly CSIRO Division 
of Radiophysics publication.  However, Hazard was actually employed by the
University of Sydney in the Chatterton Astronomy Department within the School
of Physics.  The Department had recently been opened in 1961 to commission and
operate the Narrabri Stellar Intensity Interferometer.
Allegedly \citep{1993PASAu..10..355H}, an editorial blunder had affiliated
Hazard with CSIRO rather than with the
University of Sydney. However, it has since been noted that CSIRO 
changed the details of Hazard's affiliation \citep{2018PASA...35....6H}. }

%


British radio astronomer {\bf Cyril Hazard} obtained his Ph.D.\ at Jodrell Bank,
where he mastered the timing of radio sources disappearing behind the Moon,
and then reappearing, as a means to measure their position to arcsecond
accuracy.  He moved to the University of Sydney in 1961 to supervise the
construction of the Hanbury Brown-Twiss Stellar Interferometer at Narrabri.
Nowadays, Prof.\ Cyril Hazard has a long-term visitor status at the University
of Cambridge.  For his 3C~273 observation, 
the Parkes engineer and young physicist {\bf M.\ Brian Mackey} set up the observing system.
More than that, Mackey performed much of the data reduction and 
accounted for the Moon's rough, irregular circumference.  He left radio astronomy a
few years later, and little information could be found on him. 
{\bf Albert John Shimmins} 
(1921--2007), the then deputy director of Parkes in 1963, 
was responsible for the telescope's wonderful pointing accuracy, and 
according to Miller 
Goss\footnote{\url{ftp://ftp.aoc.nrao.edu/staff/mgoss/schmidt.cohen.4july09.txt} (accessed 15 February 2015).}, it
was J.G.\ Bolton who politely suggested to Hazard that Shimmins be included on the
article rather than Bolton himself.  Previously,    
from 1953 to 1961, Melbourne-born Shimmins had been employed by the
organisation that became the Weapons Research Establishment at Salisbury and
Woomera, where he worked on the electronic guidance for the Blue Streak
intercontinental missile and rocket tracking systems.  In 1961 he moved to
CSIRO's radiophysics division at Parkes and installed their tracking computer
and data recorder/processor. He was in the control room during the Apollo 11
lunar broadcast, and he led and co-authored many valuable radio source
catalogs.\footnote{These included works with 
  Beverley June Harris (later Wills) who obtained her Ph.D.\ from The ANU in
  1969, Margaret E.\ Clarke, Ronald (Ron) David Ekers, Richard (`Dick') Norman
  Manchester, and others.} 
Upon retiring in 1981, 
after having served as the Officer-in-Charge at Parkes from 1971 to 1981, the
University of Melbourne bestowed on them a Doctor of Science degree. 
He passed away\footnote{To quote from the considerable obituary in The Age
  newspaper (29/11/2007), written by Dr Robert Shanks,  ``... to observe him
  [Albert J.\ Shimmins] 
  walking to the shops in Albert Park or Windsor in his later life---an
  elderly man wih a walking stick and cap, careless of dress, carrying a bag
  with his few needs---conjured the visage of an eccentric.  However, Shimmins
  was far more.  He was a distinguished scholar and scientist.} 
in East Malvern, Melbourne, in 2007.  
`John' Shimmins is honoured through 'The Albert Shimmins Fund' that
he very generously established (by donating over \$14.5M) 
at Melbourne University, from where he had obtained
a Masters degree in electrical engineering in 1952.  The fund supports scholarships and
research activities in the Faculty of Science.
 


\textls[-15]{During a visit to Parkes\footnote{Permission to observe was approved by
  Radiophysics Chief Bowen (1911--1991: \citet{HanburyBrown:1992}), but this was
  apparently unbeknown to the Parkes Director Bolton.}, R.N.\ Bracewell (then
at Stanford) discovered that the 10 cm wavelength emission from Centaurus-A
was linearly polarised \citep{1962Natur.195.1289B, 2002JAHH....5..107B}, just as
it was at 3-cm \citep{1962AJ.....67Q.581M}. The day after Bracewell left the
Telescope, an enthusiastic R.M.\ Price repeated the observations using the 21
cm receiver, resulting in his first publication \citep{1962Natur.195.1084C,
  1984sdra.conf..300P, 2012arXiv1210.0986P} coming out in Nature before
Bracewell's article appeared.  Subsequent mapping of 
the polarisation at three different positions, and more importantly, at
different wavelengths, further revealed Faraday rotation from uniform
large-scale magnetic fields in either our Galaxy or Centaurus-A\footnote{
A historical article on Centaurus-A can be found in
\citet{2010PASA...27..402R}, with more recent results in
\citet{2019Galax...7...44D}, and future studies of cosmic magnetism 
reviewed in \citet{2020Galax...8...53H}.}~\citep{1965AuJPh..18..589C, 2019Galax...8....4B, 2019Galax...7...43O, 2019Galax...7...47S}. }

{\bf Brian F.C.\ Cooper} (1917--1999) obtained his B.Sc.\ and
B.Eng.\ (Electrical) from the University of Sydney in 1939 and 1941, after
which he joined the CSIR Radiophysics wartime effort to develop radar. 
Helping J.H.\ Piddington, he played
a major role introducing radar at Darwin after the Japanese air
raid in early 1942.  He later played a key role in helping to pioneer the
manufacture of transistors in Australia \citep{DaviesCooper:1953} before
developing receivers and correlators for the Parkes radio
telescope~\citep{1994ptyr.conf...44C}.  Cooper became the Head of the CSIRO
Division of Radiophysics' receiver group and has been described as something
of an unsung hero in the development of Australian radio astronomy and
transistors.

US-born {\bf Richard Marcus Price} arrived at the Radiophysics Division in
1961 as a Fullbright Scholar with a B.Sc.\ degree and graduated with a
radio-based Ph.D.\ from The ANU in 1966 under the supervision of Bart Bok
(Mount Stromlo) and J.G.\ Bolton (Parkes).  By 1969 Price had returned to the
USA, working at the Massachusetts Institute of Technology before becoming the
Head of the Astronomy Department at the Albuquerque's University of New
Mexico, where he worked from 1979 and is currently an emeritus professor.
During this time, he did, however, return to Australia and was the
Officer-in-Charge at the Parkes Observatory from 1994 to 1999.

The 1965 article with Cooper and Price appears to be the first from Engineer
{\bf Douglas (Doug) J.\ Cole} (deceased), who continued to publish under the
CSIRO affiliation until at least 1978.  In 1965,
Cole used the Parkes radio telescope to track the position and Doppler shift
of NASA's Mariner IV mission to Mars. Added to NASA's data, this
collectively provided the first close-up images of Mars, which proved not to
be the welcoming place that many had expected from earlier ground-based
observations.
For example, the drawings from Australia by Walter Frederick Gale (1865--1945: 
\citet{1946MNRAS.106Q..29.}) in 1892 suggested a land of oases.\footnote{The NASA-run 
Tidbinbilla Deep Space Communication Complex in Canberra subsequently played a key role in 2012
regarding the successful landing of the Mars rover Curiosity in `Gale Crater'.}
The pursuit of water on Mars continues to this day
\citep{2020Galax...8...40N}. 


{\bf Francis (Frank) Fredrick Gardner} (1924-2002: \citet{2005JAHH....8...33M}) 
was another bright spark from Sydney who graduated in 1946 from the University
of Sydney and then went to Cambridge University in the UK, where he obtained
his Ph.D.\ in 1953.  He had already started working at CSIRO's Division of
Radiophysics in 1950, where he retired in 1988.  He was a receiver
engineer who also became a radio astronomer through the development of
the Parkes radio telescope and specialising in polarisation
studies \citep{1966ARAA...4..245G} and astro-chemistry (molecules in space).
The paper by \citet{1964AuJPh..17..340B} titled ''The
Parkes catalogue of radio sources, declination zone $-$20$^\circ$ to $-$60$^\circ$'' served as a reference for the paper by \citet{1965AuJPh..18..627B} 
titled ``Identification of extragalactic radio sources
between declinations $-20^{\rm o}$ and $-44^{\rm o}$'', which just missed out
on entering into Table~\ref{Tab61-65}.  Both Ronald D.\ Ekers (on the
preceding paper with Margaret E.\ Clarke\footnote{Working with Tony Hewish,
  M.E.\ Clarke obtained her Ph.D.\ from Cambridge, where she had
  discovered the phenomenon of Interplanetary Scintillations.  She subsequently
  came to Parkes in the mid-1960s.})  and Jennifer A.\ Ekers\footnote{Prior to
  becoming an astronomer at Parkes, J.A.\ Ekers had obtained a degree in
  chemistry and conducted medical research.} (married to R.D.\ Ekers)
contributed to these ongoing radio survey papers \citep{1966AuJPh..19..559B,
  1969AuJPA...7....3E}. 


{\bf Frank John Kerr} (1918-2000: \citet{2000BAAS...32.1674W}) was born in St Albans,
England, to Australian parents who returned to Australia after World War I
had finished.  
He obtained his physics degree from the University of Melbourne, and if 
J.L.\ Pawsey is the father of radio astronomy in Australia, then Kerr is 
the father of 21 cm astronomy.  He established the Southern Hemisphere 21-cm line
program in the late 1940s and made sure that the surface of the Parkes 
radio telescope was built smoothly enough that observations down to 
10-cm wavelengths could be performed \citep{2009cnhe.book.....S}.  Well before
Parkes came 
online, Kerr made the first detection of a radio spectral line in an external
galaxy, namely HI in the Magellanic Clouds.  He also worked with G\'erard
de Vaucouleurs (who was at the Commonwealth Observatory) and others to 
determine the rotation and masses of the 
Magellanic Clouds~\citep{1955AuJPh...8..508K}. 
Kerr additionally mapped the Milky Way at 21-cm \citep{1962MNRAS.123..327K} and coined
the term ``Galactic Warp'' in 1956 to describe the gravitational effect of the
Magellanic Clouds on our Galaxy \citep{1957AJ.....62...93K}. 
Having helped to precisely determine the 
plane of the hydrogen gas in our Galaxy, his work became the basis of
our new galactic coordinate system \citep{1960MNRAS.121..132G}, 
while his other work strengthened the view that the Milky Way is a spiral
galaxy \citep{1958MNRAS.118..379O}.  He joined the Radiophysics
Lab in 1940, and from 1966 to 1987 he worked at the University of
Maryland, USA.  From 1986--1990 he discovered many optically 'hidden'
galaxies, due to obscuring dust, in the Milky Way's ``zone of
avoidance''.



{\bf Bengt E.\ Westerlund} (1921--2008: \citet{2008Msngr.133...58D}) obtained
his Ph.D.\ at the Uppsala Observatory, Sweden, in 1953.  After a teaching
position in France, from 1957 to 1967 he operated the new Schmidt telescope at
the ``Uppsala Southern Station'' at Mount Stromlo Observatory.\footnote{Due to
 Canberra's light pollution, the Uppsala Schmidt telescope was moved to
  Siding Spring Observatory in 1982.}  In 1967 he commenced work at the
Steward Observatory in the USA, and he later became the European Southern
Observatory's director in Chile in 1970.  Although his most cited article
(included in Table~\ref{Tab61-65}) pertains to Population I stars in the Large
Magellanic Cloud \citep{1961UppAn...5....1W}, he is also well known for the
article \citet{1961PASP...73...51W} titled ``A Heavily Reddened Cluster in
Ara'' in which he discovered what has come to be called ``Westerlund 1''.
Furthermore, known as the Ara Cluster, this is the most massive compact young star
cluster in our Galaxy and the Local Group.  Some think that it may evolve into
a Globular Cluster \citep{2002IAUS..207..745G}.  The cluster also contains
``Westerlund 1-26'': a red supergiant that is one of the largest stars we
know. 
The year 2015 marked the silver jubilee (25 years) for the Hubble Space
Telescope, and an HST image of ``Westerlund 2'' (another giant cluster
of some 3000 stars, \citet{1961ArA.....2..419W}) was recently used as an
official Hubble 25th Anniversary image (see
Figure~\ref{Fig_West}).\footnote{This obscured cluster resides in the Gum~29
  nebula, in the constellation Carina, also known as RCW~49.}  Moreover, the
\citet{1997macl.book.....W} textbook about the Magellanic Clouds is regarded
as something of a bible.  Upon his retirement, asteroid ``2902 Westerlund''
was named in his honour, and in 2004 he inaugurated the ``Westerlund
Telescope'' at the Uppsala Astronomical Observatory.
\begin{figure}[H]
\includegraphics[height=5.0cm,angle=0]{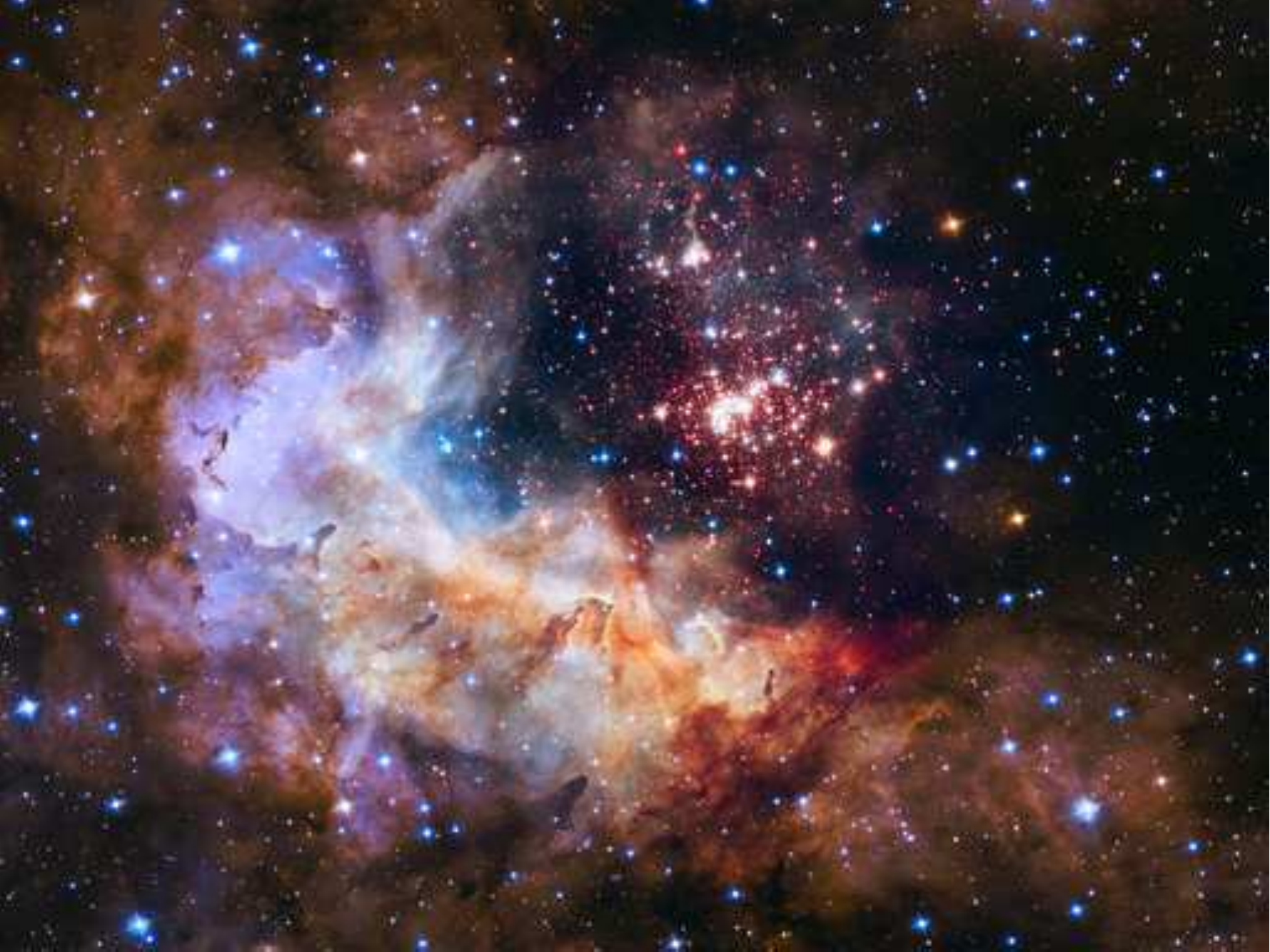}
\caption{Westerlund 2 in the Carina Nebula. 
Credit: NASA, ESA, the Hubble Heritage Team (STScI/AURA),
A. Nota (ESA/STScI), and the Westerlund 2 Science Team
\label{Fig_West} 
}
\end{figure}

Having met J.A.\ Roberts in Section~\ref{Sec_51-55} (and A.E.\ Ringwood in
Section~\ref{Sec_56-60}), we are left with {\bf Max M.\ Komesaroff} (deceased
1988), who reported on the radio polarisation of
Jupiter in \citet{1965Icar....4..127R}, exploring Jupiter's radiation belt.
Komesaroff joined the CSIRO group at Dapto in 1953, and he is also well known
for his many CSIRO-badged Nature articles about pulsars, such as: ``Spectral
Fine Structure in Pulsar Radiation''~\citep{1968Natur.220..358K}; ``Measurements
on the Period of the Pulsating Radio Source at 1919 +
21'' \citep{1968Natur.218..229R}; ``Evidence in Support of a Rotational Model
for the Pulsar PSR 0833-45'' \citep{1969Natur.221..443R}; and his theoretical
work on the ``polar cap'' model titled ``Possible Mechanism for the Pulsar Radio
Emission'' \citep{1970Natur.225..612K}.  This latter work built on that by 
\citet{1969Natur.221...25G} and \citet{1969ApJ...157..869G}, 
see \mbox{Section~\ref{Sec_66-69}}.  
\citet{1968Natur.220..358K} involved the pulsating radio source at
1919$+$21 (known as CP 1919, it was the first pulsar to be
discovered: \citet{1968Natur.217..709H}). A pen-chart recording of its pulses was
obtained at Parkes by Brian John Robinson (1930--2004:~\citet{Whiteoak:2006}) on 8
March 1968 \citep{1968Natur.218.1143R, 2003ASPC..302...23M}, and in 1973 this
appeared on the Australian \$50 paper note along with an image of the Parkes
radio telescope.


\subsection{1966--1969}\label{Sec_66-69}
The brilliant North American {\bf Peter Goldreich} (1939--) obtained his
Ph.D.\ from Cornell in 1963, under Thomas Gold (1920--2004).  In 1966 he
joined Caltech as an Associate Professor, and he was made a full professor in the
year that \citet{1969ApJ...157..869G} was
published, along with his sole-author Australian
publication \citet{1969PASAu...1..227G} on this same topic.  His most cited
article, 
Goldreich \& Julian was
written during an extended visit to the University of Sydney's School of
Physics in 1969.  It explored the physics of a rotating neutron star whose
magnetic dipole moment is aligned with the axis of rotation, and the
associated slowing of the rotation due to magnetic torques and energy losses
to surrounding particles.  {\bf William H.\ Julian} is a fellow North American
astronomer who arrived at Caltech the year after Goldreich, and after having
penned the important \citet{1966ApJ...146..810J} article 
regarding the dynamics and stability of thin, differentially rotating disc
galaxies in response to lumpy perturbers and the onset of spiral
structure\footnote{Their model suggests that spiral patterns are transient
  responses to a lumpy disc, at odds with the standing density wave theory of
  \citet{1964ApJ...140..646L} which is likely responsible for the
  `grand design' spirals.}.  Julian has since relocated and is now a Professor
Emeritus at the New Mexico State University.

{\bf Leo John Gleeson} (deceased 1979: \citet{1979PASAu...3Q.406W}) was a key
contributor to the theory of cosmic ray modulation.  He published a
tremendously influential paper \citep{1968ApJ...154.1011G} with W.I.\ Axford
just as he arrived in the Mathematics Department at Monash University.
Although in 1968 they were both affiliated with the University of California,
San Diego, Gleeson's ``Present address'' when the paper came out was Monash
University, hence its inclusion here (and on the Monash University's Applied
Mathematics web-pages) which some may debate.  The pair had two other very
well-cited articles in 1967 and 1968.  The first is titled ``Cosmic Rays in the Interplanetary
Medium'' \citep{1967ApJ...149L.115G} and has 300 citations but a Cornell
University affiliation in the USA, albeit at their ``Cornell-Sydney University
Astronomy Center''.  The second is titled ``The Compton-Getting Effect''
\citep{1968ApSS...2..431G} amd has a Monash University and a University of
California, San Diego, affiliation, but without enough citations (107) to make it
into Table~\ref{Tab66-69}.  Together with Prof.\ Robert Street (1920--2013), Dr
Denis Walter Coates, and Mr Robert Luke Bryant (Monash Physics Department),
Gleeson helped establish the Mount Burnett
Observatory in the Dandenong Ranges 
near Melbourne using a 16-inch reflector purchased from the Bendigo estate of
Mr L.\ Jeffree.

\textls[-15]{From Dannevirke, New Zealand, {\bf William Ian Axford, Sir}
(1933--2010:~\citet{2010AG....51c..37V}) 
%
%
was one of the pioneers of space plasma physics.  In addition to his
work on the origin and acceleration of cosmic rays by shocks, he explained
phenomena related to the solar heliosphere, explained how the solar wind couples
with the Earth's magnetosphere and ionosphere, and explained how the solar wind 
interacts with the interstellar medium \citep{1961CaJPh..39.1433A}. 
He became the Director at the Max Planck Institute for Aeronomy (later MPI for
Solar System Research) in 1974.  He was the Vice-Chancellor of Victoria
University, Wellington, from 1982 to 1985, before returning to MPI for
Solar System Research, where he retired in 2001.  New Zealand's Ian Axford
[exchange] Fellowship in Public Policy is named in his honour. }

{\bf Venkataraman Radhakrishnan (Rad)} (1929--2011) was from India, as were Twiss
and Hanbury Brown.  He was the son of the 1930 Nobel laureate physicist Sir
Chandrasekhara Venkata Raman (1888--1970)\footnote{The 1983 Nobel physicist
  Subrahmanyan Chandrasekhar was the nephew of Sir Chandrasekhara Venkata
  Raman.}, who is famous for ``Raman Scattering''.  After working at Caltech as a
Senior Research fellow from 1959 to 1964 (without a Ph.D.), Radhakrishnan
sailed across the Pacific in a 35-foot trimaran called\footnote{Cygnus~A was the first radio galaxy discovered \cite{1944ApJ...100..279R, 1946Natur.158..234H}.} Cygnus~A and
joined the CSIRO Division of Radiophysics as a Senior Research Scientist in
1965.  At Parkes, he found the radio pulse from pulsar PSR 0833-45 to be
$95\pm5$ percent polarised, and that the plane of polarisation rotated close to
90$^\circ$ during the pulse \citep{1969Natur.221..443R}, with his observations
at different frequencies \citep{1969ApL.....3..225R} leading him and D.J.~Cooke
to correctly hypothesise about the rotating magnetic fields and poles of the
pulsar.  This research followed similar work with rotating magnetic neutron stars by
\citet{1968Natur.218..731G} at Cornell University, and 
\citet{1969ApJ...157..869G} from Caltech and later Monash, and
Radhakrishnan's own work on the magnetic field and internal rotation of
Jupiter \citep{1960PhRvL...4..493R}. In another work, 
\citet{1969Natur.222..228R} reported on a `change of state' in this
pulsar, with the period abruptly decreasing by two parts per million.
Radhakrishnan is perhaps equally well known for his series of articles in 1972
titled ``The Parkes Survey of 21-Centimeter Absorption in Discrete-Source
Spectra''~\citep{1972ApJS...24..161R}.  This built on his 21-cm research portfolio
that he started at the Onsala observatory, Sweden, in the late1950s and
which he continued with Bolton in the early-1960s while they were at Caltech together.
After Radhakrishnan's father died in 1970, he left the Division of
Radiophysics and spent some of 1971--1972 at the Meudon Observatory, France,
before returning to India to become the Director of the Raman Research
Institute, which had been established by his father in Bangalore.  He held this
role from 1972 to 1994, where he successfully continued his research on pulsar
astronomy and liquid crystals.  He was also the IAU Vice-President from 1988
to 1994. Still without a Ph.D., in 1996 the University of Amsterdam awarded
him an honorary doctorate degree.

\textls[-15]{{\bf David (Dave) J.\ Cooke}, a co-author of pulsar
papers with Radhakrishnan 
\mbox{\citep{1969ApL.....3..225R, 1969Natur.221..443R, 1968Natur.218..229R}}, 
graduated with an engineering degree from The University of
Adelaide.  He worked at the Weapons Research Establishment at Salisbury in
South Australia (the base for the Woomera rocket range), where A.J.\ Shimmins
had worked, and at the Radar Research Establishment at Malvern in the UK.
Cooke joined the staff at Parkes in 1967, and in 1969 he was the Senior Radio
Receiver Engineer on duty during the Apollo 11 lunar landing broadcast around
the world, for which he received an award in Houston, Texas, in 2007 (along
with an Australia Day ``Stars of Australia'' award).  He was the
Officer-in-Charge of the Parkes Observatory from 1988 to 1993.}

{\bf Lindsey Fairfield Smith} attended the Women's College at the University
of Sydney---as had Joan Freeman (Section~\ref{Sec_45-50})---where she
obtained her B.Sc.\ degree (1958--1961) before then obtaining her Ph.D.\ from
The ANU.  She completed her thesis in 1966 and graduated in 1967.  In addition
to her two 1968 papers \citep{1968MNRAS.138..109S, 1968MNRAS.140..409S}  
from Mount Stromlo (listed in Table~\ref{Tab66-69}), in 
that same year, while at the University of Colorado, she published her third
important Wolf-Rayet star paper regarding the distribution of those stars in
our Galaxy \citep{1968MNRAS.141..317S}. 
In 1978 she returned to Australia as a Senior Lecturer in Physics at the
University of Wollongong.  She then closed the circle by returning to Mount
Stromlo in 1989--1991, and by 1993 she was back at the University of Sydney,
where she developed a useful three-dimensional classification for
nitrogen-dominant Wolf-Rayet stars \citep{1996MNRAS.281..163S}. 

{\bf Betty `Louise' Turtle, n\'ee Webster} (1941-1990:
\citet{1991PASAu...9....6S}) 
was another graduate student at Mount Stromlo during the Bok years, who was
present at the same time as L.F.\ Smith and supervised by B.\ Westerlund.
After graduating, and while at the University of Wisconsin, Webster authored
her own (similarly themed and) popular survey paper titled ``The masses and
galactic distribution of southern planetary nebulae'' \citep{1969MNRAS.143...79W} which
has been cited 115 times.  Webster subsequently worked with Sir Richard
Woolley (former Director at Mount Stromlo Observatory) at the Royal Greenwich
Observatory, including time at the South African Astronomical Observatory,
before helping to commission the AAT and working at the newly formed
Anglo-Australian Observatory, and then working at the University of New South
Wales from 1978 onwards. She has many papers of note, including work on
chemical abundance gradients in galaxies and identifying the X-ray
binary source Cygnus X1 as a possible black hole \citep{1972Natur.235...37W}. 
She is largely responsible for having established the Bok Prize, and she
herself is honoured through the Louise Webster Prize, which 
the Astronomical Society of Australia also administer.

{\bf Richard (Dick) Xavier McGee} (1921--2012: \citet{2013:Sim}) was a clerk in
the Customs Department when he was called up for military service in 1941.  He
was posted to Darwin, where he was very nearly killed during the 1942 air raid
when bombs fell on either side of the truck that he was driving.  It is not
known to us if he met J.H.\ Piddington at that point.  He later trained as a
navigator in Canada and flew in Lancasters in the UK.  
After the war, in 1947, he quit
his job as a clerk (which he had initially returned to) and went to
the University of Sydney on a Commonwealth scholarship.  He obtained a 
first-class honours degree in Physics and subsequently worked for CSIRO's
radiophysics laboratory from 1950 until his retirement in 1986.  From 1971 to
1988, he was the editor of the Publications of the Astronomical Society of
Australia.  He is known for his contribution to our 
knowledge of the Milky Way's chemical abundance
gradient \citep{1983MNRAS.204...53S}, and his detailed HI
survey \citep{1966AuJPh..19..343M} which revealed the
distribution and motion of hydrogen gas in the LMC, including a primitive
spiral pattern and local fuel supply for the stellar nurseries scattered
around the galaxy\footnote{The survey was so important that it was repeated
  at Parkes in 1984 \citep{1984AA...137..343R}.}.  {\bf Miss Janice A.\ Milton}
(later Mrs J.A.\ Weedon) and McGee also published a survey of neutral hydrogen
across the southern sky, in addition to their focus on the LMC, which was
repeated with the Parkes telescope when its instrumentation improved. 
\startlandscape
\begin{specialtable}[H]
\widetable
\setlength{\tabcolsep}{7.1mm}
\caption{The ten most cited articles from 1966 to 1969
  are numbered in column 1, and involve 12 distinct authors.  Ten distinct
  authors can be found above the horizontal line.\label{Tab66-69}} 
\begin{tabular}{llllclc}
\toprule
\textbf{\#} & \textbf{Title}  &  \textbf{Affiliation}  &  \textbf{Author(s)}  &  \textbf{Year}  &  \textbf{Journal}  &  \textbf{Cites} \\  
   &        &               &             &        & \textbf{Vol.\ Page} &         \\
\midrule
1. & Pulsar Electrodynamics                                      & Univ.\ Sydney                              &  Goldreich, P.                &  1969  &  ApJ      & (1660) \\
   &                                                             & \hspace{2mm} \& Caltech                    & \hspace{2mm} \& Julian, W.H.\ &        & 157, 869  &        \\
2. & Solar Modulation of Galactic Cosmic Rays                    &  UCSD (USA)                                &  Gleeson, L.J.                &  1968a &  ApJ      & (518)  \\ 
   &                                                             & \hspace{2mm} \& Monash University          & \hspace{2mm} \& Axford, W.I.  &        & 154, 1011 &        \\ 
3. & Magnetic Poles and the Polarization                         &  CSIRO-Radiophysics                      &  Radhakrishnan, V.            &  1969  &  ApL      & (498)  \\  
   & \hspace{5mm} Structure of Pulsar Radiation                  &                                            & \hspace{2mm} \& Cooke. D.J.   &        & 3, 225    &        \\  
4. & A Revised Spectral Classification System \&                 &  MSO                                       &  Smith, L.F.                  &  1968a & MNRAS     & (297)  \\  
   & a new catalogue for galactic Wolf-Rayet stars               &                                            &                               &        & 138, 109  &        \\  
5. & Absolute Magnitudes and Intrinsic Colours                   &  MSO                                       &  Smith, L.F.                  &  1968b & MNRAS     & (275)  \\  
   & \hspace{5mm} of Wolf-Rayet Stars                            &                                            &                               &        & 140, 409  &        \\  
6. & 21 cm hydrogen-line survey of the Large                     &  CSIRO-Radiophysics                      &  McGee, R.X.                  &  1966  & AuJPh     & (272)  \\
   & \hspace{5mm} Magellanic Cloud.\ II...                       &                                            & \hspace{2mm} \& Milton, J.A.  &        & 19, 343   &        \\
7. & Chemical evolution of the terrestrial planets               &  The ANU:                                  &  Ringwood, A.E.               &  1966  & GeCoA     & (242)  \\  
   &                                                             & Geophys. \& Geochem.                       &                               &        &  30, 41   &        \\  
\midrule 
8. & New subdwarfs. II. Radial velocities,                       &  Mount Wilson (USA)                        &  Sandage, A.                  &  1969b & ApJ       & (222)  \\ 
   & photometry and preliminary space motions                    & \hspace{2mm} \& MSO                        &                               &        & 158, 1115 &        \\ 
   & for 112 stars with large proper motion                      &                                            &                               &        &           &        \\ 
9. & The Double Cepheid CE Cassiopeiae in                        &  MSO                                       &  Sandage, A.\ \&              &  1969  & ApJ       & (209)  \\  
   & \hspace{5mm} NGC 7790...                                    & \hspace{2mm} \& Univ.\ Basel (Switz.)      & \hspace{2mm} Tammann, G.A.    &        & 157, 683  &        \\  
10.& The Reddening, Age Difference, and Helium                   &  MSO                                       &  Sandage, A.                  &  1969a & ApJ       & (189)  \\  
   & \hspace{5mm} Abundance of the Globular Clusters...          &                                            &                               &        & 157, 515  &        \\  
\bottomrule
\end{tabular}
\end{specialtable}

\finishlandscape





%
%




\textls[-15]{{\bf Allan Rex Sandage} (1926--2010: \citet{2010Natur.468..898T,
  2011BAAS...43..038D, 2011Obs...131..109K, 2012BMFRL..58..245L,
  2011JRASC.105...38V}) was a hugely influential North American astronomer who
obtained his Ph.D.\ from Caltech in 1953---the year that Edwin Hubble, to
whom Sandage was a student assistant, passed away.  Sandage spent a year at
Mount Stromlo in 1969, and in that time, he wrote three of Australia's most
cited astronomy articles from the latter half of the 1960s
\citep{1969ApJ...157..515S, 1969ApJ...158.1115S, 1969ApJ...157..683S}.
Prolific as always, his articles in Table~\ref{Tab66-69} discuss stars with
large proper motions, calibrating the cepheid period-luminosity relation
(recall the work of~\citep{1952PASP...64..196G}), and globular cluster helium
abundances and ages \citep{1953AJ.....58...61S, 1970ApJ...162..841S,
  1982ApJ...252..553S}.  In one of astronomy's longer running
debates\footnote{Tension over the precise value of the Hubble constant
  continues to this day (e.g., \citep{2018PhRvD..97j3529B}).}, Sandage
famously advocated \citep{1975ApJ...196..313S} a `Hubble constant' of around
55 km s$^{-1}$ Mpc$^{-1}$, while at the same time, de Vaucouleurs (who had
been at Mount Stromlo 12--18 years earlier), along with Sidney van den Bergh
(1929--), were advocating a higher value around 80--100 km s$^{-1}$ Mpc$^{-1}$.
The disparity led to a ``Key Project'' on the Hubble Space Telescope, which
reported a value of 71 $\pm$ 6 km s$^{-1}$ Mpc$^{-1}$, with a corrected value of
68 $\pm$ 6 km s$^{-1}$ Mpc$^{-1}$ reported by \citet{2000ApJ...529..786M}.
Sandage also worked on other topics such as cepheid variables and globular
clusters as a means to probe the distance scale of the Universe
\citep{2013IAUS..289...13T}.  He is, of course, also known for his research
into observational cosmology, the ``monolithic collapse'' model for galaxies
\citep{1962ApJ...136..748E}\footnote{Olin Eggen wrote this paper while at
  Caltech, but he was later the Director at Mount Stromlo Observatory from
  1966--1977.  He was reportedly something of a character who drove a Bolwell
  fibreglass sports car with a Ford V8 motor. He died while visiting Canberra
  in 1998.}, contributions to the colour-magnitude relation for early-type
galaxies~\citep{1978ApJ...223..707S}\footnote{{\bf Natarajan Visvanathan
    (Vis)} (1932--2001: \citet{2002BAAS...34.1386F}) was a particularly
  friendly staff member at Mount Stromlo Observatory from 1975 to 2001, often
  seen driving his yellow Porsche sports car up and down the mountain.}, and
his many galaxy catalogs \citep{1961hag..book.....S, 1981rsac.book.....S,
  1985AJ.....90.1681B}. }

{\bf Gustav Andreas Tammann} (1932--2019: \citet{2019Msngr.176...58L}) 
was the director of the University of
Basel's Astronomical Institute in Switzerland.  His work with supernovae, the
extragalactic distance scale, and the expansion of the Universe has earned him
several international medals, and asteroid 18872 Tammann is named after him.






\section{Concluding Remarks}\label{Sec_Con}

By the second half of the 1940s, the more cited astronomical studies
(Table~\ref{Tab45-50}) focussed on the Sun at radio wavelengths and
explored its influence on Earth's magnetosphere and ionosphere.
Indeed, eight of the top-10 and twelve of the top-15, most-cited
articles from this period relate to this.  Notable at this time were
the contributions from several women who did not just capably fill the
void that had arisen with so many men away fighting in WWII but who
excelled in their roles.  In particular, Rachel E.B.\ Makinson tutored
radiophysics at The University of Sydney to RAAF airmen and
pre-(radio astronomers) during the war.  She eventually went on to
study the physics of fibres and advance the wool industry for
Australia, following in the footsteps of Elizabeth and John Macarthur, 
who helped establish that very industry.  One of the first radio
astronomers to emerge after the war was Ruby Payne-Scott, who had
co-developed radar during the war with Joan Jelly, n\'ee Freeman, and
others.  While Joan Freeman switched slightly from radiophysics and
went on to become one of the world's leading nuclear physicists, Ruby
Payne-Scott is the author of one of Australia's most cited astronomy
articles from the years 1945 to 1950 (Table~\ref{Tab45-50}).  As discussed, her
work built on that performed by another remarkable woman, Elizabeth
Alexander, in New Zealand.

One of the few exceptions to the 1945--1950 articles that were predominantly
related to solar phenomena is {\em the} most-cited article from this
period.  Written by Prof.\ `Walter' Stibbs, it pertains to magnetic
variability in another star.  Together with an article about the
detection of three radio-sources beyond our solar system, it signalled how
astronomy was starting to branch out.  
As technology developed over the years, the research expanded to include yet
more stars, galactic and extragalactic radio sources, and various other
aspects of our Galaxy and neighbouring galaxies.  Reflecting this
development, the ten most-cited articles from the latter half of the 1960s
(Table~\ref{Tab66-69}) pertain to eight different topics, including planets,
stars, pulsars, globular clusters, cosmic rays, and HI in the Large Magellanic
Cloud.


Among the articles from 1966 to 1969 is not one but two articles from Lindsey
Fairfield Smith, who had attended the Women's College at the University of
Sydney where Joan Freeman had been two decades earlier.  In one of her papers
from 1968, Lindsey Smith presented a catalogue for the distribution of
Wolf-Rayet stars in our Galaxy\footnote{Still a popular subject, \citet{2019Galax...7...74N} provide an
  update for our Local Group.}, as seen from the Southern Hemisphere.  In the
following year, Betty `Louise' Turtle, n\'ee Webster---another Ph.D.\ student
at Mount Stromlo Observatory under Bart J.\ Bok's Directorship---published
on the Galactic distribution of southern planetary nebulae.\footnote{Following
  Webster's pioneering work, Gaia has
  enabled distances to over 1000 planetary nebulae \citet{2020Galax...8...29G}.}  It should be
remembered that contributing to the apparent 20-year hiatus of women in our
Tables was society in general.  Married women were expected to look after the
house; until 1966, they were denied permanent jobs in the Australian Public
Service and were fired if they became pregnant.  Provisions for (unpaid)
maternity leave within the Public Service did not arrive until 1973, and the
concept of equal pay for equal work did not come into practice until the 1970s
(since 1950, it had been at 75 percent of the male basic wage).  However, helping to
turn the tide was the husband and wife team of Bart J.\ Bok (Director)
and Priscilla F.\ Bok (astronomer), known for increasing the number of
female Ph.D.\ astronomy students at Mount Stromlo Observatory.  They arrived
the year that the wife and husband team of Antoinette and Gerlad H.\ de
Vaucoulers left (after six years at Mount Stromlo Observatory).  More recently, 
this century, there has been another couple at the Observatory, but with a
nice reversal. There was a female Director (Penny D.\ Sackett) with her partner 
astronomer (Frank H.\ Briggs). 

Although working as either paid professionals or as funded students, the individuals
appearing in Tables~\ref{Tab45-50}--\ref{Tab66-69} almost certainly did not
achieve their success by only working from 9 to 5. They undoubtedly, and
literally, worked day and night to render their valuable service for the
credit of Australia and the advancement of astronomical science.  It is
appropriate that their labours are remembered here and elsewhere.  Of
exceptional distinction are three names in the preceding Tables.  J.P.\ Wild
appears in the first four tables spanning two decades, while A.E.\ Ringwood
appears in the last three tables.  Based on his analysis of lunar rocks
returned by Neil Armstrong and the Apollo missions, Ringwood later 
advanced the theory that our Moon was split-off from an early-Earth.
Finally, E.E.\ Salpeter has four very highly cited articles, taking out the first and
second, and first and third, spots in two different tables\footnote{Had we
  extended the tables slightly further, we note that J.H.\ Piddington would
  also have featured in every Table.}.

\textls[-15]{While not included in the above tables of scientific research articles, the
year 1969 is, of course, also memorable for the valuable role that Australia
played in receiving and broadcasting the Apollo 11 lunar landing to the
world (see Figure~\ref{Fig_Radio}). 
This historic event will likely be remembered for centuries to come; it
was when humankind's exploratory nature culminated in us leaving our planet
and walking on another astronomical body.  Neil Armstrong's stepping onto the
Moon, footage of which was captured by the engineering team at the 26 m dish
at the Honeysuckle
Creek\footnote{\url{http://members.pcug.org.au/$\sim$mdinn/TheDish/}  (accessed 29 March 2021).} NASA tracking
station in Canberra\footnote{This historic telescope was later relocated to
  Tidbinbilla, Canberra, where NASA's 70 m dish 
  operates \citep{1994AuJPh..47..497W} (pp. 501--503).}, 
along with the subsequent footage (after the initial 8 minutes) taken by the
team at the Parkes Observatory, was relayed around the world.  The large 
64~m Parkes radio telescope continues to conduct world-leading research
today. Examples include: 
the Parkes Pulsar Timing Array (PPTA)
to detect the presence of long-wavelength 
gravitational radiation~\citep{2010CQGra..27h4013H, 2013PASA...30...17M, 
2015Sci...349.1522S, 2018PASA...35...13T}; 
the search for fast radio bursts (FRB: \citep{2015MNRAS.447..246P,
  2018MNRAS.475.1427B}); 
a survey of methanol masers at 6.7 and 12.2 GHz 
\citep{2010MNRAS.401.2219B, 2010MNRAS.404.1029C}; and 
the exciting `Breakthrough Listen' project 
\citep{2020AJ....159...86P} expanding upon the 1995 
Search for Extra-Terrestrial Intelligence (SETI) `Project Phoenix'
\citep{1996SPIE.2704...24T} in which then-PhD student 
A.Graham was one of the Official Observers. }
\begin{figure}[H]
\hfill
{\includegraphics[height=5.0cm]{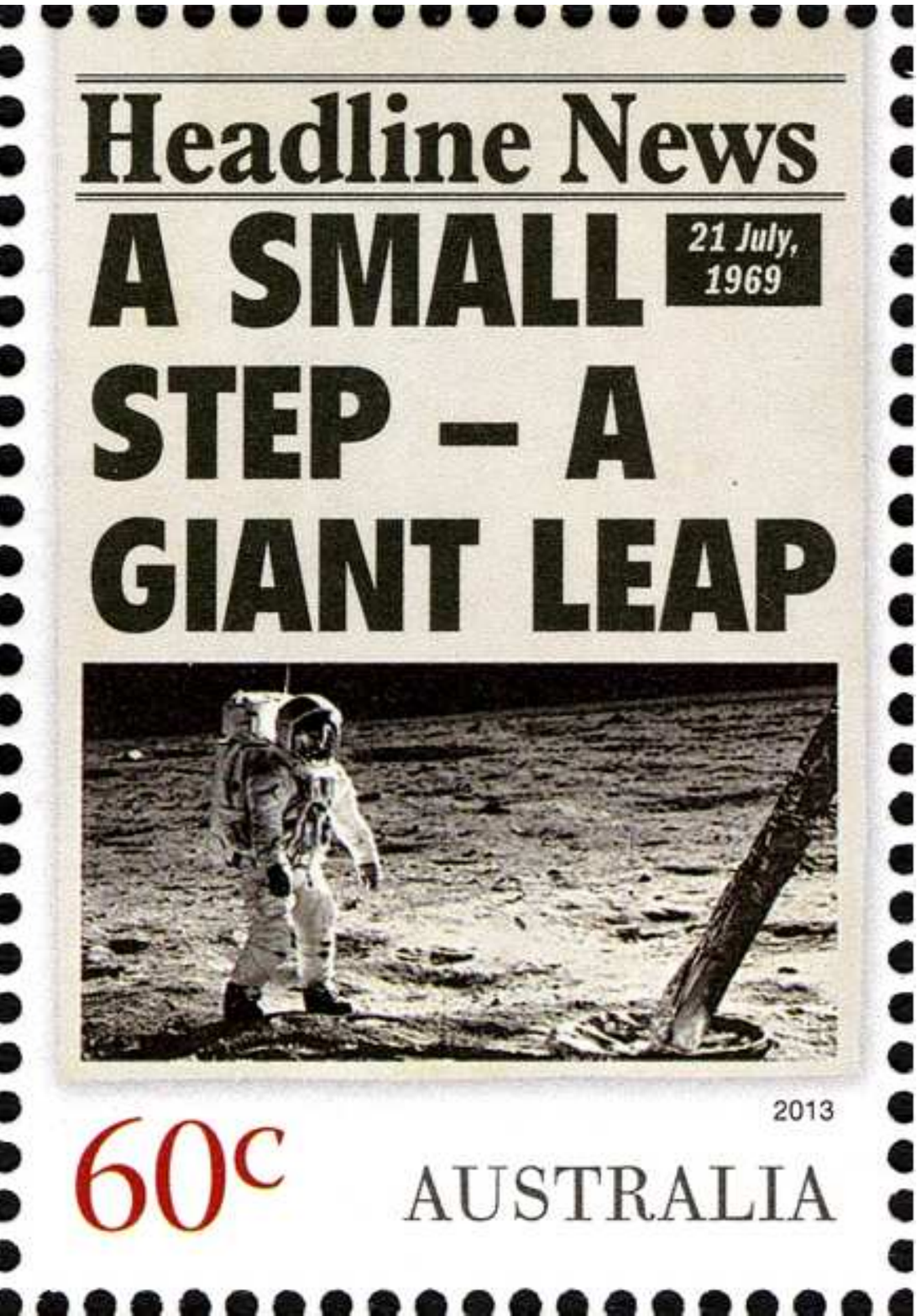}}
\hfill
{\includegraphics[height=5.0cm]{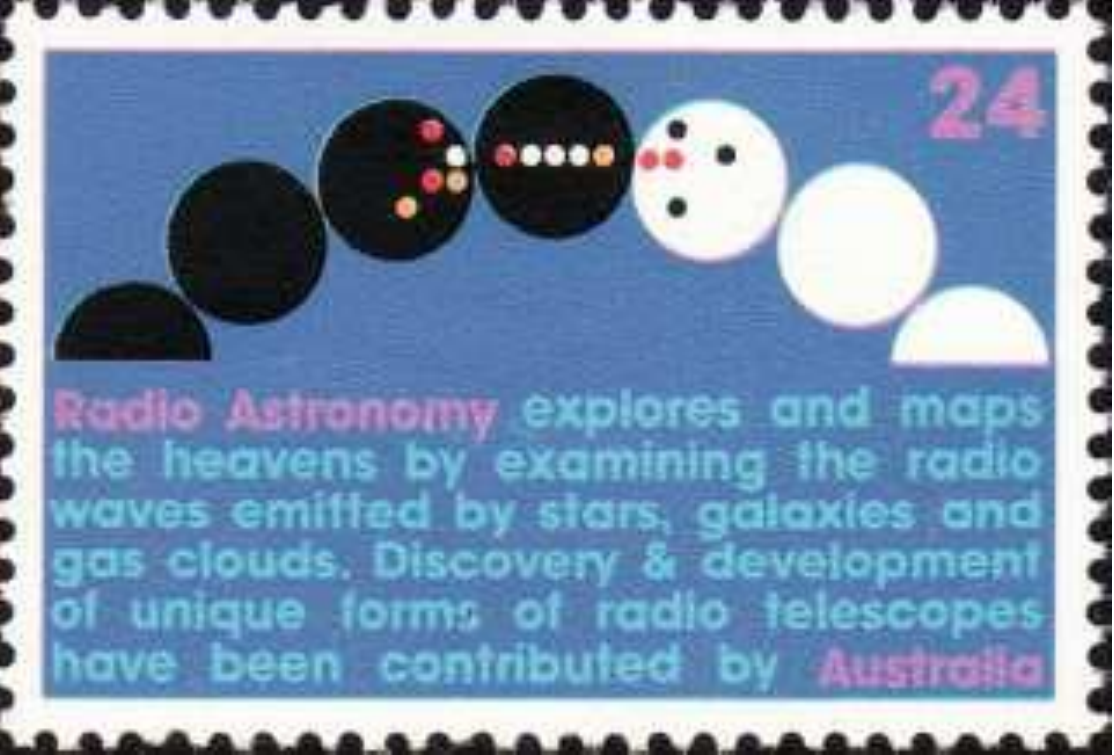}}
\hfill
\caption{
Left panel: Australia Post stamp from the 23 July 2013 ``Headline News''
series, showing the lunar footage taken from Australia on 20 July 1969. 
Right panel: Australia Post stamp from the 14 May 1975 
``Scientific Development'' series. 
\label{Fig_Radio}
}
\end{figure}


\subsection{Looking Forward} 

From 1955 to 1974, 
until the 3.9 m telescope opened at Siding Spring Observatory, 
the 74-inch (1.9 m) telescope at Mount Stromlo Observatory, 
together with the 1950-opening of the 74-inch Radcliffe
Telescope \citep{1951ASPL....6..170K} in Pretoria, South Africa, were the largest optical
telescopes in the Southern Hemisphere, giving these two countries a unique
vantage point.  With the ~Square Kilometer
  Array (SKA)\footnote{\url{www.skatelescope.org} (accessed 29 March 2021).} to be located in both Australia
and South Africa, these two countries will again host a large telescope. 
This time a radio facility two orders of magnitude more sensitive
and rapid in sky surveys than present radio facilities.\footnote{The Chinese
  have constructed the behemoth ``Five hundred meter Aperture
%
%
  Spherical Telescope (FAST: \citep{2011IJMPD..20..989N}) in
  Pingtang County, Guizhou Province, which will also contribute tremendously
  to furthering our knowledge of the Universe.}  Due to Australia's lack of
high mountains, it is not an optimal site for a large, next-generation 
optical/infrared telescope \citep{1995PASA...12...97W}.  Ongoing access to
such a telescope, in particular the Giant Magellan Telescope 
(GMT)\footnote{\url{www.gmto.org} (accessed 29 March 2021).} in Chile \citep{2012SPIE.8444E..1HJ}, 
is, of course, recognised as key to the ongoing success of 
astronomy in Australia.  As such, Australia is a partner in this grand
venture.  Indeed, not one but three such giant telescopes are under
development by different groups worldwide due to the wide-spread desire
for access to such a facility given the ground-breaking scientific questions
that they can tackle.  

Investment today, in new facilities and instruments for existing telescopes, 
promises to secure Australia's astronomical future, such as the international 
  Murchison Widefield Array 
(MWA\footnote{\url{http://mwatelescope.org} (accessed 29 March 2021).}) \citep{2013PASA...30....7T,
  2018PASA...35...33W, 2019PASA...36...30O} in Western Australia and the
Australian Square Kilometre Array Pathfinder (ASKAP: \citep{2007PASA...24..174J,
  2011PASA...28..215N, 2013PASA...30....3D, 2013PASA...30....6M, 2019PASA...36...24L}) also
sited at the Murchison Radio-astronomy Observatory (MRO) in Western Australia.
The long-wavelength MWA operates at frequencies of \mbox{70--300~MHz}, and 
can study the Sun \citep{2005SPIE.5901..124S, 2011ASInC...2..397C}, 
as previously done at The University of Western Australia after
WWII \citep{1947Natur.160..708W}.  It will additionally enable studies of the
Earth's ionosphere, variable radio sources, neutral atomic hydrogen at the
epoch of reionisation, and more~\citep{2013PASA...30...31B,
  2018PASA...35...43R, 
  2019PASA...36....4F, 2019PASA...36...23T, 2019PASA...36....2M}. This follows the
successes of earlier low-frequency antennae in Sydney, 
Tasmania \citep{2015JAHH...18....3O, 2015JAHH...18...14G}, and also near
Adelaide.  The University of Adelaide's Buckland Park Array, which at 1 square
km in size, was the world's largest low-frequency telescope when it was
finished in 1969.  In the northern hemisphere, the Caltech-operated Long
  Wavelength Array (LWA\footnote{\url{http://www.tauceti.caltech.edu/lwa} (accessed 29 March 2021).}) at the
Owens Valley Radio Observatory, California, currently operates at 28 MHz to 88
MHz and has the same science goals as the MWA, as does the SKA-pathfinder
Low-Frequency Array (LOFAR\footnote{\url{http://www.lofar.org} (accessed 29 March 2021).}), built by the
Netherlands Institute for Radio Astronomy,~ASTRON.

Although not national facilities, smaller aperture, optical telescopes such as
The ANU's automated 1.35 m SkyMapper
telescope \citep{2005AAS...206.1509S, 
  2007PASA...24....1K, 2018PASA...35...10W} at Siding Spring Observatory, the 1 m fast-slew robotic
Zadko Telescope\footnote{\url{http://www.zt.science.uwa.edu.au} (accessed 29 March 2021).} at the Gingin
Observatory near Perth \citep{2010PASA...27..331C}, and the Hungarian-made
Automated Telescope Network-South (HATSouth) unit looking for exo-planets from
Siding Spring Observatory \citep{2013PASP..125..154B}, are also expected to
deliver many exciting discoveries.  Interestingly, 60-odd years ago, a 13 cm
patrol telescope operated daily at the Culgoora Solar Observatory near
Narrabri as a part of a worldwide network watching for solar flares.  This
supported the USA space programme and was funded by the USA National
Oceanic and Atmospheric Administration.  The observations were to support
NASA’s Skylab, which, ironically, crash-landed in Australia because of solar
flares that had heated our upper atmosphere and which subsequently expanded
and thus increased the wind drag on 
Skylab.   
Regular solar observations for monitoring space weather and ionospheric
conditions are still made today from Culgoora, and many other places, such as
Learmonth in Western Australia.  Learmonth additionally hosts one of the six Global
Oscillations Network Group (GONG) instruments that provide helioseismic
doppler data, magnetograms, and H$\alpha$ images of the Sun.  Radar is also
used to study the ionosphere and aurora from Tasmania via the Tasman International Geospace
Environment Radar (TIGER\footnote{\url{http://www.tiger.latrobe.edu.au/} (accessed 29 March 2021).})
team within the Space Physics Group at La Trobe
University \citep{2013RaSc...48..722C}. 

Of course, Australia's reputation for astronomy is no
longer just based on optical and radio (including
millimetre) observations.  Much of Australia's
research is now enabled by access to significant computing power, 
off-shore facilities, and space-based satellites that have genuinely opened up astronomy. 
This research encompasses areas such as: 
wide-field, high-spatial-resolution optical studies not feasible from the
ground; 
infrared astronomy\footnote{In the 1960s, the Australian Government's 
  Department of Supply was involved with infrared mappings of the sky 
  using balloon-bourne detectors.}; 
ultra-violet research\footnote{Ultra-violet observations of the Sun were made by The University of 
  Adelaide in collaboration with the Weapons Research Establishment at
  Salisbury using rockets and satellites.  
An ultra-violet telescope named Endeavour was built at
  Mount Stromlo Observatory by Auspace Ltd.  It flew aboard the space
  shuttle Discovery (flight STS~42) in January 1992 and took observations
  from the space shuttle Endeavour (STS~67) in March 1995, along with the USA's
  ASTRO-2 UV program.}; 
X-ray studies\footnote{In the late 1960s, X-ray
  astronomy was conducted by the University of Tasmania and The University of
  Adelaide from sounding rocket and balloon payloads launched from Mildura.}; 
and the gamma-ray universe.  In addition, atmospheric
particle showers from cosmic rays\footnote{Such studies have been conducted at
  The University of Adelaide since the early 1970s.} \citep{2004ApJ...606L.115T,
  2008arXiv0801.3886R} also contribute to Australia's broad and rich astronomical
research portfolio.

Rather like radio physics in the late 1940s, gravitational wave physics in
Australia has expanded from physics into astronomy, especially after the
detection of gravitational radiation from coalescing black
holes \citep{2016PhRvL.116f1102A}. This area of research holds much
promise, and many Australian researchers now have involvement with the large overseas ground-based
facilities searching for gravitational radiation \citep{2006CQGra..23S..41M} 
and any associated electromagnetic counterpart \citep{2017PASA...34...69A, 2019PASA...36...19D}. 
Work is currently conducted at The Australian International
Gravitational Research Centre based at The University of Western Australia,
within the OzGrav Centre of Excellence\footnote{\url{https://www.ozgrav.org} (accessed 29 March 2021).} funded
by the Australian Research Council, and elsewhere. 
There are also indirect searches occurring in Australia through the Parkes 
Pulsar Timing Array mentioned earlier, and many other astronomers in Australia
have written theoretical papers on the subject. 



Further afield, the gold mine\footnote{Arete Capital Partners own the Stawell
  Gold Mine.} in the town of 
Stawell---where the unique sundial of astronomer 
Charles James Merfield (1866--1931) can be found---has 
connections with astronomy via the search for the Universe's alleged but 
elusive {\it dark matter} \citep{2006isdm.book.....F, 
2016arXiv160503299U, 
2019Galax...7...68A, 
2019Galax...7...16B, 
2019Galax...7...81Z}. 

While the astronomical connections with the esteemed
Fairfax and Packer families have largely disappeared\footnote{The prestigious
  Fairfax Prize still exists; it is awarded to first-year students at The University of Sydney
  who receive the highest possible entrance score.},
new connections seem hopeful with the Australian Government's recent (1 July 2018) creation 
of the Australian Space Agency headquartered in Adelaide.  
Business leaders
come philanthropists, such as Stanley Chatterton who co-founded Woolworths
(sponsor of the Stawell Gift foot race) and enabled the Chatterton
  Astronomy Department at The University of Sydney,
have in the past supported scientific investigations of discovery.
%
%
While most astronomical facilities in Australia are still simply 
named after their location---just a few are 
named after pioneering astronomers, such as the Paul Wild Observatory in 
Culgoora---, Australia is starting to see naming rights awarded to
individuals recognised for their support of astronomy.  
The Zadko Telescope\footnote{The 1 m Zadko 
  Telescope was made possible through a generous donation from
%
%
  businessman James Zadko to The 
  University of Western Australia.}, which opened in 2009, is perhaps the first such facility. 
Another example is the 
W.M.\ Keck Observatory's\footnote{\url{http://www.keckobservatory.org/} (accessed 29 March 2021).}
Remote Observing Facility at the Swinburne University of Technology, affectionally known
as the Baker Control Room after it was partially funded through a generous donation
from the Eric Ormond Baker Charitable fund.  While more such 
partnerships would be heartily welcomed, it is clear that 
Australia has the drive, talent, and potential to 
continue in the footsteps of its pioneering astronomers. 




\vspace{6pt}
\authorcontributions{While the text was written by A.W.G., most of the
  table entries were discovered by the remaining authors. In 2014, the latter
  three authors started the survey as Year 10 highschool ``work experience'' 
  students.  In 2015, K.H.K.\ double-checked the entries as an undergrdaute
  ``summer student'', and started to research the article's
  authors. A.W.G.\ subsequently triple-checked for missing entries and
  completed researching the article's authors. Writing---original draft preparation: A.W.G., K.H.K., L.J.B., V.C.L.D., K.K. All authors have read and agreed to the published version of the manuscript.
}


\funding{This research received no external funding.}

\institutionalreview{Not applicable.}

\informedconsent{Not applicable.}

\dataavailability{This research has made use of the Astrophysics Data System (ADS),
operated by the Smithsonian Astrophysical Observatory (SAO) under a National
Aeronautics and Space Administration (NASA) grant. 
This article has additionally benefited from the National Library of
Australia's (free and wonderful) online and digitised {\it Trove} database of
old newspapers and more (\url{http://trove.nla.gov.au/}, accessed 15 February 2015).}

\acknowledgments{We thank Ron Ekers, Richard Hunstead, Nick Lomb, Jeremy Mould and John Norris for their comments, either in 2015 or 2021.}



\conflictsofinterest{The authors declare no conflict of interest.}

\appendixtitles{yes}
\appendixstart
\appendix

\section{Further Reading}
\label{AppA}

During our research, we encountered several informative books about astronomy
in Australia over the past century.  This led us to search for more books, and
while we do not confess to having read or reviewed the following, we
nonetheless felt that we could list and acknowledge them here for the benefit
of readers who may wish to explore this topic~further.
\begin{itemize}
\item {\bf 1987}, C.B.\ Schedvin, {\it Shaping Science and Industry: A History
  of Australia's Council for Scientific and Industrial Research 1926--1949},
  Allen \& Unwin. 
\item {\bf 1990}, S.C.B.\ Gascoigne, K.M.\ Proust \& M.O.\ Robins, {\it The
  Creation of the Anglo-Australian Observatory}, Cambridge University Press.
\item {\bf 1991}, R.S.\ Bhathal \& Graeme White, {\it Under the Southern
  Cross: a brief history of astronomy in Australia}, Kangaroo Press,
  Kenthurst, NSW.
\item {\bf 1992}, Peter Robertson, {\it Beyond Southern Skies: Radio Astronomy
  and the Parkes Telescope}, Cambridge University Press.
%
%
\item 1993, Kerrie Dougherty \& Matthew L. James, \emph{Space Australia: The Story of Australia's Involvenment In Space, Powerhouse Publishing}.
\item {\bf 1994}, D.E. Goddard \& D.K. Milne (Eds.) {\it Parkes: Thirty Years of Radio Astronomy},
  CSIRO Publishing.
\item {\bf 1996}, Raymond Haynes, Roslynn Haynes, David Malin \& Richard
  McGee, {\it Explorers of the Southern Sky: A History of
  Australian Astronomy}, Cambridge University Press. 
\item {\bf 1996}, Ragbir Bhathal, {\it Australian Astronomers: Achievements at
  the Frontiers of Astronomy}, National Library of Australia Publishing, Canberra.
\item {\bf 2003}, T.\ Frame \& D.\ Faulkner, {\it Stromlo: An Australian
  Observatory}, Sydney, Allen \&~Unwin. 
\item {\bf 2004}, Nick Lomb, {\it Transit of Venus: the Scientific Event that
  Led Captain Cook to Australia}, Sydney, Powerhouse Publishing. 
\item {\bf 2008}, Roslyn Russell, {\it Two People \& a Place: The Family Who
  Lived in Sydney Observatory}, Yarralumla, A.C.T., Roslyn Russell Museum Services. 
\item {\bf 2009}, Woodruff T.\ Sullivan III, {\it Cosmic Noise: A History of
  Early Radio Astronomy}, Cambridge University Press. 
\item {\bf 2010}, Miller Goss \& Richard McGee, {\it Under the Radar: The First Woman in
  Radio Astronomy: Ruby Payne-Scott}, Springer. 
\item {\bf 2011}, Richard Gillespie, {\it The Great Melbourne Telescope},
  Read How You Want. 
\item {bf 2011}, Nick Lomb, {\it Transit of Venus: 1631 to the Present},
  University of New South Wales Press. 
\item {\bf 2013}, Miller Goss, {\it Making Waves: The Story of Ruby Payne-Scott:
  Australian Pioneer Radio Astronomer}, Springer. 
\item {\bf 2013}, Ragbir Bhathal, Ralph Sutherland \& Harvey Butcher, {\it Mt Stromlo
  Observatory: From Bush Observatory to the Nobel Prize}, CSIRO Publishing. 
\item {\bf 2013}, David P.D.\ Munns, {\it A Single Sky: How an International
  Community Forged the Science of Radio Astronomy}, The MIT Press.
\item {\bf 2017}, R.H.\ Frater, W.M.\ Goss, H.W.\ Wendt, {\it Four Pillars of Radio Astronomy: Mills, Christiansen, Wild, Bracewell}, Springer International Publishing AG.
\item \textbf{2017}, P. Robertson, \emph{Radio Astronomer John Bolton and a New Window on the Universe}, NewSouth Publishing.

\item {\bf 2021}, W.\ Goss, C.\ Hooker, R.\ Ekers, {\it From the Sun to the Cosmos - Joseph Lade Pawsey, Founder of Australian Radio Astronomy}, in preparation.
\end{itemize}

\section{Update}
\label{AppB} 

Based on (a medley of) metrics, rankings of articles, 
journals (e.g., \url{https://www.scimagojr.com}, accessed 29 March 2021), 
Universities (e.g., \url{http://www.shanghairanking.com}, accessed 29 March 2021), 
and even the world's top 2\% of scientists \citep{Ioannidis2020} will change with
time.  Since this research was conducted, the various papers have
jostled around slightly in the Tables.  Of note are the following changes.

From the 1945--1950 time period,
%
%
three `new' publications
have appeared on the horizon, the most prominant of which is
(`Relative Times of Arrival of Bursts of Solar Noise on
Different Radio Frequencies' \citet{1947Natur.160..256P}), coming in at
number 10.  One new name from this three-author publication is {\bf 
Donald (Don) E.\ Yabsley}, who was involved in many pioneering,
multi-frequency, solar radio astronomy experiments around Sydney.  He
also led the Division of Radiophysics' 1949 expedition in Tasmania to
observe a partial eclipse of the Sun \citep{2008JAHH...11...71W}.
The two other `new' publications are (\citep{1948MNRAS.108..163G}, `Chromospheric
Flares') and (\citep{1948Natur.161..312B}, `Variable Source of Radio
Frequency Radiation in the Constellation of
Cygnus'), which now enter the tail (10 to 15) of
the list, and are written by authors whom we have already encountered.


From the 1951--1955 time period, citations to the top-ranked paper
\citep{1955ApJ...121..161S} have increased by nearly a half, while
citations to \citet{1951PRCO....2...85O} have more than doubled,
catapulting it from 12th to 7th.

From the 1956--1960 time period, citations to
\citet{1956AuJPh...9..198B} have increased by nearly a half,
moving it from 9th to 4th.  Three `new' publications have also
appeared on the radar, the most prominant of which is (\citet{1958MNRAS.118..379O}, `The
galactic system as a spiral nebula (Council
Note)'), entering the list at number 7.  This is
a curious publication in that the three authors (J.H.\ Oort, F.J.\
Kerr \& G.\ Westerhout) provide no affiliation.  It is the authorship
of F.J.\ Kerr and the contribution of radio data from Australia which
reveals this is partially an Australian publication.  Jan Hendrik Oort
(1900--1992: \citep{1993JApA...14....3W, 1993PASP..105..681B, 2019ASSL..459.....V}) was an
eminent Dutch astronomer, and Gart Westerhout 
(1927--2012: \citep{2012PhT65online}) was a Dutch-American
astronomer.  Furthermore, now {\em just} outside the top-ten are
(\citet{1959AuJPh..12..327R}, `Solar Radio Bursts of Spectral Type II') and 
(\citet{1957AJ.....62...93K}, `A Magellanic effect on the galaxy') by authors we have met.
From 1961 to 1965, citations to the `new' publication (\citep{1963AuJPh..16..570H}, `A Low
  Resolution Hydrogen-line Survey of the Magellanic System. II. Interpretation
  of Results') are up over 40 percent and it is now equal
tenth with the prior entries 10 and 11.  We have already met the co-authors
F.J.\ Kerr and R.X.\ McGee, while J.V.\ (Jim) Hindman is new.  Hindman started
work at the CSIRO Division of Radiophysics with Jack H.\ Piddington and as an
assistant to W.N.\ (Chris) Christiansen \citep{2011arXiv1109.4880F,
  2014bea..book..421E, 2017fpra.book.....F} in the late 1940s and early 1950s
\citep{2002ASPC..276...19R}. Hindman went on to co-publish 17 refereed
articles up until 1970.

From 1966 to 1969, citations to the `new' publication (\citet{1967AuJPh..20..147H}, `A high
  resolution study of the distribution and motions of neutral hydrogen in the
  Small Cloud of Magellan') are up 10 percent and it is now
equal tenth.

All of the 8 `new' entries mentioned above were authored at the
CSIR/CSIRO Divison of Radiophysics, except for the publication from
Giovanelli at the Divison of Physics.  This is a testament to the longevity of
their pioneering research in what was then the nascent field of radio astronomy.
\startlandscape
\appendix
\begin{specialtable}[H]
\widetable
\setlength{\tabcolsep}{7.5mm}
\caption{New Entries.\label{Update}}
\begin{tabular}{lllcl}
\toprule
\textbf{Title}  &  \textbf{Affiliation}  &  \textbf{Author(s)}  &  \textbf{Year}  &  \textbf{Journal}  \\
\midrule
 Relative Times of Arrival of Bursts of Solar Noise     &  CSIR, Div.\ Radiophysics  &  Payne-Scott, R., Yabsley, D.E.  &  1947  &  Nature, 160, 256 \\
                                                        &  &  \& Bolton, J.G.   &   &    \\
 Variable Source of Radio Frequency Radiation in        &  CSIR, Div.\ Radiophysics &  Bolton, J.G.\ \& Stanley, G.J.              &  1948  &  Nature, 161, 312   \\
 \hspace{5mm} the Constellation of Cygnus               &                     &      &    &     \\
 Chromospheric Flares                                   &  CSIR, Div.\ Physics       &  Giovanelli, R.G.  &  1948  &  MNRAS, 108, 163  \\
 The galactic system as a spiral nebula                 &        ...                 &  Oort, J.H., Kerr, F.J. &  1958  & MNRAS, 118, 379   \\
                                                        &  &  \& Westerhout, G.   &   &    \\
 Solar Radio Bursts of Spectral Type II                 &  CSIRO, Div.\ Radiophysics &  Roberts, J.A.                      &  1959  &  AuJPh, 12, 327   \\
 A Magellanic effect on the galaxy                      &  CSIRO, Div.\ Radiophysics &  Kerr, F.J.                         &  1957  &  AJ, 62, 93      \\
 A Low Resolution Hydrogen-line Survey of               &  CSIRO, Div.\ Radiophysics &  Hindman, J.V., Kerr, F.J.  &  1963  &  AuJPh, 16, 570   \\
 \hspace{5mm} the Magellanic System.\ II...             &    &   \& McGee, R.X.   &    &     \\
 A high resolution study of the distribution and        &  CSIRO, Div.\ Radiophysics &  Hindman, J.V.                     &  1967  &  AuJPh, 20, 147   \\
 \hspace{5mm} motions of neutral hydrogen in the [SMC]  &    &      &    &     \\
\bottomrule
\end{tabular}
\end{specialtable}

\finishlandscape

\end{paracol}
\reftitle{References}

\end{document}